\documentclass[useAMS,usenatbib]{mn2e}

\usepackage{graphicx}
\usepackage{hyperref}
\usepackage{enumerate}
\usepackage{xcolor}
\usepackage{longtable}
\usepackage{deluxetable}
\usepackage{txfonts}
\usepackage{times}
\def\apj{ApJ}
\def\aap{A\&A}
\def\apjl{ApJL}
\def\apjs{ApJS}
\def\mnras{MNRAS}
\def\na{New Astronomy}
\def\aj{AJ}

\newcommand{\hi}{{\sc H\,i}}
\newcommand{\mhi}{$M$(\hi)}

\newcommand{\lk}{{$L_\mathrm{K}$}}

\newcommand{\mhil}{$M$(\hi)/\lk}
\newcommand{\mhim}{\mhi$/M_\star$}

\newcommand{\msun}{{M$_\odot$}}
\newcommand{\lsun}{{L$_\odot$}}

\newcommand{\ltsima} {$\; \buildrel < \over \sim \;$}
\newcommand{\gtsima} {$\; \buildrel > \over \sim \;$}
\newcommand{\lta} {\lower.5ex\hbox{\ltsima}}
\newcommand{\gta} {\lower.5ex\hbox{\gtsima}}

\newcommand{\atlas}{ATLAS$^{\rm 3D}$}

\defcitealias{cappellari2011a}{Paper I}
\defcitealias{krajnovic2011}{Paper II}
\defcitealias{emsellem2011}{Paper III}
\defcitealias{young2011}{Paper IV}
\defcitealias{davis2011a}{Paper V}
\defcitealias{bois2011}{Paper VI}
\defcitealias{cappellari2011b}{Paper VII}
\defcitealias{khochfar2011}{Paper VIII}
\defcitealias{duc2011}{Paper IX}
\defcitealias{davis2011b}{Paper X}
\defcitealias{serra2012a}{Paper XIII}
\defcitealias{cappellari2013a}{Paper XV}
\defcitealias{krajnovic2013a}{Paper XVII}
\defcitealias{cappellari2013b}{Paper XX}
\defcitealias{krajnovic2013b}{Paper XXIII}
\defcitealias{naab2013}{Paper XXV}

\title[\hi\ discs in fast and slow rotators]{The \atlas\ project -- XXVI. \hi\ discs in real and simulated fast and slow rotators}
\author
[Paolo Serra et al.]{\parbox{\textwidth}{Paolo Serra,$^{1,2}$\thanks{E-mail:\texttt{paolo.serra@csiro.au}}
Ludwig Oser,$^{3,4}$
Davor Krajnovi\'c,$^{5}$
Thorsten Naab,$^{3}$
Tom Oosterloo,$^{1,6}$
Raffaella Morganti,$^{1,6}$
Michele Cappellari,$^{7}$
Eric Emsellem,$^{8,9}$
Lisa M. Young,$^{10}$
Leo Blitz,$^{11}$
Timothy A. Davis,$^{8}$
Pierre-Alain Duc,$^{12}$
Michaela Hirschmann,$^{13}$
Anne-Marie Weijmans,$^{14,15}$\thanks{Dunlap Fellow}
Katherine Alatalo,$^{11,16}$
Estelle Bayet,$^{7}$
Maxime Bois,$^{8,9}$
Fr\'ed\'eric Bournaud,$^{12}$
Martin Bureau,$^{7}$
Roger L. Davies,$^{7}$
P. T. de Zeeuw,$^{8,17}$
Sadegh Khochfar,$^{18,19}$
Harald Kuntschner,$^{8}$
Pierre-Yves Lablanche,$^{8,9}$
Richard M. McDermid,$^{20}$
Marc Sarzi,$^{21}$
and Nicholas Scott$^{22,23}$}\vspace{0.4cm}\\ 
\parbox{\textwidth}{$^{1}$Netherlands Institute for Radio Astronomy (ASTRON), Postbus 2, 7990 AA Dwingeloo, The Netherlands\\
$^{2}$CSIRO Astronomy and Space Science, Australia Telescope National Facility, PO Box 76, Epping, NSW 1710, Australia\\
$^{3}$Max-Planck-Institut f\"ur Astrophysik, Karl-Schwarzschild-Str. 1, 85741 Garching, Germany\\
$^{4}$Department of Astronomy, Columbia University, New York, NY 10027, USA\\
$^{5}$Leibniz-Institut f\"ur Astrophysik Potsdam (AIP), An der Sternwarte 16, D-14482 Potsdam, Germany\\
$^{6}$Kapteyn Astronomical Institute, University of Groningen, Postbus 800, 9700 AV Groningen, The Netherlands\\
$^{7}$Sub-Dept. of Astrophysics, Dept. of Physics, University of Oxford, Denys Wilkinson Building, Keble Road, Oxford, OX1 3RH, UK\\
$^{8}$European Southern Observatory, Karl-Schwarzschild-Str. 2, 85748 Garching, Germany\\
$^{9}$Universit\'e Lyon 1, Observatoire de Lyon, Centre de Recherche Astrophysique de Lyon and Ecole Normale Sup\'erieure de Lyon, 9 avenue Charles Andr\'e, F-69230 Saint-Genis Laval, France\\
$^{10}$Physics Department, New Mexico Institute of Mining and Technology, Socorro, NM 87801, USA\\
$^{11}$Department of Astronomy, Campbell Hall, University of California, Berkeley, CA 94720, USA\\
$^{12}$Laboratoire AIM Paris-Saclay, CEA/IRFU/SAp -- CNRS -- Universit\'e Paris Diderot, 91191 Gif-sur-Yvette Cedex, France\\
$^{13}$INAF - Astronomical Observatory of Trieste, via G.B. Tiepolo 11, I-34143 Trieste, Italy\\
$^{14}$Dunlap Institute for Astronomy \& Astrophysics, University of Toronto, 50 St. George Street, Toronto, ON M5S 3H4, Canada\\
$^{15}$School of Physics and Astronomy, University of St Andrews, North Haugh, St Andrews KY16 9SS, UK\\
$^{16}$Infrared Processing and Analysis Center, California Institute of Technology, Pasadena, California 91125, USA\\
$^{17}$Sterrewacht Leiden, Leiden University, Postbus 9513, 2300 RA Leiden, the Netherlands\\
$^{18}$Max-Planck Institut f\"ur extraterrestrische Physik, PO Box 1312, D-85478 Garching, Germany\\
$^{19}$Institute for Astronomy, University of Edinburgh, Royal Observatory, Blackford Hill, Edinburgh EH9 3HJ, UK\\
$^{20}$Gemini Observatory, Northern Operations Centre, 670 N. A`ohoku Place, Hilo, HI 96720, USA\\
$^{21}$Centre for Astrophysics Research, University of Hertfordshire, Hatfield, Herts AL1 9AB, UK\\
$^{22}$Sydney Institute for Astronomy (SIfA), School of Physics, The University of Sydney, NSW 2006, Australia\\
$^{23}$ARC Centre of Excellence for All-sky Astrophysics (CAASTRO), 44 Rosehill Street, Redfern, NSW 2016, Australia
}}

\begin{document}

\date{Accepted .... Received ...; in original form ...}

\pagerange{\pageref{firstpage}--\pageref{lastpage}} \pubyear{2013}

\maketitle

\label{firstpage}

\clearpage

\begin{abstract}
\rm One quarter of all early-type galaxies (ETGs) outside Virgo host a disc/ring of \hi\ with size from a few to tens of kpc and mass up to $\sim10^9$ \msun. Here we investigate whether this \hi\ is related to the presence of a stellar disc within the host making use of the classification of ETGs in fast and slow rotators (FR/SR). We find a large diversity of \hi\ masses and morphologies within both families. Surprisingly, SRs are detected as often, host as much \hi\ and have a similar rate of \hi\ discs/rings as FRs. Accretion of \hi\ is therefore not always linked to the growth of an inner stellar disc. The weak relation between \hi\ and stellar disc is confirmed by their frequent kinematical misalignment in FRs, including cases of polar and counterrotating gas. In SRs the \hi\ is usually polar. This complex picture highlights a diversity of ETG formation histories which may be lost in the relative simplicity of their inner structure and emerges when studying their outer regions.

We find that $\Lambda$CDM hydrodynamical simulations have difficulties reproducing the \hi\ properties of ETGs. The gas discs formed in simulations are either too massive or too small depending on the star formation feedback implementation. Kinematical misalignments match the observations only qualitatively. The main point of conflict is that nearly all simulated FRs and a large fraction of all simulated SRs host corotating \hi. This establishes the \hi\ properties of ETGs as a novel challenge to simulations.\rm 

\end{abstract}


\section{Introduction}
\label{sec:intro}

The observation of both neutral hydrogen (\hi) and molecular gas (H$_2$, traced by CO) has shown that early-type galaxies (E/S0s, hereafter ETGs), while on average gas-poorer than spirals, can in some cases host a significant mass of cold gas (for early \hi\ studies see, e.g., \citealt{gallagher1975,knapp1977,bieging1977,knapp1985,wardle1986}; for CO see \citealt{wiklind1989,lees1991}). What has remained unclear until recently is whether gas-rich ETGs are  peculiar, rare objects or represent a significant fraction of the overall population \citep[e.g.,][]{vangorkom1997}. This question has now been answered by large \hi\ and H$_2$ surveys of morphologically-selected ETGs carried out as part of the \atlas\ project \citep[][hereafter Paper I]{cappellari2011a}.

In \citet[hereafter Paper XIII]{serra2012a} we show that $\sim40$ per cent of all ETGs outside the Virgo cluster and $\sim10$ per cent inside Virgo host \hi\ down to a mass of $\sim10^7$ \msun\ and a column density of $\sim3\times10^{19}$ cm$^{-2}$, placing recent, previous results \citep{morganti2006,diseregoalighieri2007,grossi2009,oosterloo2010} on a strong statistical basis. We find that in $\sim2/3$ of all detections (and $1/4$ of all ETGs outside Virgo) the \hi\ is distributed in a low-column-density disc or ring with size from a few to tens of kpc, and mass from $\sim10^7$ to more than $10^9$ \msun\ \citep[see also][]{vandriel1991,oosterloo2007}. The remaining detected ETGs host \hi\ on an unsettled configuration which usually reveals recent dynamical interaction within a group.

Concerning molecular gas, \citet[hereafter Paper IV]{young2011} detect CO in 22 per cent of all ETGs independent of galaxy luminosity or environment, with $M$(H$_2$) between $\sim10^7$ \msun\ (the typical sensitivity of their data) and $\sim10^9$ \msun\ \citep[for recent, previous surveys of smaller samples see][]{combes2007,sage2007}. Follow-up interferometric observations show that this gas is mostly distributed on discs and rings with typical radius $\sim1$ kpc, just a fraction of the galaxy effective radius \citep{alatalo2013,davis2013}.

Taken together, these studies establish that about half of all ETGs host between $\sim10^7$ and $\sim10^9$ \msun\ of cold gas (atomic and/or molecular) usually settled within the galaxy potential. \rm Very sensitive observations are required to reach such a detection rate (for example, this is significantly higher than in the recent surveys of massive galaxies by \citealt{catinella2010} and \citealt{saintonge2011}) but when these are performed we find that ETGs with a significant mass of cold gas are not peculiar, rare systems\rm. The obvious implication is that in order to understand the assembly of ETGs as a family we need to understand how they get \rm or retain \rm their gas and what impact this has on the host, on its star formation history and on its structure. Here we focus on the latter aspect.

In most previous studies of this kind ETGs are simply divided into ellipticals and lenticulars. When this is done, molecular gas is almost exclusively found in lenticulars (\rm \citealt{lees1991}, \citealt{welch2010}, \rm \citetalias{young2011}), consistent with the fact that CO is invariably associated with dust lanes. Furthermore, \hi\ seems to be more abundant in lenticular than in elliptical objects at a typical \mhi\ sensitivity of $\sim10^9$ \msun\ \citep{roberts1975,wardle1986,bregman1992,sadler2001} although more sensitive observations of smaller samples do not reveal any significant difference \citep{grossi2009,oosterloo2010}. One complication is that the distinction between ellipticals and lenticulars can be ambiguous and is affected by projection effects \citep{jorgensen1994}. A more robust and physically meaningful approach is to classify ETGs on the basis of their internal kinematics \citep[e.g.,][]{davies1983,cappellari2007,emsellem2007}.






In this context a key result of the \atlas\ project is that the vast majority of all ETGs ($\sim85$ per cent) are structurally similar to spirals. They rotate like discs \citep[][hereafter Paper II]{krajnovic2011}, can be described as a family of oblate rotators on the $\lambda_\mathrm{R}$-$\epsilon$ diagram \citep[][hereafter Paper III]{emsellem2011}, and cover the same range of intrinsic flattening and bulge-to-disc ratio of Sa to Sc spirals (\citealt{cappellari2011b,cappellari2013b}, hereafter Paper VII and Paper XX, respectively; \citealt{krajnovic2013a}; Weijmans et al., submitted; see also \citealt{laurikainen2011}; \citealt{kormendy2012}). We call these ETGs fast rotators (FRs). In practice, FRs include \rm nearly \rm all classical lenticulars plus a large number of galaxies morphologically misclassified as ellipticals because of unfavourable viewing angle \citepalias{emsellem2011}. The evolutionary history of most FRs must include the growth and survival of a stellar disc following accretion of cold gas from the surrounding medium as well as gas-rich minor and major mergers (\citealt{khochfar2011}, hereafter Paper VIII; \citealt{naab2013}, hereafter Paper XXV).

The remaining $\sim15$ per cent of all ETGs do not exhibit the disc-like stellar rotation observed in FRs and occupy a region of the $\lambda_\mathrm{R}$-$\epsilon$ diagram corresponding to intrinsically lower specific angular momentum and flattening \citepalias{emsellem2011}. We call these galaxies slow rotators (SRs). We find that SRs form a heterogeneous family including systems with no rotation at all, non-disc-like rotation and kinematically distinct cores (KDCs). Massive SRs are thought to grow via (a small number of) major mergers at $z>1$, which lower the system angular momentum and may create a KDC, followed by many minor mergers with low gas fraction, which preserve the KDC and make the stellar body rounder (\citealt{naab2009}; \citealt{bois2011}, hereafter Paper VI; \citetalias{khochfar2011}; \citetalias{naab2013}). This sequence of events is most likely to occur in a dense environment, in agreement with the increased fraction of SRs in the centre of the Virgo cluster \citepalias{cappellari2011b}.

Within this picture one may naively think that SRs should lack a significant cold-gas phase, unlike FRs. Indeed, \citetalias{young2011} shows that the CO detection rate is much lower in SRs than in FRs. However, the situation is not necessarily so simple. For example, the detection of kinematically-misaligned ionized gas \citep[][hereafter Paper X]{davis2011b}, cold dust \citep{smith2012} and stellar nuclear cusps \citep[][hereafter Paper XXIII]{krajnovic2013b} in SRs suggests that they too can accrete some gas during their life; and simulations show that at least some SRs may have formed in gas-rich mergers (\citetalias{bois2011}; \citetalias{naab2013}). Furthermore, the mere fact that FRs are found at basically all environment densities \citepalias{cappellari2011b} implies that there must be a large variety of evolutionary histories leading to their formation, and therefore potentially a large variety of resulting cold-gas properties.

\rm All that is required to mark the visual difference between a FR and a spiral at $z=0$ is that the former should have lower star-formation-rate surface density on the stellar disc. In other words, they should be quenched owing to a lower $M_\mathrm{gas}/M_\star$ and/or star-formation efficiency \it within the stellar body\rm. Observations indicate that such quenching may be related to the growth of the stellar bulge: the cold gas content of the stellar body decreases as the bulge becomes more dominant (\citealt{catinella2010}; \citealt{saintonge2011}; \citetalias{cappellari2013b}) and, additionally, the bulge may stabilize the remaining gas against star formation \citep{kawata2007,martig2009,saintonge2012,martig2013}. Yet it is still possible for a FR to host a large mass of cold gas well outside the stellar body and at low surface density (e.g., \citealt{morganti2006}; \citealt{oosterloo2007}; \citetalias{serra2012a}). Such gas distributions are indicative that there are many ways of making a FR and that this information, lost in the apparent homogeneity of their shape and stellar kinematics, may emerge when observing their cold-gas phase. \rm Furthermore, these tenuous gas systems are only weakly bound to the host and hence are easily perturbed. Their actual morphology is therefore a sensitive probe of the past dynamical history of the host, and indeed in \citetalias{serra2012a} we show that the \hi\ morphology varies systematically with the density of the environment. \rm
 
In this spirit we study here the \hi\ properties of ETGs as a function of their structure within the context of the \atlas\ project. The few previous attempts in this direction reveal no clear relation between \hi\ and ETG structure but are based on significantly smaller samples \citep{morganti2006,oosterloo2010}. Here we aim at clarifying the situation by studying the large \atlas\ \hi\ sample. We describe the sample and the \hi\ data in Sec. \ref{sec:data}, discuss \hi\ detection rate, mass and morphology as a function of galaxy structure in Sec. \ref{sec:lambdaeps}, analyse the kinematical misalignment between \hi\ and stars in Sec. \ref{sec:misal}, perform for the first time a comparison to the predictions of $\Lambda$CDM hydrodynamical simulations in Sec. \ref{sec:sims}, and summarize the results in Sec. \ref{sec:summary}.

\section{Sample and \hi\ survey}
\label{sec:data}

The \atlas\ \hi\ sample and the radio observations which this study is based on are described in detail in \citetalias{serra2012a}. Here we provide a short summary. \rm We study a subset of the full \atlas\ sample of ETGs (\citetalias{cappellari2011a}; $\mathrm{distance}<42$ Mpc, $M_\mathrm{K}<-21.5$). This subset includes all galaxies with declination $\delta\geq10\,^{\circ}$ except the four objects closest to Virgo A, resulting in a volume-limited sample of 166 ETGs (127 outside the Virgo cluster)\rm. We obtain Westerbork Synthesis Radio Telescope \hi\ data for most of them either as part of the \atlas\ project or from previous projects \citep{morganti2006,jozsa2009,oosterloo2010}. We make use of Very Large Array data for 1 galaxy with \hi\ \citep{chung2009} and of Arecibo spectra for 20 galaxies \citep[all undetected;][]{giovanelli2005}.

As mentioned in Sec. \ref{sec:intro}, our Westerbork observations allow us to detect \hi\ down to a column density of a few times $10^{19}$ cm$^{-2}$ and a mass of $\sim10^7$ \msun. We detect \hi\ associated with the target ETG in 53 galaxies (49 outside Virgo). In \citetalias{serra2012a} we classify detected ETGs in the following classes on the basis of the \hi\ morphology: $D$ = large discs (14 objects); $R$ = large rings\footnote{In fact, in \citetalias{serra2012a} we group $D$s and $R$s in a single class. In table 1 of that paper we indicate galaxies where the \hi\ is distributed on a ring. Here we use that table as a basis for the division into $D$ and $R$ objects.} (10); $d$ = small discs (10); $u$ = unsettled distributions (14); $c$ = gas clouds (5). Gas distributions characterized by ordered rotation ($D$, $R$ and $d$ classes) are therefore the most common ones among \hi-rich ETGs, amounting to $2/3$ of all detections and $1/4$ of all ETGs outside Virgo. As discussed in \citetalias{serra2012a}, the above \hi\ classes are a way to simplify the observed continuum of gas morphologies, with a number of objects being intermediate (or in transition) between classes. We make use of the \hi\ classes $D$, $R$, $d$, $u$ and $c$ in the rest of this paper.

In \citetalias{serra2012a} we show \hi\ images of all detections. Each image is obtained after creating a mask including all emission in the corresponding \hi\ cube. The images shown in \citetalias{serra2012a} are the zero-th moment of the masked cubes. In the present paper we show and analyse for the first time the \hi\ velocity fields obtained as the first moment of the same masked cubes.

\section{Relation between \hi\ and ETG inner structure}
\label{sec:lambdaeps}

In this section we investigate whether the \hi\ detection rate, mass and morphology of ETGs are related to their stellar kinematics as well as position on the $\lambda_\mathrm{R}$-$\epsilon$ diagram\footnote{We adopt the $\lambda_\mathrm{R}$ and $\epsilon$ measurements obtained within a $1\ R_\mathrm{e}$ aperture in \citetalias{emsellem2011}. Note that not all galaxies have their stellar kinematics measured out to $1\ R_\mathrm{e}$, as discussed in that paper.} and mass-size plane. In particular, we make use of the kinematical classification presented in \citetalias{krajnovic2011}. In that paper ETGs are divided in the following classes based on a quantitative analysis of their stellar velocity field\footnote{Note the addition of a $\star$ subscript to the kinematical classes name to avoid confusion with the \hi\ classes defined in Sec. \ref{sec:data}.}: $a_\star$ = no rotation; $b_\star$ = some rotation but with a velocity field not consistent with that of a rotating disc; $c_\star$ = KDC; $d_\star$ = two counterrotating stellar discs; $e_\star$ = velocity field consistent with that of a rotating disc. These classes are important to understand the classification of ETGs into FRs and SRs discussed in Sec. \ref{sec:intro}. In \citetalias{emsellem2011} we define FRs and SRs as galaxies above and below the line $\lambda_\mathrm{R}=0.31\sqrt{\epsilon}$ on the $\lambda_\mathrm{R}$-$\epsilon$ diagram, respectively. Fast rotators and $e_\star$ ETGs are essentially the same family and represent $\sim85$ per cent of all objects. On the other hand, the SR family is very heterogeneous and includes galaxies in all other groups ($a_\star$ to $d_\star$).

\subsection{Fast versus slow rotators}

\begin{table}
\centering
\caption{Number of galaxies in the various \hi\ morphological classes \citepalias{serra2012a} as a function of stellar kinematics class \citepalias{emsellem2011} and SR/FR classification \citepalias{krajnovic2011} for the full \atlas\ \hi\ sample.}
\begin{tabular}{rrrrrrrrrr}
\hline
\multicolumn{2}{c}{\rm Full\rm}& \multicolumn{8}{c}{Stellar kinematics} \\
\multicolumn{2}{c}{\rm Sample\rm}                   &All     & $a_\star$  & $b_\star$  & $c_\star$    & $d_\star$ & $e_\star$ & SR & FR\\
 \hline
& All               & 166 & 4  & 8  & 14 & 4 & 136 & 22 & 144\\ 
& Undet.         & 113 & 3  & 5 &  7   & 4 & 94 & 13 & 100\\
& Det.              & 53   & 1  & 3  &  7  & 0 & 42 & 9 & 44\\
\\
\hi\ & $D$              & 14  & 0  &1   & 1  & 0 & 12 & 2 & 12\\
& $R$                   & 10 & 0  & 0  & 2   &0 & 8   & 2 & 8\\
& $d$               & 10 & 0  & 0  & 0  & 0 & 10 & 0 & 10\\
& $u$               & 14  & 1  & 2  & 2  & 0 & 9 & 3 & 11\\
& $c$               & 5    & 0  & 0 & 2  & 0 & 3 & 2 & 3\\
\hline
\end{tabular}
\label{tab:summary1}
\end{table}

\begin{figure*}
\includegraphics[width=18cm]{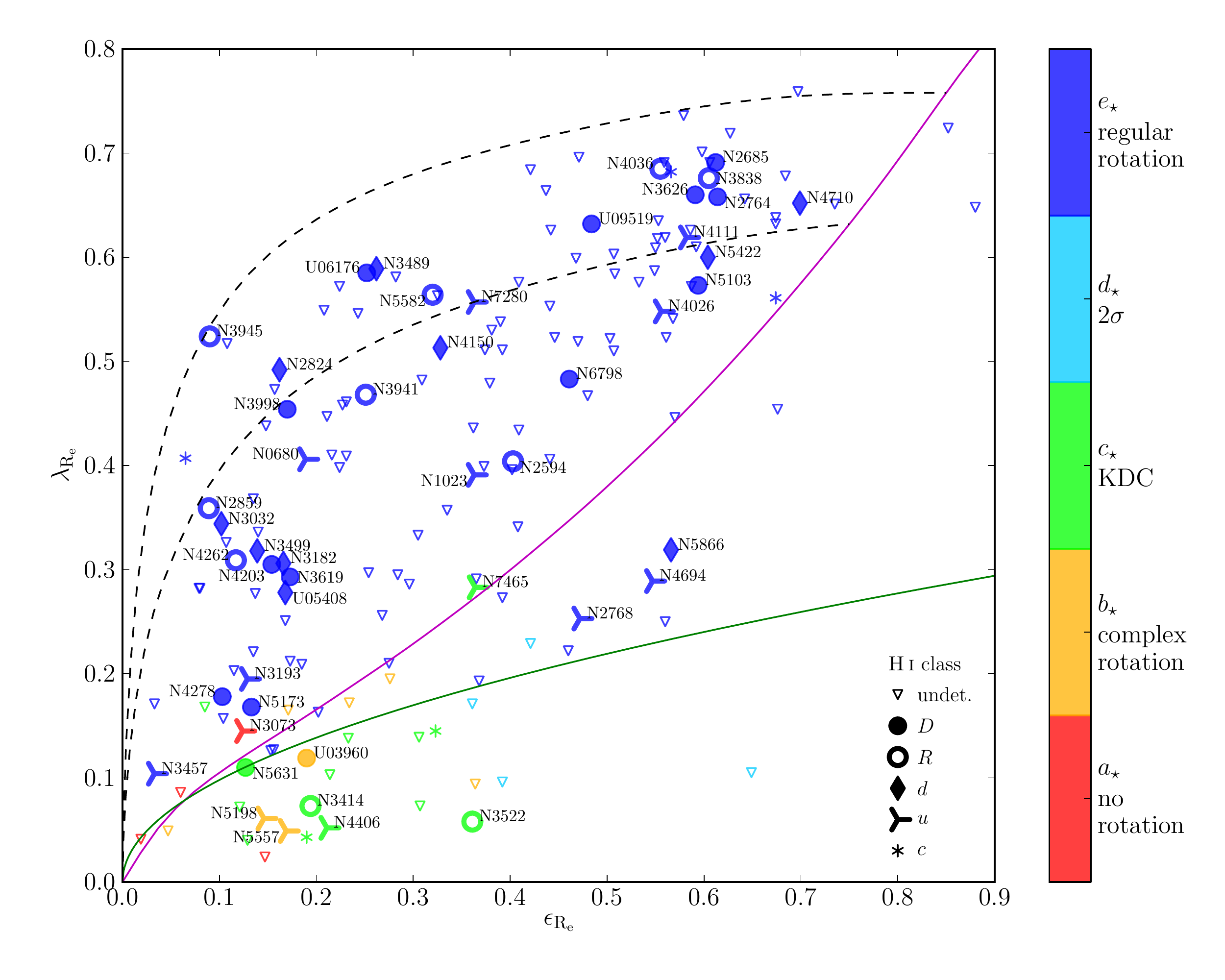}
\caption{\hi\ on the $\lambda_\mathrm{R}$-$\epsilon$ diagram. The colour scheme represents the kinematical classification of ETGs presented in \citetalias{krajnovic2011}. The markers shape represent the \hi\ morphology as in \citetalias{serra2012a} with the small modification described in Sec. \ref{sec:data}. The magenta line represents the edge-on view of anisotropic, oblate models with $\beta=0.65\times\epsilon$ \citep{cappellari2007}, and the black dashed lines represent the models with $\epsilon=0.75$ and 0.85 viewed under all possible inclinations from edge-on (top right) to face-on (bottom left). The green line represents the empirical separation between FRs and SRs $\lambda_\mathrm{R}=0.31\sqrt{\epsilon}$ proposed in \citetalias{emsellem2011}. We indicate the name of all \hi\ detections except that of $c$ galaxies.}
\label{fig:le_kinclass}
\end{figure*}

Figure \ref{fig:le_kinclass} shows all galaxies on the $\lambda_\mathrm{R}$-$\epsilon$ diagram. In this figure we indicate galaxies with different stellar kinematics and \hi\ morphology using different colours and markers, respectively. To first order Fig. \ref{fig:le_kinclass} confirms the results of \citet{morganti2006} and \citet{oosterloo2010}: the relation between \hi\ properties and galaxy structure is not a simple one.

We detect \hi\ in galaxies at all locations on the $\lambda_\mathrm{R}$-$\epsilon$ diagram where there are ETGs, both FRs and SRs. Many of the \hi\ discs and rings ($D$, $R$ and $d$) are hosted by flat, nearly edge-on FRs (top-right of the diagram) and their lower-inclination counterparts (at lower values of $\lambda_\mathrm{R}$ and $\epsilon$). This is reasonable as these galaxies have a pronounced stellar disc component and therefore must have experienced a considerable level of dissipation during their assembly. Their current \hi\ discs/rings may be a remnant of that process. The more detailed analysis of Sec. \ref{sec:misal} shows that this is indeed the case for about half of all gas-rich FRs, but also that in the remaining half the gas accretion episode traced by \hi\ does not seem to be related to the formation of the current stellar disc.

What might be more surprising is to find \hi\ also among SRs ($a_\star$, $b_\star$ and $c_\star$ objects only) and that in some of them the detected \hi\ is distributed on a large disc or ring. This indicates that SRs too can experience gas accretion and  dissipative processes during their formation, and gas-poor evolutionary paths are not the only way to make a galaxy with this internal structure. \rm A similar conclusion is reached by \citetalias{davis2011b} based on the misalignment between ionized gas and stellar kinematics in these galaxies within $\sim 1\ R_\mathrm{e}$, by \cite{smith2012} based on the frequent detection of cold dust in SRs, by \citetalias{krajnovic2013b} based on the existence of SRs with a cuspy nuclear light profile, and by both idealized and $\Lambda$CDM hydrodynamical simulations (\citetalias{bois2011}; \citetalias{naab2013}). Additional observations such as very deep optical imaging may provide further clues about the formation of SRs in presence of a cold gas component \citep{duc2011}.\rm

These qualitative results are illustrated in more quantitative terms in Table \ref{tab:summary1} for the full sample and in Table \ref{tab:summary2} for non-Virgo galaxies only. Within the limited statistics available for SRs (and for classes $a_\star$ to $d_\star$), their \hi\ detection rate is the same as that of FRs (and $e_\star$ objects). Namely, we detect \hi\ in $41\pm14$ per cent of all SRs and $31\pm5$ per cent of all FRs, while outside Virgo the detection rate becomes $50\pm18$ per cent for SRs and $37\pm6$ per cent for FRs (error bars assume a binomial distribution). The fraction of galaxies hosting a rotating \hi\ distribution (\hi\ classes $D$, $R$ and $d$) is $18\pm9$ per cent for SRs and $21\pm4$ per cent for FRs over the full sample, while outside Virgo it becomes $25\pm13$ per cent for SRs and $25\pm5$ per cent for FRs.

\begin{table}
\centering
\caption{Number of galaxies in the various \hi\ morphological classes \citepalias{serra2012a} as a function of stellar kinematics class \citepalias{emsellem2011} and SR/FR classification \citepalias{krajnovic2011} for non-Virgo ETGs in the \atlas\ \hi\ sample.}
\begin{tabular}{rrrrrrrrrr}
\hline
\multicolumn{2}{c}{\rm Outside\rm}& \multicolumn{8}{c}{Stellar kinematics} \\
\multicolumn{2}{c}{\rm Virgo\rm}                   &All     & $a_\star$  & $b_\star$  & $c_\star$    & $d_\star$ & $e_\star$ & SR & FR\\
 \hline
& All               & 127 & 2   & 7   & 11 & 1   & 106          & 16 & 111\\ 
& Undet.         & 78  & 1    & 4  &  5   & 1  & 67             & 8 & 70\\
& Det.              & 49   & 1   & 3  &  6  & 0   & 39            & 8 & 41\\
\\
\hi\ & $D$        & 14  & 0  &1   & 1  & 0      & 12          & 2 & 12\\
& $R$              &  9   & 0  &0   & 2  &0        & 7            & 2 & 7\\
& $d$               & 9    & 0   & 0  & 0  & 0      & 9            & 0 & 9\\
& $u$               & 12  & 1  & 2   & 1  & 0     & 8            & 2 & 10\\
& $c$               & 5    & 0   & 0  & 2  & 0      & 3            & 2 & 3\\
\hline
\end{tabular}
\label{tab:summary2}
\end{table}

The above paragraph describes the \hi\ properties (detection rate and morphology) as a function of stellar kinematics. A complementary approach is to examine the stellar kinematics as a function of \hi\ properties, but also in this case Tables \ref{tab:summary1} and \ref{tab:summary2} show that there is no clear trend. Within each \hi\ class (including that formed by all undetected ETGs) the fraction of SRs and FRs is consistent with the global values of $\sim15$ and $\sim85$ per cent, respectively. The same is true considering all galaxies with a settled \hi\ distribution together ($D$, $R$ and $d$). We also compare the $\lambda_\mathrm{R}/(0.31\sqrt{\epsilon})$ distribution of $D+R+d$ ETGs to that of $\mathrm{undetected}+c$ objects\footnote{In this context $c$ ETGs can be grouped together with non-detections since \mhi\ is very low and the detected gas does not appear obviously associated with the host \citepalias{serra2012a}.} and find that they are indistinguishable -- according to a two-sample KS test the probability that the two samples are not drawn from the same parent distribution is 78 per cent, insufficient to claim a significant difference.

The only possible exception to this lack of trends is that small \hi\ discs ($d$) include no SRs. This is consistent with the fact that \hi\ in these galaxies seems to be strongly linked to the presence \rm of molecular gas and star formation within a stellar disc \rm \citepalias{serra2012a}. However, the number of $d$s is low and only $\sim1$ SR would be expected among them based on the global SR fraction. Better statistics would therefore be needed to confirm this result.

\begin{figure*}
\includegraphics[width=18cm]{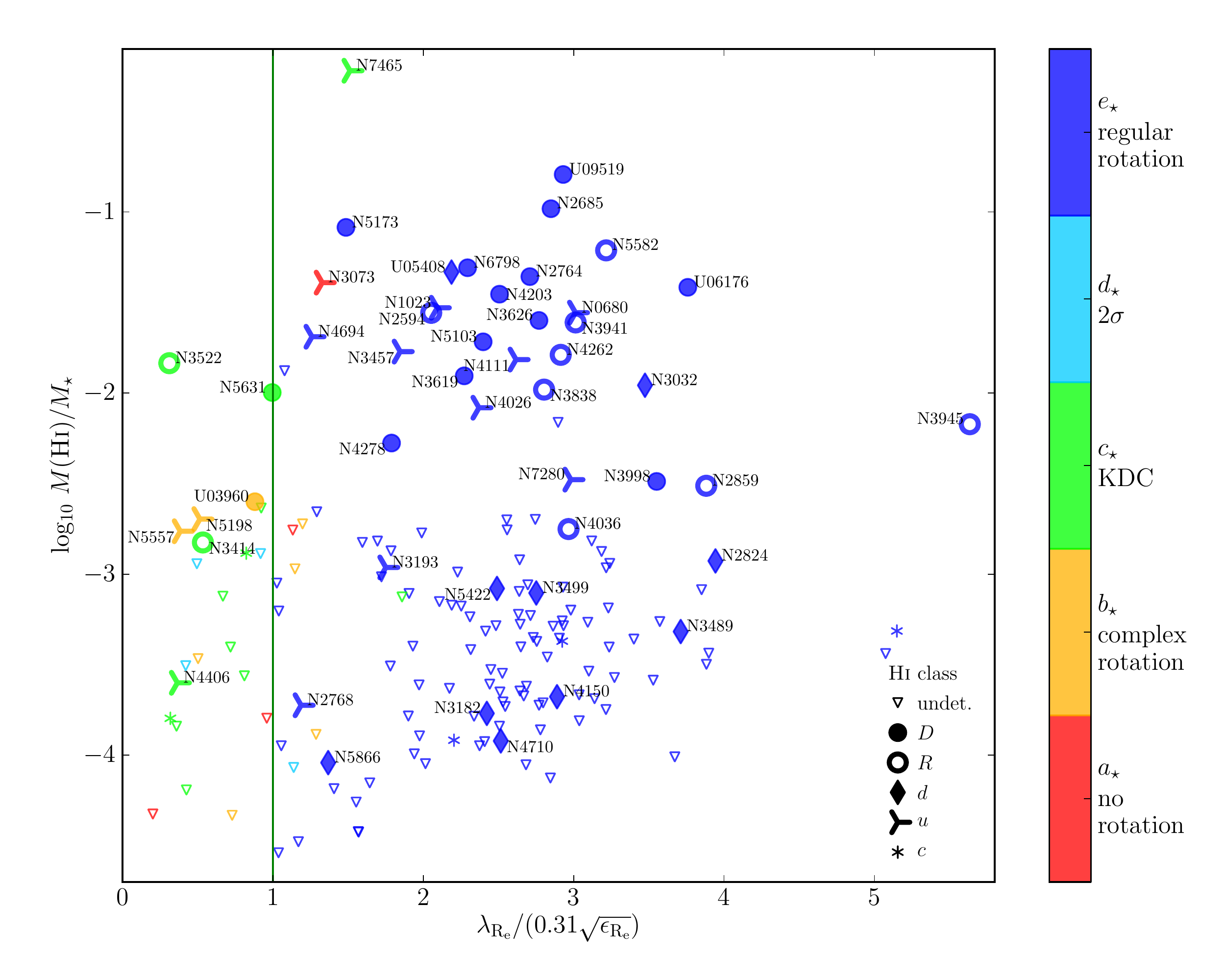}
\caption{\mhim\ plotted against the ratio $\lambda_\mathrm{R}/(0.31\sqrt{\epsilon})$, such that SRs and FRs are found to the left and right of the green, vertical line, respectively. Colours and markers are as in Fig. \ref{fig:le_kinclass}.}
\label{fig:mhille}
\end{figure*}

Another important quantity to analyse is the \hi\ mass. Figure \ref{fig:mhille} shows the \mhim\ ratio plotted against the $\lambda_\mathrm{R}/(0.31\sqrt{\epsilon})$ ratio. \rm Here and in the rest of this paper we adopt for $M_\star$ the value derived by \citet[][hereafter Paper XV]{cappellari2013a} based on dynamical modelling -- $M_\mathrm{JAM}$ in that paper\footnote{The results discussed in this section do not change if we normalize \mhi\ to $L_\mathrm{K}$. This follows from the fact that the distribution of $M_\star/L_\mathrm{K}$ ratio is very narrow for ETGs, peaking at 1.2 \msun/\lsun\ and with an rms of 0.4 \msun/\lsun. This scatter is insufficient to significantly alter a plot like the one showed in Fig. \ref{fig:mhille}, where galaxies spread over $\sim3$ orders of magnitude in \mhim.}. One galaxy, PGC~071531, has no reliable $M_\star$ measurement and is excluded from this analysis. \rm Figure \ref{fig:mhille} shows that \hi-detected SRs are found over a wide range of \mhim. They seem not to reach the large \mhim\ values allowed for FRs but the overall low number of SRs makes this difference statistically insignificant. For example, the fraction of galaxies with \mhim\ $\geq 0.002$ is $23\pm10$ per cent within the SR family and $24\pm4$ per cent for FRs. These fractions remain comparable to one another if we consider ETGs outside Virgo only, \rm and these (and the following) results do not change if we adopt a different \mhim\ threshold\rm . Conversely, we define two ETG subsamples with \mhim\ above and below 0.002, including 39 and 126 galaxies respectively. We find that they contain a similar fraction of SRs consistent with the global value of $\sim15$ per cent, both considering all galaxies or non-Virgo galaxies only. Finally, according to a two-sample KS test the probability that the $\lambda_\mathrm{R}/(0.31\sqrt{\epsilon})$ distributions of the two subsamples are not drawn from the same parent distribution is just 4 per cent.


\begin{figure}
\includegraphics[width=8.5cm]{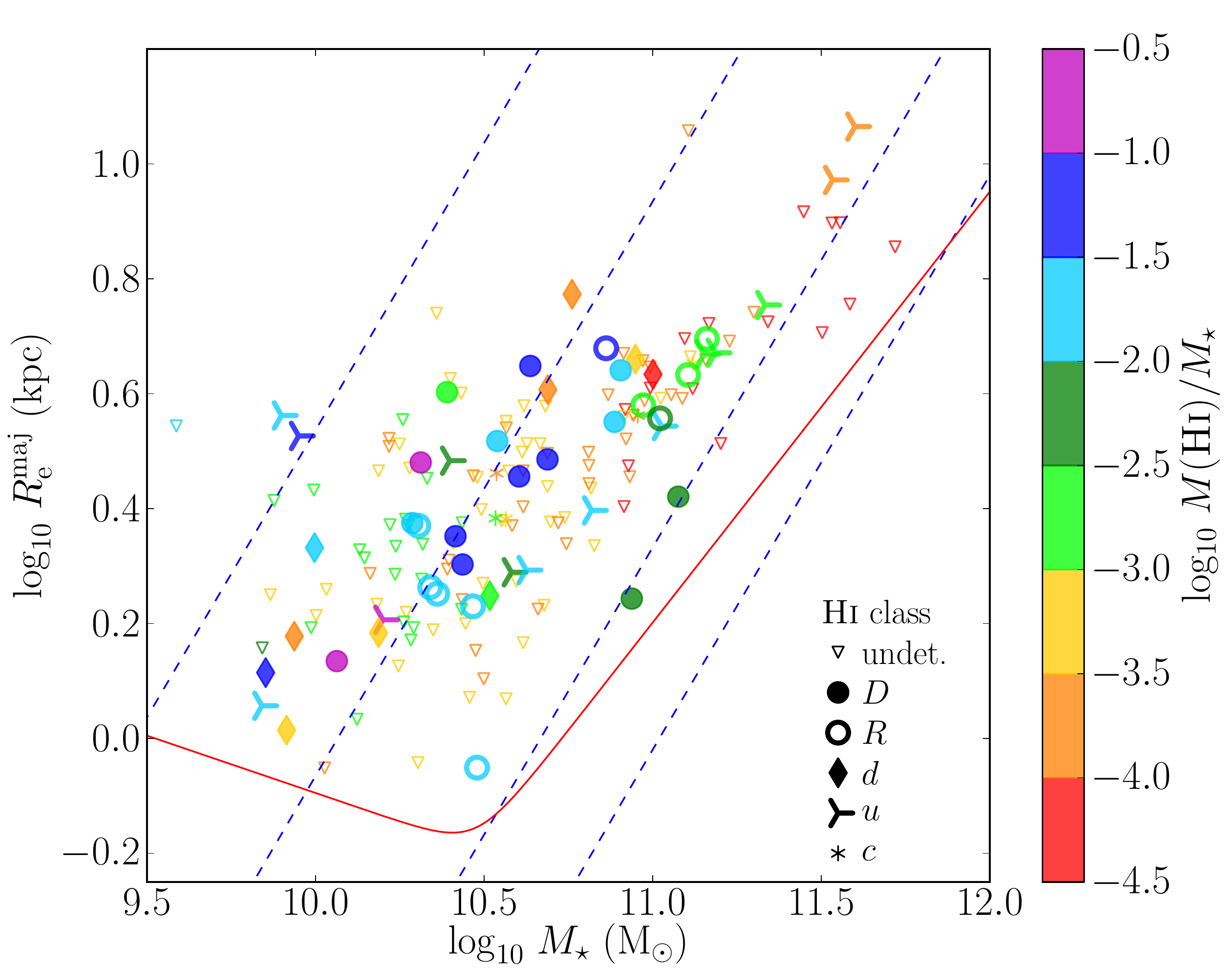}
\caption{\rm \hi\ properties on the mass-size plane. Different \hi\ morphologies are indicated as in previous figures (see legend on the bottom right) while markers are colour-coded according to the \mhi/$M_\star$ value. The velocity dispersion is constant along the blue dashed lines for galaxies following the virial relation ($\sigma_\mathrm{e}=50$, 100, 200 and 300 km s$^{-1}$ from top-left to bottom-right, respectively; see \citealt{cappellari2006}). The red solid line indicates the Zone of Exclusion discussed in \citetalias{cappellari2013b}.\rm }
\label{fig:msize}
\end{figure}

\subsection{\hi\ properties within the FR family}


Considering now FRs alone, we find no systematic variation of \hi\ properties as a function of position on the $\lambda_\mathrm{R}$-$\epsilon$ diagram. Specifically, we compare intrinsically flatter FRs to rounder ones. We do this by considering $e_\star$ galaxies only and splitting them in two groups above and below the line of 0.75 intrinsic ellipticity, respectively (see Fig. \ref{fig:le_kinclass}\rm; we comment below on alternative ways of defining flat and round galaxies\rm). This results in 47 `flat' and 89 `round' $e_\star$ ETGs (39 and 67 outside Virgo, respectively). We detect \hi\ in $32\pm8$ per cent of all flat objects and $30\pm6$ per cent of all round ones. The detection rate of rotating \hi\ distributions ($D$, $R$ and $d$ objects) is $26\pm7$ and $20\pm5$ per cent, respectively, with both flat and round ETGs exhibiting a mix of $D$, $R$ and $d$ objects. No significant difference emerges if we consider non-Virgo galaxies only.

Conversely, the fraction of flat and round $e_\star$ ETGs does not vary significantly as a function of \hi\ class (also when grouping together all $D$, $R$ and $d$ systems). Within the error bars, we always find $\sim35$ per cent of flat objects and $\sim65$ per cent of round galaxies. Furthermore, the $\lambda_\mathrm{R}/\lambda_\mathrm{R,\epsilon=0.75}$ distribution of galaxies with an \hi\ disc or ring ($D$, $R$ and $d$) is statistically undistinguishable from that of galaxies with no \hi\ (undetected and $c$). The two-sample KS-test probability that the two distributions are not drawn from the same parent distribution is 38 per cent.

We also investigate possible variations of \mhim\ within the FR family. We find that $19\pm6$ per cent of all flat galaxies have \mhim\ $\geq 0.002$, comparable to the $26\pm5$ per cent found for round objects. Conversely, we define an \hi-rich and an \hi-poor subsample of $e_\star$ ETGs adopting the same \mhim\ separation as above. The two samples include 32 and 103 galaxies, respectively. The fraction of flat objects in the two groups is comparable ($28\pm9$ and $37\pm6$ per cent, respectively). Finally, a two-sample KS test gives a probability of just 10 per cent that the $\lambda_\mathrm{R}/\lambda_\mathrm{R,\epsilon=0.75}$ distributions of the two subsamples are not drawn from the same parent distribution.

\rm The definition of flat and round FRs adopted above relies on a simple de-projection of galaxies' shape along model lines on the $\lambda_\mathrm{R}$-$\epsilon$ diagram. A more accurate estimate of the intrinsic ellipticity of ETGs is derived in \citetalias{cappellari2013b} based on the modelling presented in \citetalias{cappellari2013a} and on the ellipticity measurement at large radius reported in \citetalias{krajnovic2011}. However, we have verified that using these intrinsic ellipticity values does not change any of the above conclusions. Flat and round FRs keep having equal \hi\ detection rates as well as equal fractions of \hi\ discs and of galaxies with \mhim\ $\geq0.002$. Similarly, the distribution of FR intrinsic ellipticity does not change as a function of \hi\ mass or morphology.\rm

\begin{figure*}
\includegraphics[width=2.4cm]{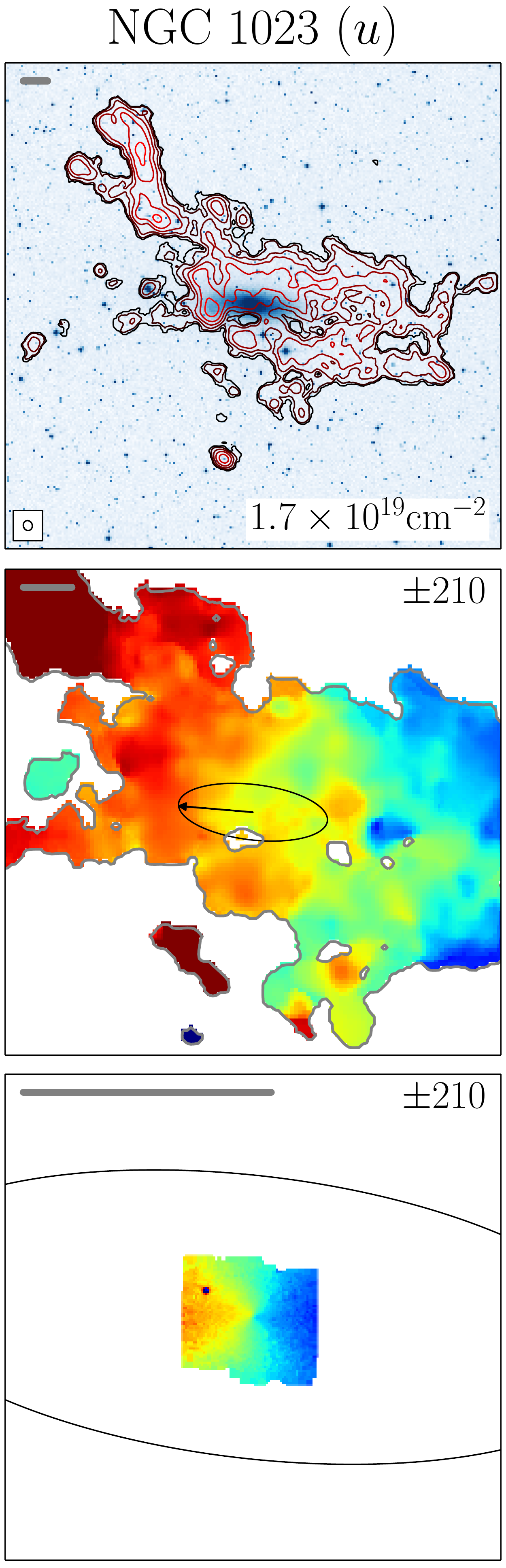}
\includegraphics[width=2.4cm]{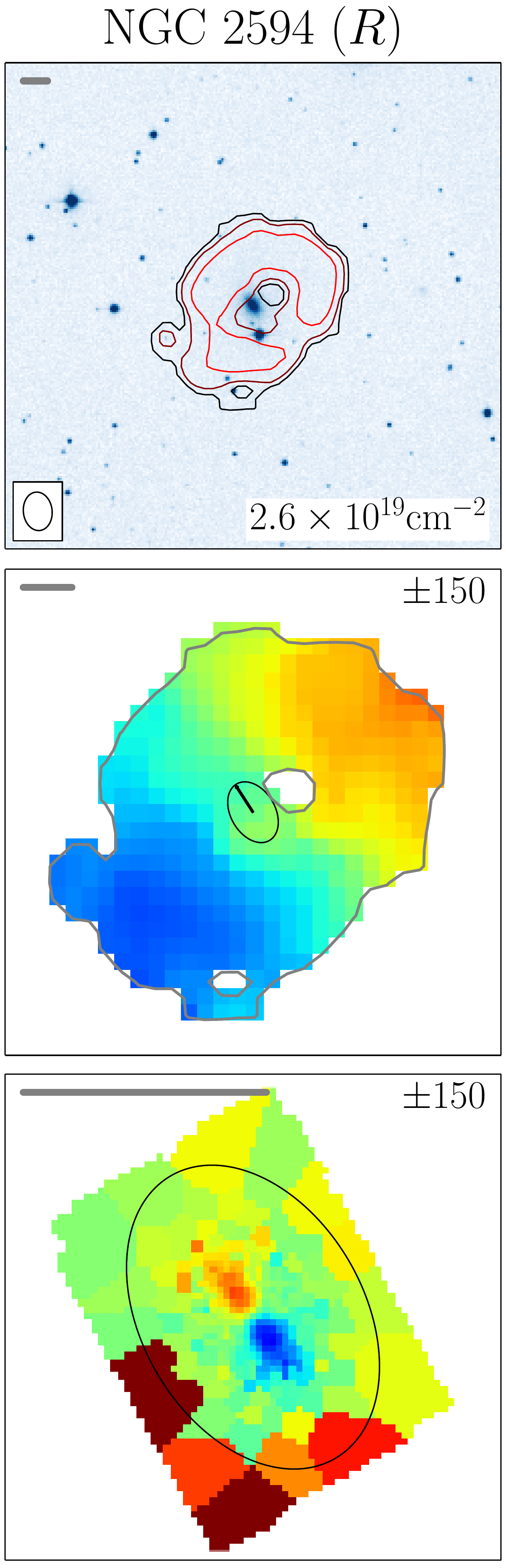}
\includegraphics[width=2.4cm]{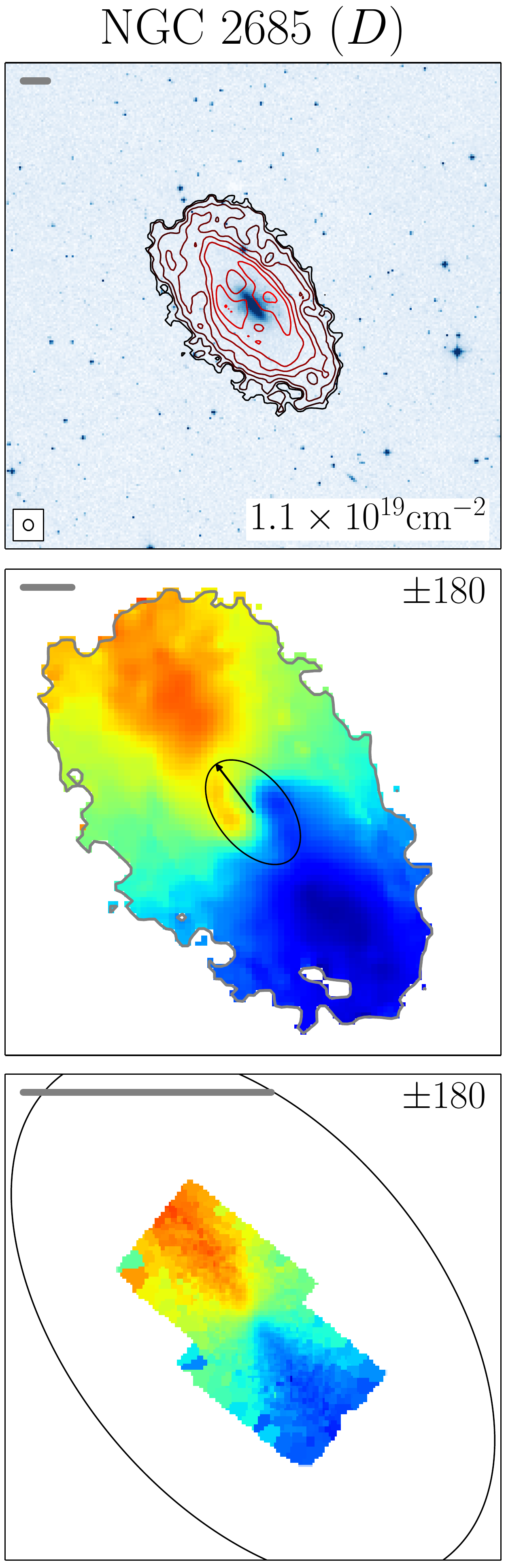}
\includegraphics[width=2.4cm]{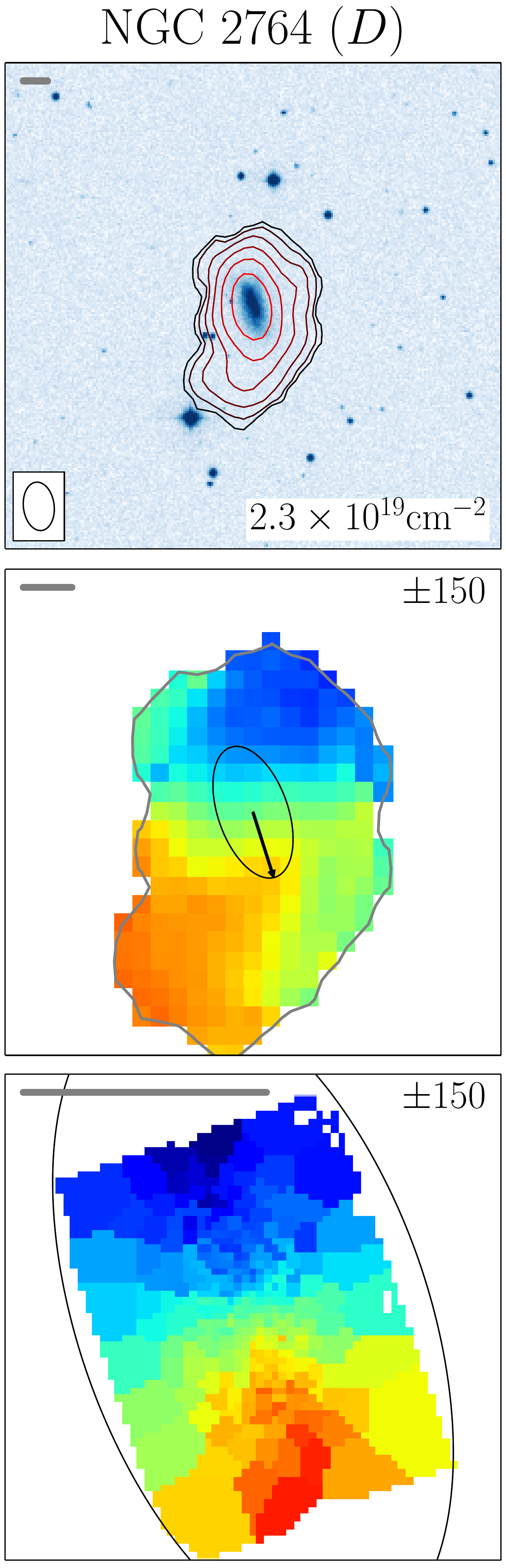}
\includegraphics[width=2.4cm]{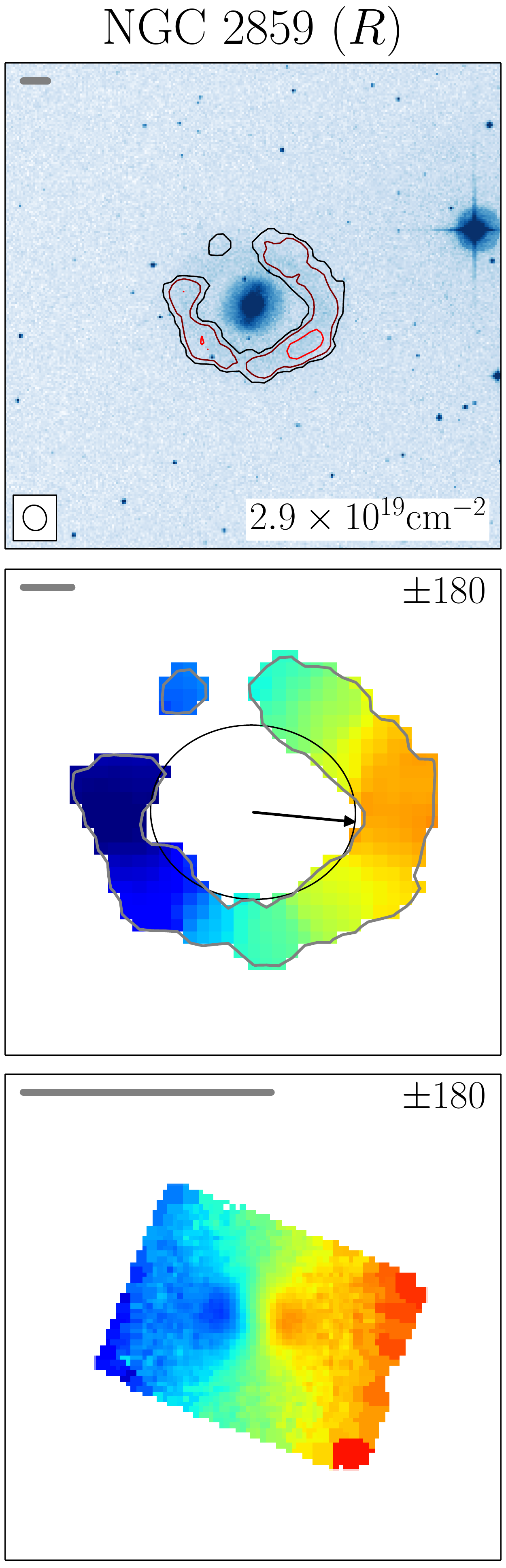}
\includegraphics[width=2.4cm]{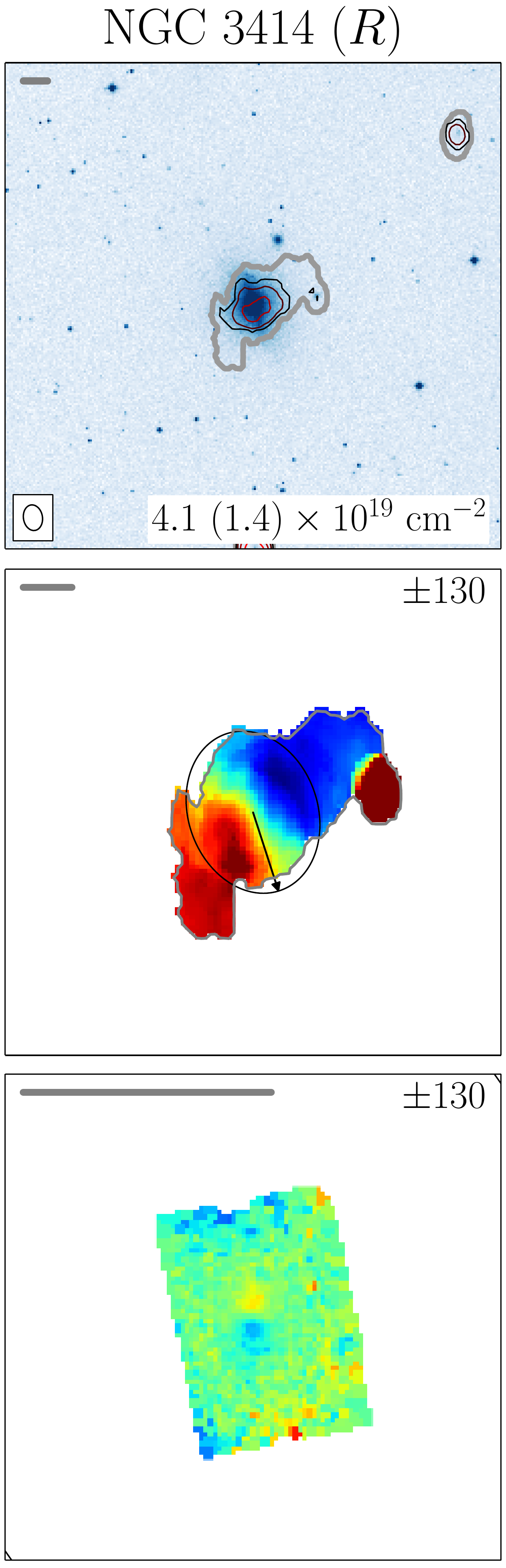}
\includegraphics[width=2.4cm]{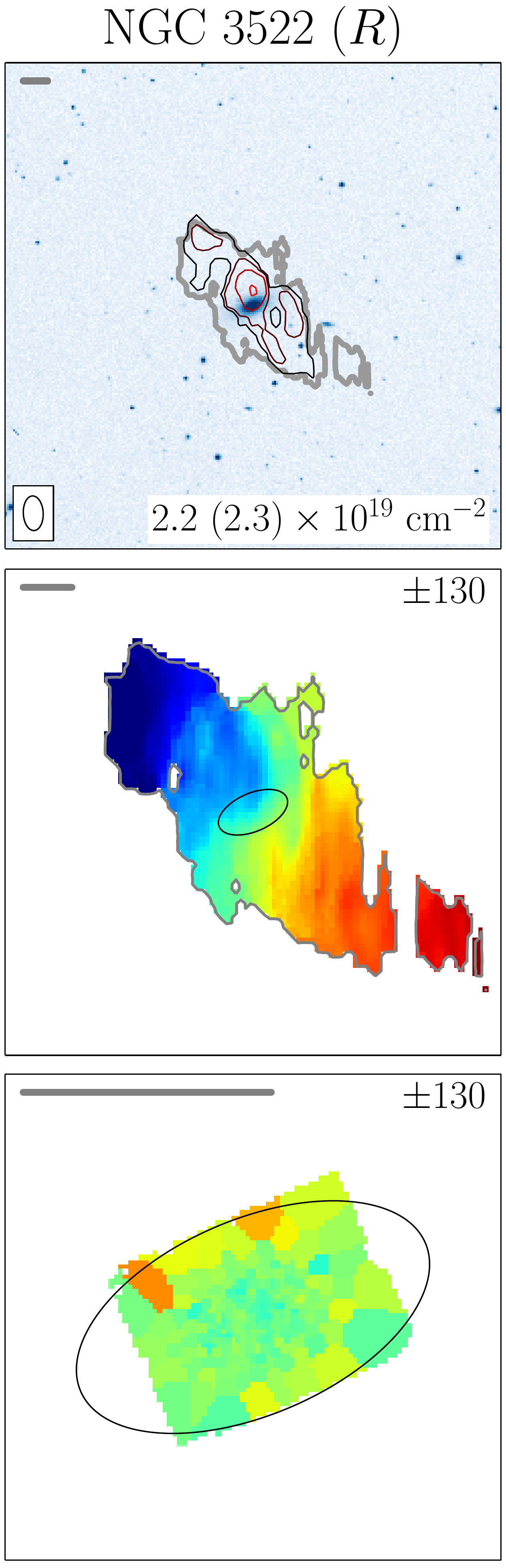}

\vspace{0.4cm}

\includegraphics[width=2.4cm]{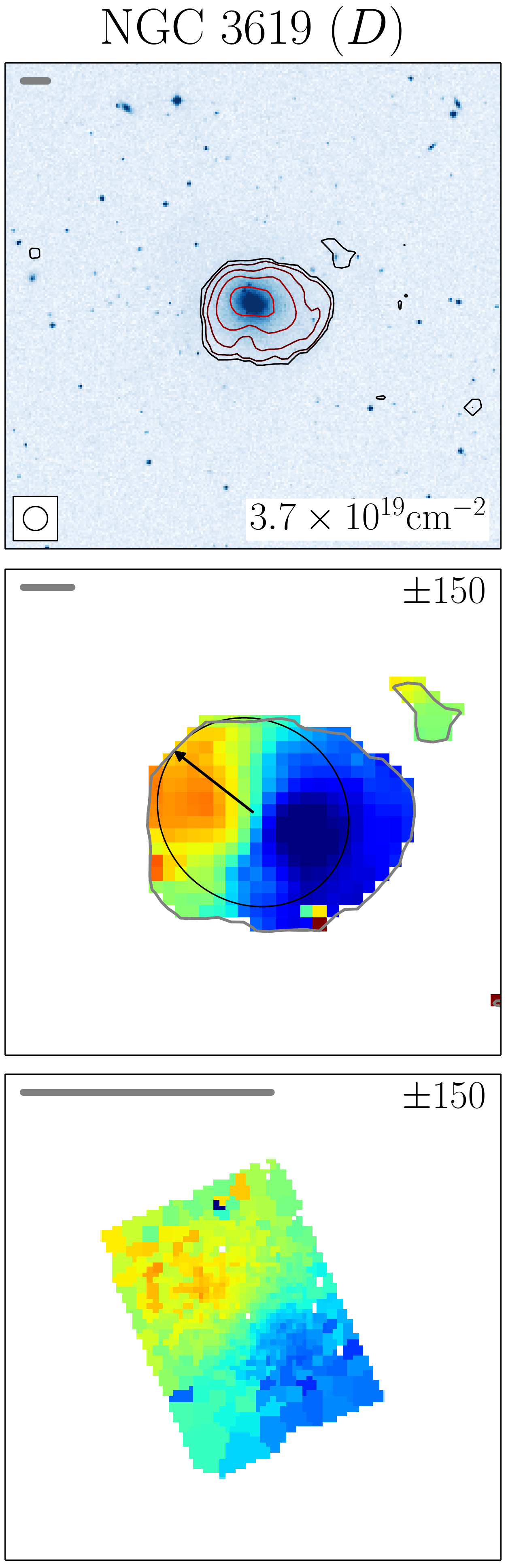}
\includegraphics[width=2.4cm]{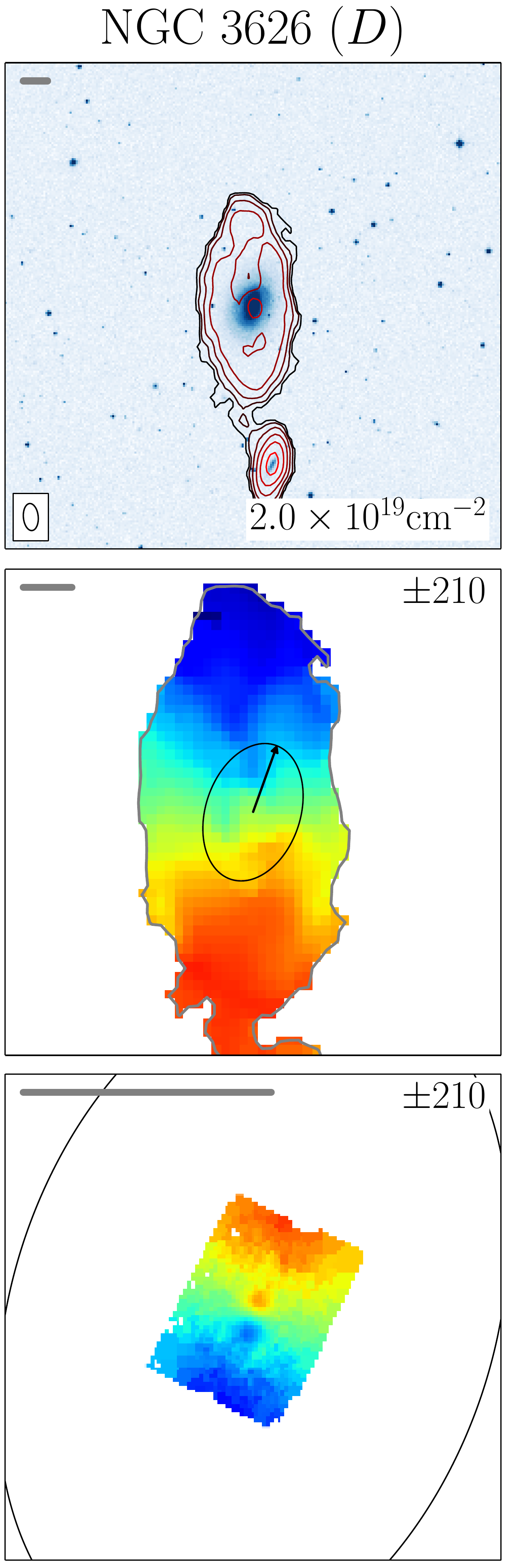}
\includegraphics[width=2.4cm]{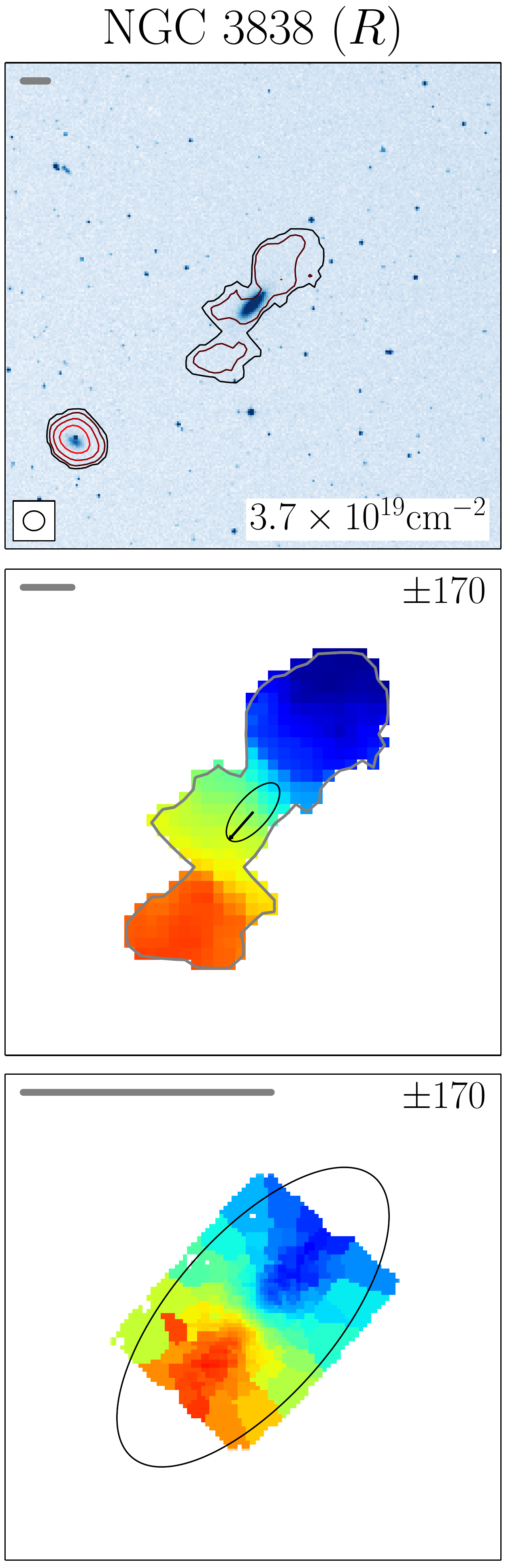}
\includegraphics[width=2.4cm]{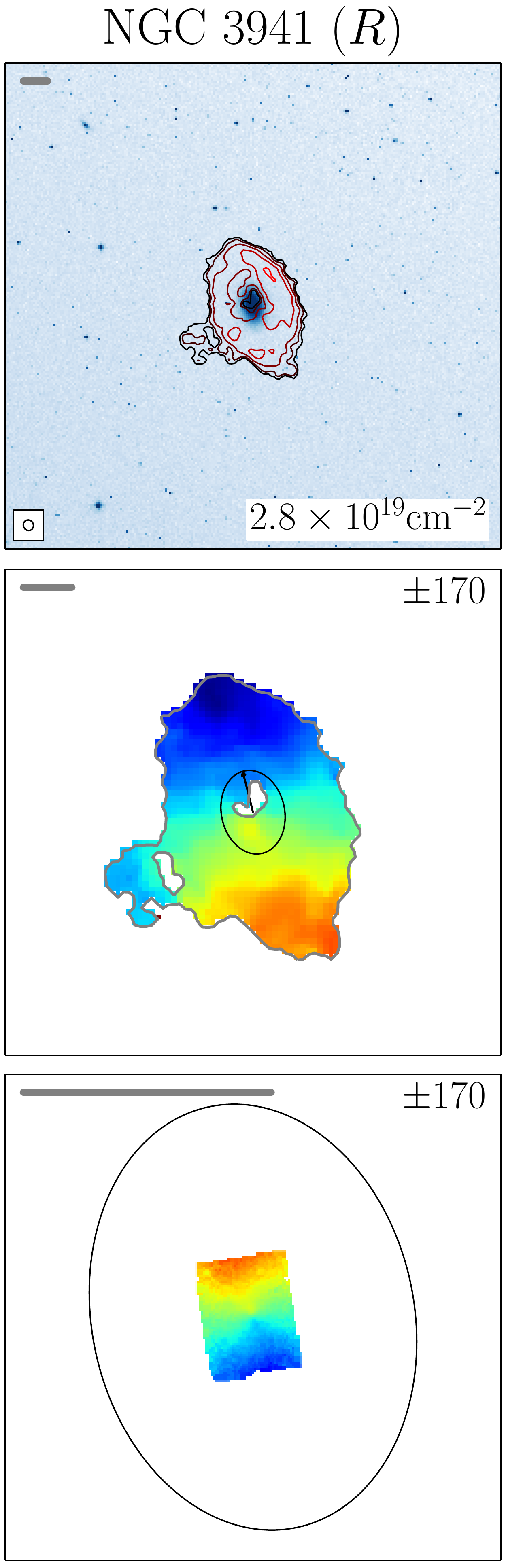}
\includegraphics[width=2.4cm]{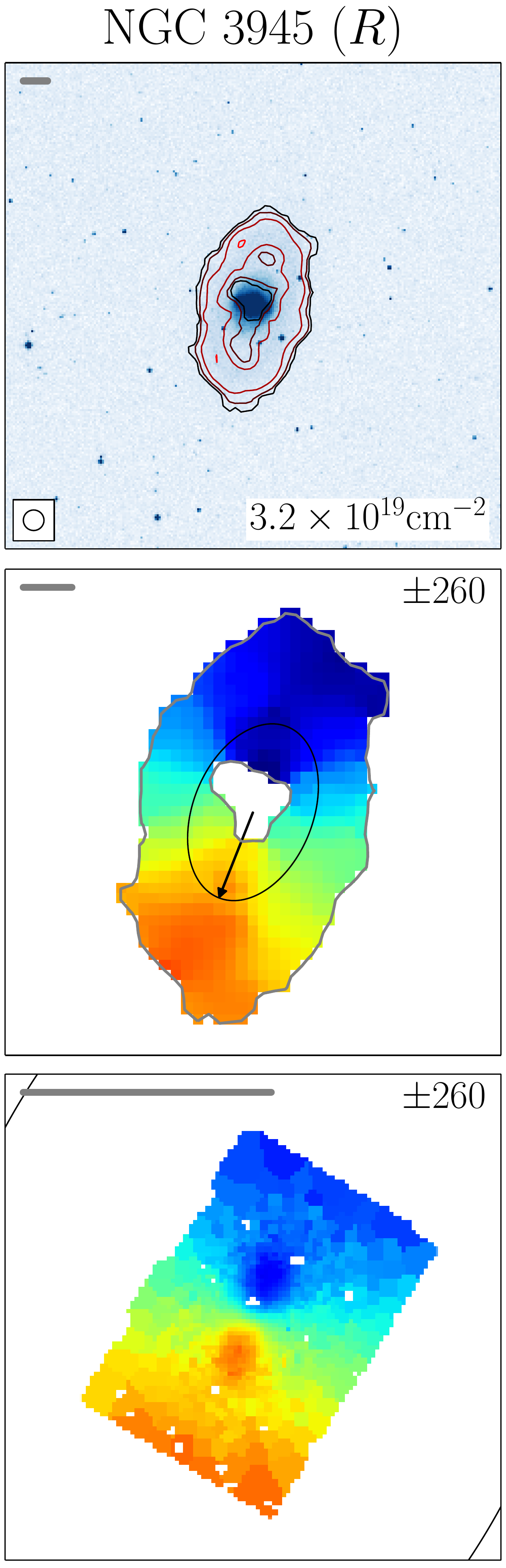}
\includegraphics[width=2.4cm]{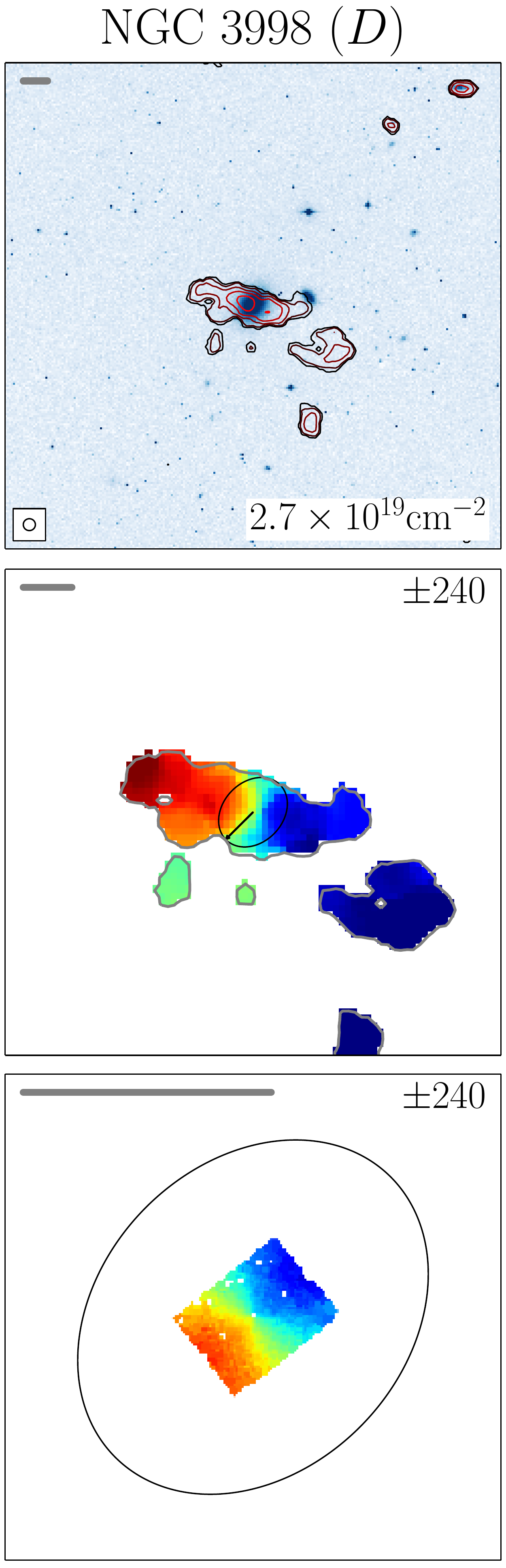}
\includegraphics[width=2.4cm]{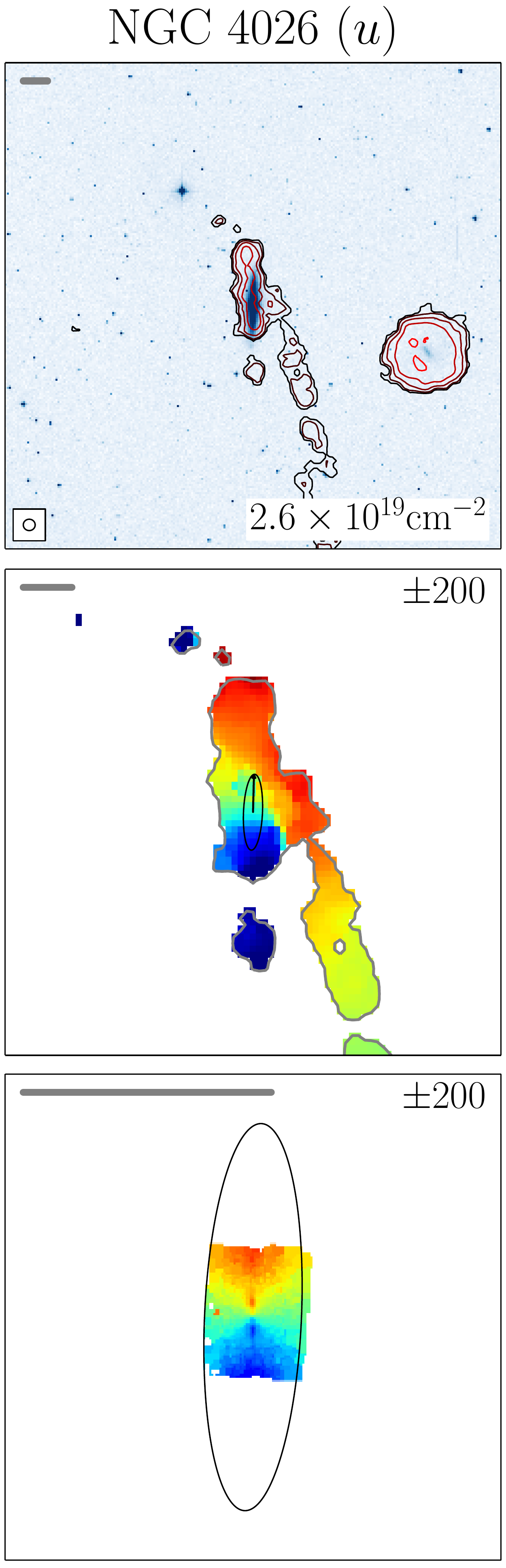}

\includegraphics[width=17.6cm]{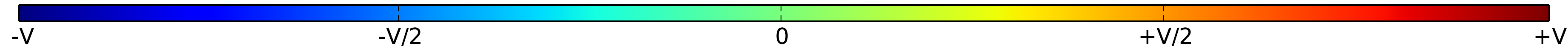}

\caption{From top to bottom (three panels per galaxy), \hi\ images ($100\times100$ kpc$^2$), \hi\ velocity fields ($50\times50$ kpc$^2$) and stellar velocity fields ($10\times10$ kpc$^2$) of all $D$ and $R$ ETGs, plus the two $u$'s showing some ordered rotation (NGC~1023 and NGC~4026). The grey scale-bar represents 5 kpc in all figures. The black ellipses overlaid on all velocity fields have semi-major axis equal to $3\ R_\mathrm{e}$ \rm and ellipticity taken from \citetalias{krajnovic2011}. The black arrow overlaid on the \hi\ velocity fields indicates the PA of the stellar kinematical major axis (receding side) from \citetalias{krajnovic2011} and scales with the size of the ellipse. In a few cases the stellar kinematical PA could not be determined reliably and no arrow is shown (see the text). \rm For all velocity fields the velocity range is indicated on the top-right. \rm This is kept fixed for the stellar and \hi\ velocity field of a same galaxy in order to enable a comparison of the relative line-of-sight rotation amplitude. \rm The colour scheme of the velocity fields is represented by the horizontal colour bar at the bottom of the figure.  \rm}
\label{fig:allvelfield}
\end{figure*}

\addtocounter{figure}{-1}
\begin{figure*}
\includegraphics[width=2.4cm]{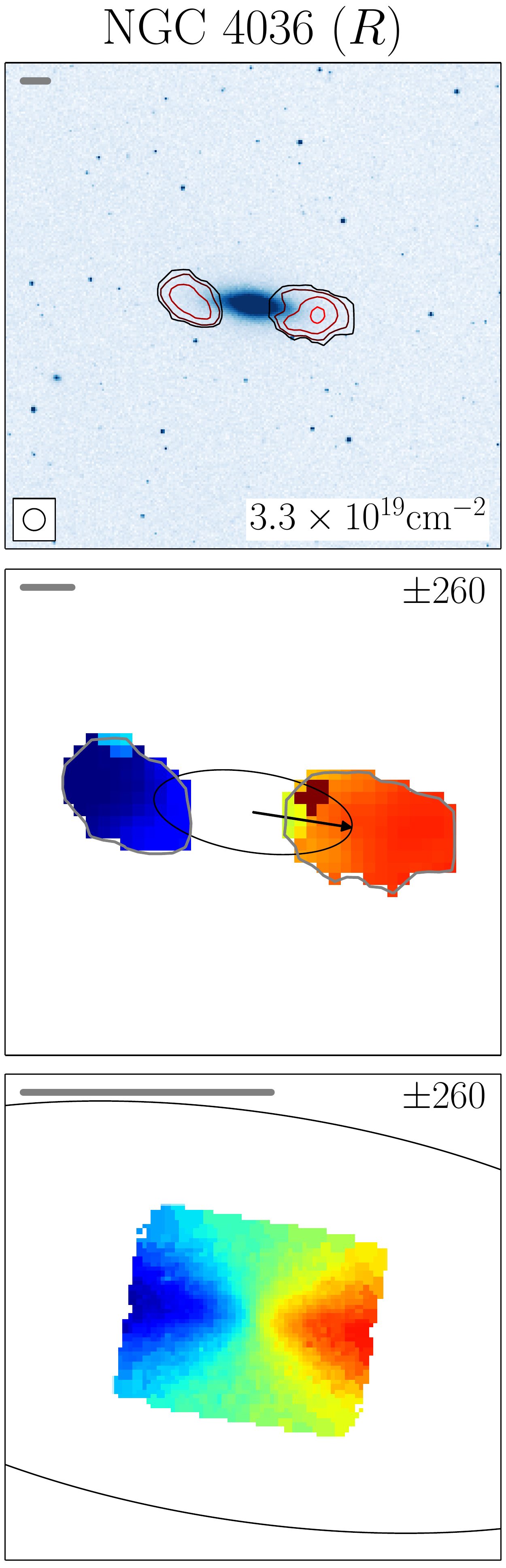}
\includegraphics[width=2.4cm]{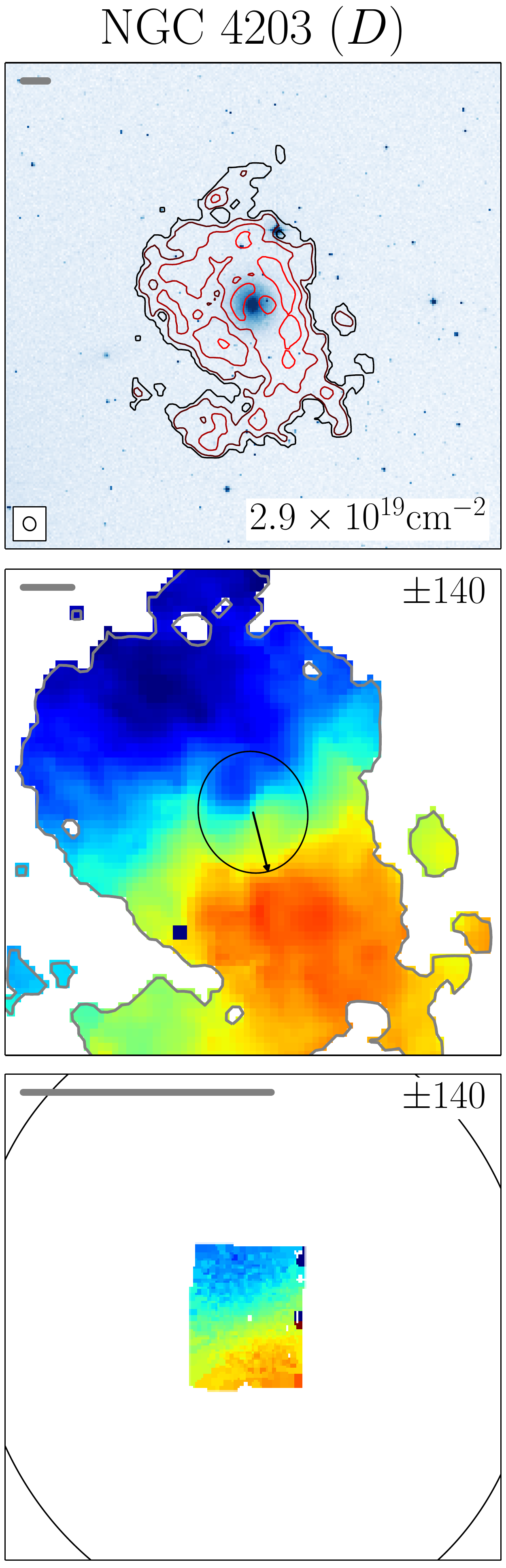}
\includegraphics[width=2.4cm]{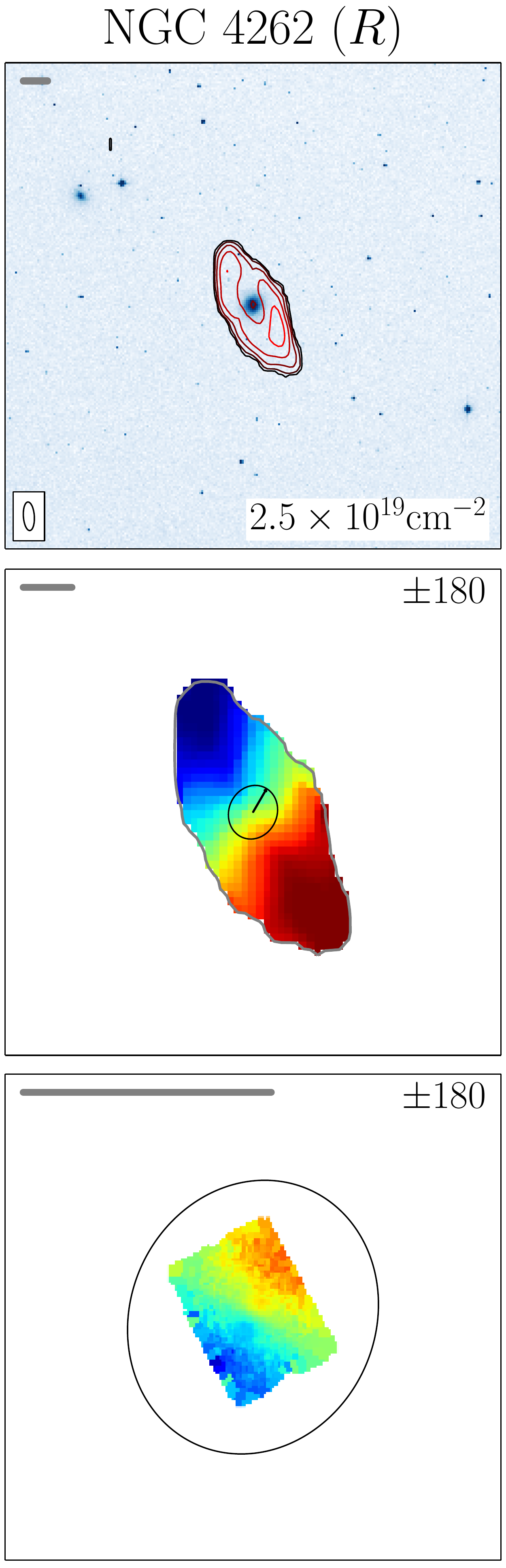}
\includegraphics[width=2.4cm]{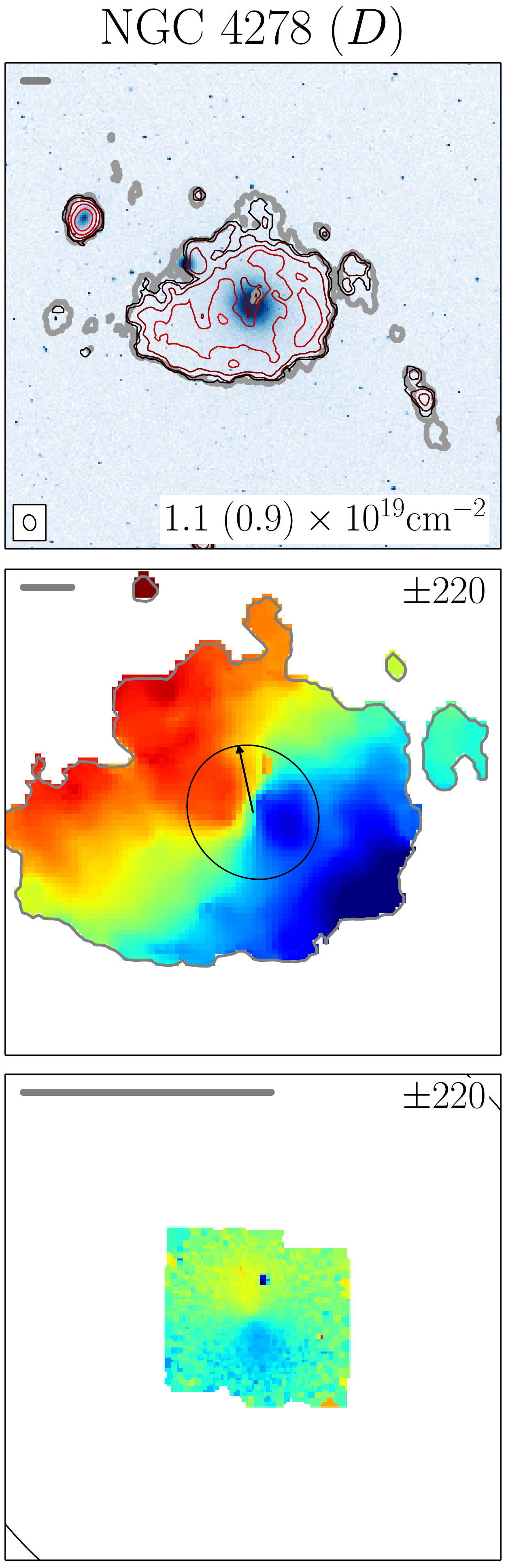}
\includegraphics[width=2.4cm]{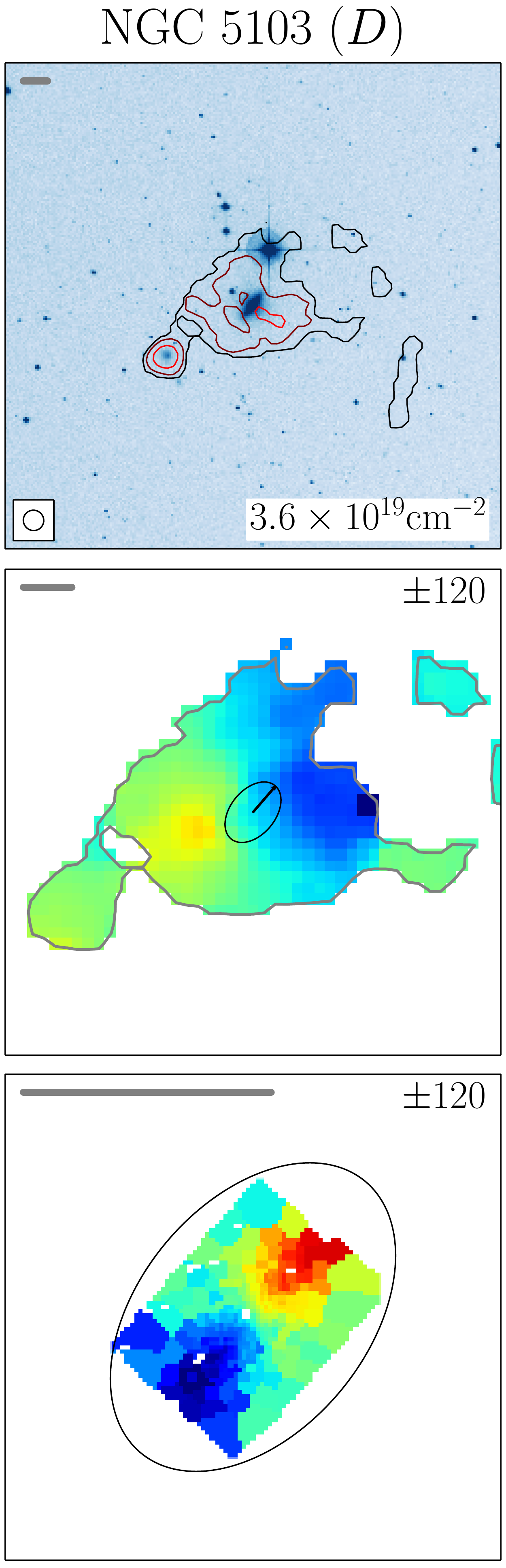}
\includegraphics[width=2.4cm]{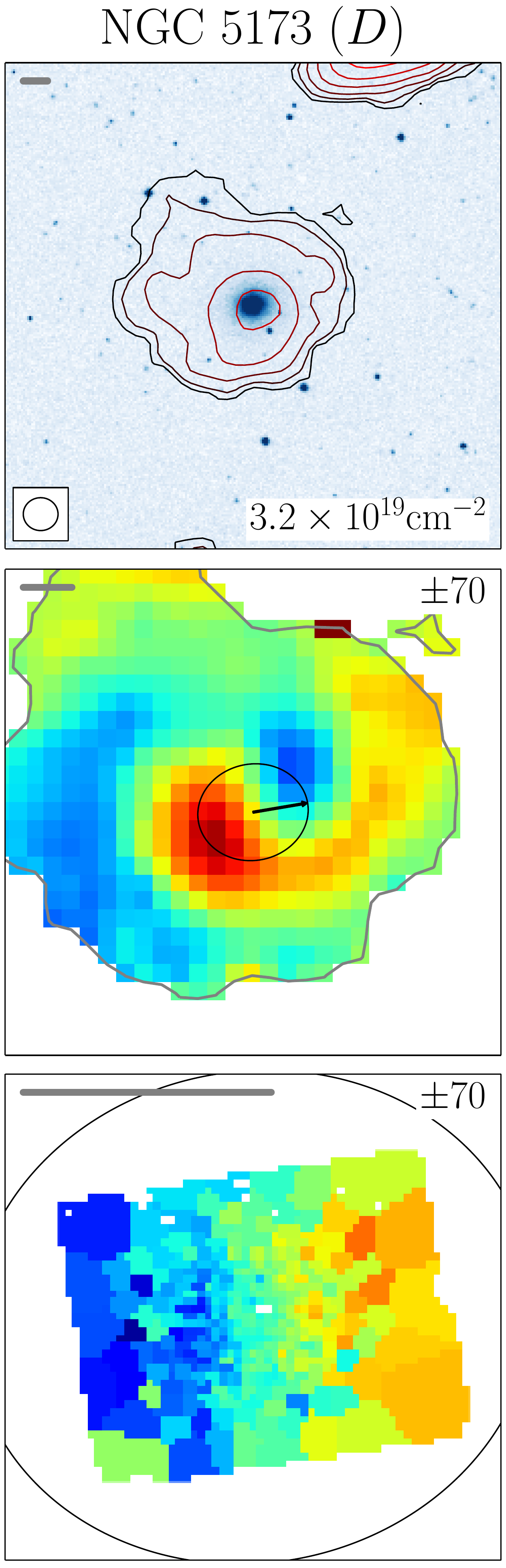}
\includegraphics[width=2.4cm]{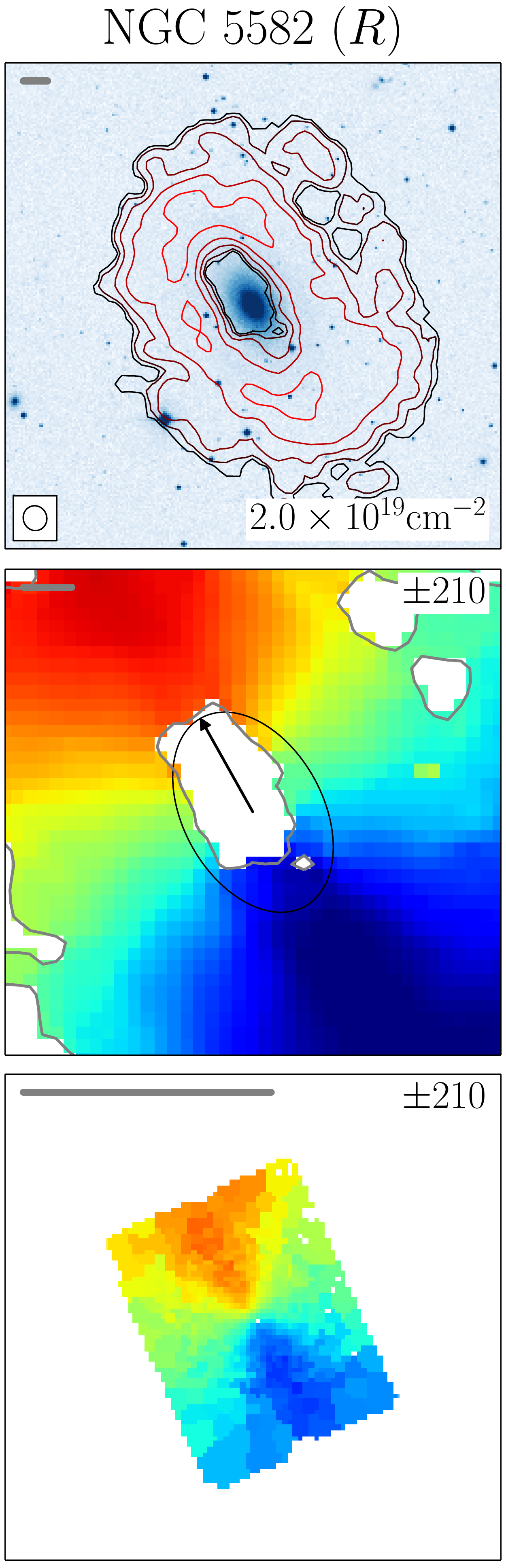}

\vspace{0.4cm}

\includegraphics[width=2.4cm]{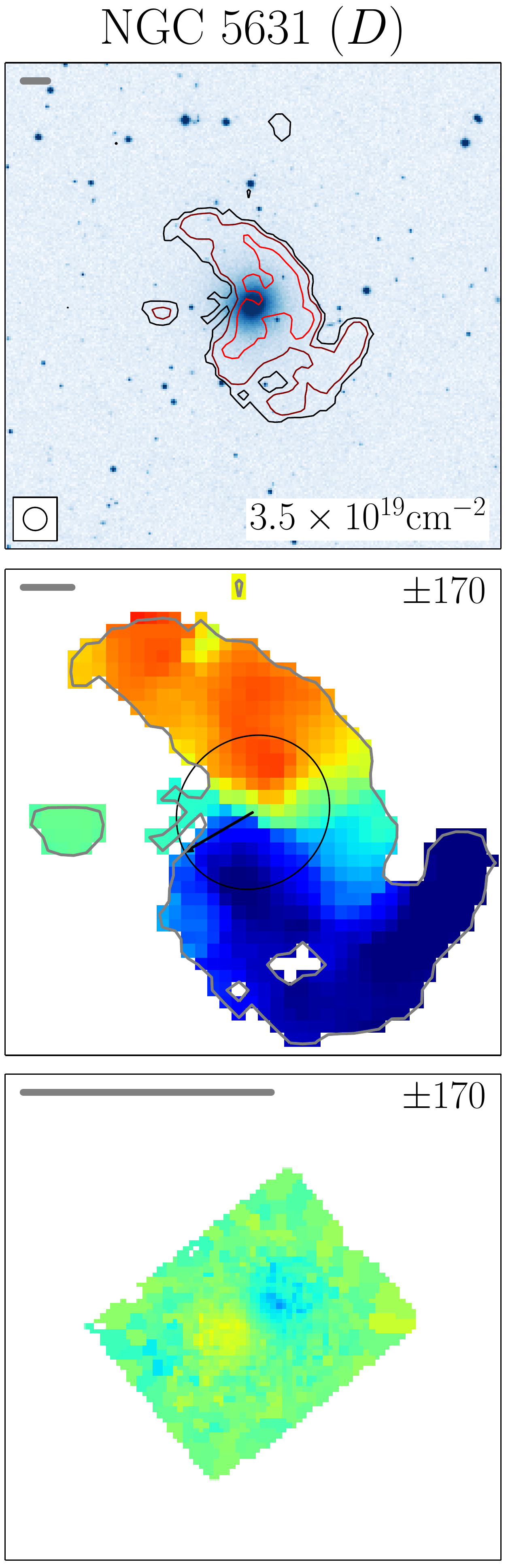}
\includegraphics[width=2.4cm]{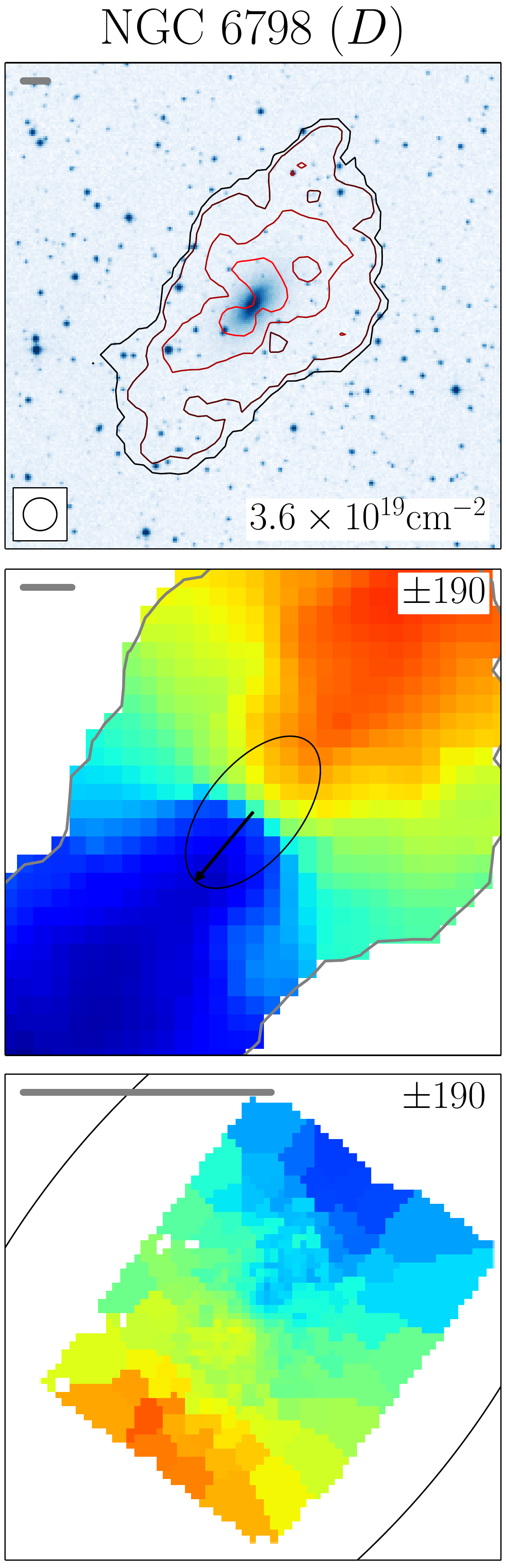}
\includegraphics[width=2.4cm]{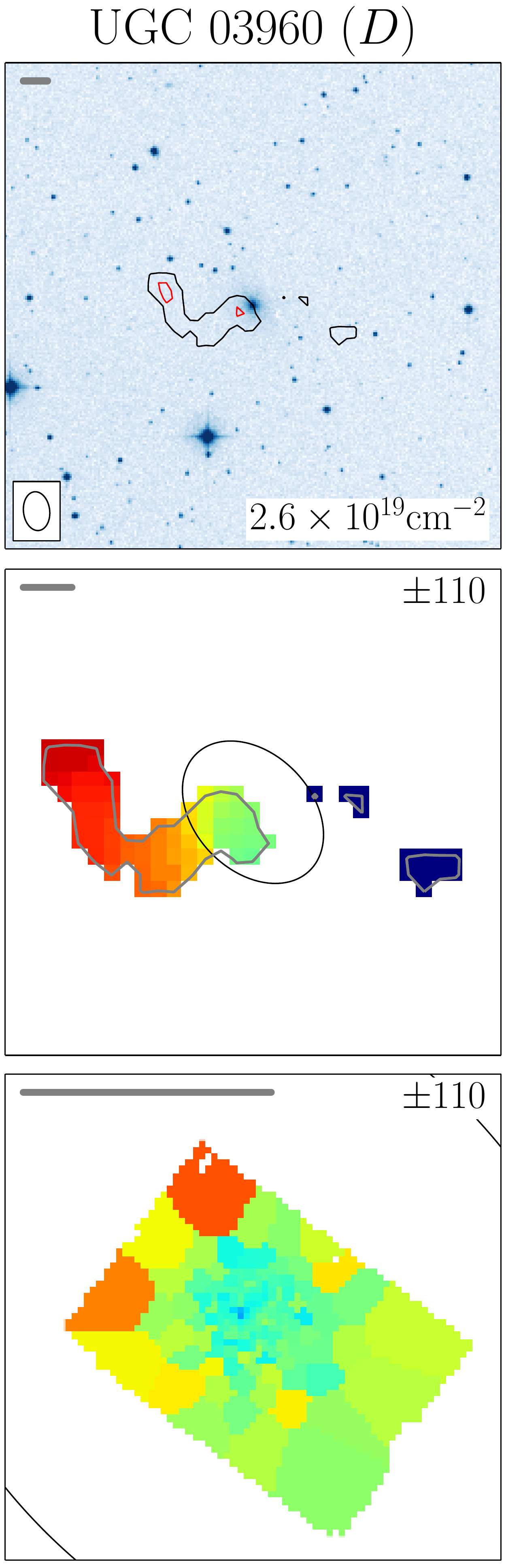}
\includegraphics[width=2.4cm]{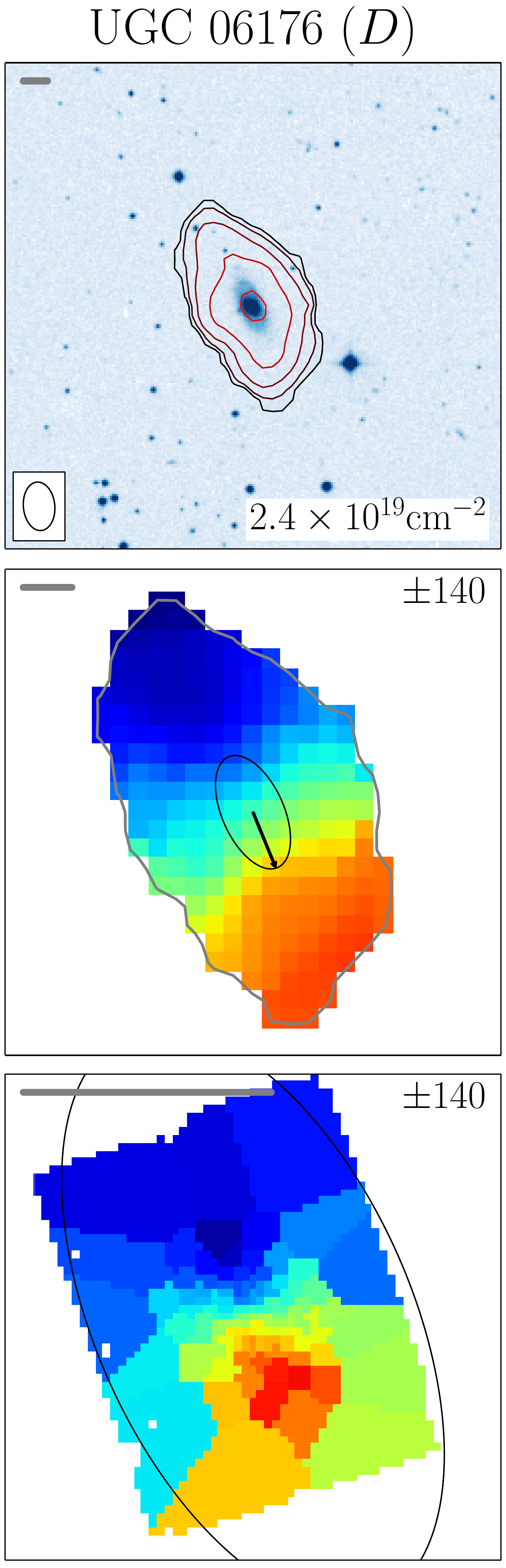}
\includegraphics[width=2.4cm]{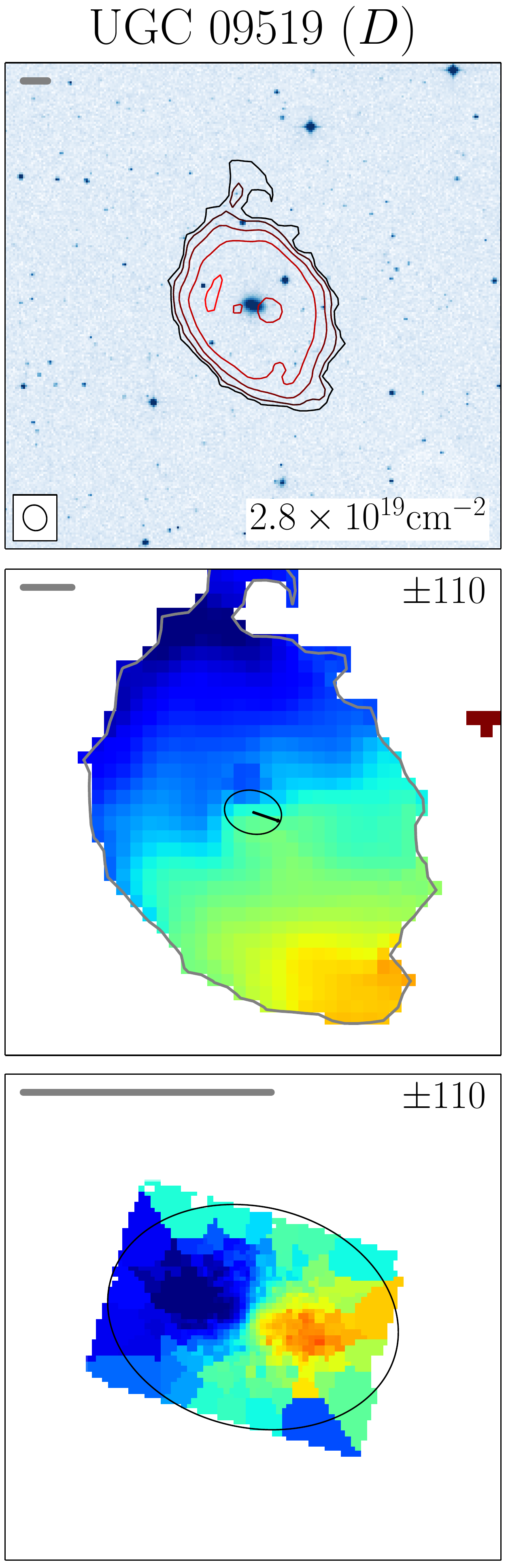}

\includegraphics[width=17.6cm]{colorbar.pdf}

\caption{\it Continued \rm}
\end{figure*}

\subsection{Mass-size plane}
\label{sec:msize}

\rm

Although the $\lambda_\mathrm{R}$-$\epsilon$ diagram is very useful to separate SRs from FRs, \citetalias{cappellari2013b} shows that a number of key ETG properties vary in a much clearer way on the mass-size plane instead. In particular, properties linked to the star formation history of galaxies such as their mass-to-light ratio, optical colour, H$\beta$ absorption and $M$(H$_2$)/$M_\star$ ratio correlate with bulge-to-disc ratio (traced by velocity dispersion) on this plane. We therefore investigate whether the ETG \hi\ properties too exhibit any systematic variation on the mass-size plane.

Figure \ref{fig:msize} shows all galaxies on the mass size plane of \citetalias{cappellari2013b}. This figure does not show any clear trend other than the decrease of the typical \mhi\ value and the prevalence of disturbed \hi\ morphologies at high galaxy mass, already reported in \citetalias{serra2012a}. As with Fig. \ref{fig:mhille}, the main observational result is a very large scatter of \hi\ properties, reinforcing our conclusion that \hi\ does not appear to be linked in any clear way to the inner galaxy structure.

\rm

\subsection{Diversity of gas accretion histories}
\label{sec:mainpoint}

The lack of a relation between ETG \hi\ properties (detection rate, mass and morphology) and ETG structure demonstrates the importance of cold gas as an independent probe of the evolution of these objects. Neutral hydrogen observations give us clues about the diversity of ETG assembly histories which are otherwise lost in the relative simplicity of their inner shape and kinematics. Our \hi\ survey demonstrates that there are very diverse evolutionary paths for FRs, which include the formation of both gas-poor objects (undetected in \hi\ or with very low \mhim) as well as galaxies with a considerable mass of \hi\ spread over a large area (and therefore with low mean column density; see \citetalias{serra2012a}), but that this diversity is not reflected in  the structure of ETGs within the FR family. Furthermore, we find that SRs too can accrete cold gas during their life and that they do so at a similar rate as FRs.

It is important to realise that SRs have a fundamentally different internal structure than FRs despite having similar \hi\ properties (in a statistical sense). This means that the accretion of cold gas has not altered their structure, which remains characterized by a relatively round shape and low specific angular momentum within $\sim 1\ R_\mathrm{e}$. In other words, gas accretion has not resulted in the growth of a significant stellar disc component in the central galaxy region. The same might be true for FRs: the \hi\ discs and rings in FRs do not need to be causally related to the stellar discs revealed by integral-field spectroscopy inside $\sim 1\ R_\mathrm{e}$, although they might well be in at least some cases. To investigate this aspect in more detail we now turn to the study of the kinematical misalignment between \hi\ and stars in these objects.

\section{Kinematical misalignment between \hi\ and stars}
\label{sec:misal}

In order to study the kinematical misalignment between \hi\ and stars we define a sample made of all galaxies for which we can reliably measure the position angle (PA) of the \hi\ kinematical major axis, PA(\hi). This includes all galaxies in the $D$, $R$ and $d$ classes plus two objects hosting \hi\ on an overall unsettled configuration ($u$) but where some of the gas shows ordered rotation -- NGC~1023 and NGC~4026. The total sample studied in this section includes therefore 36 ETGs.

\subsection{Polar, corotating, counterrotating and warped \hi}
\label{sec:misal1}

Figure \ref{fig:allvelfield} shows \hi\ images (from \citetalias{serra2012a}), \hi\ velocity fields (shown for the first time here) and stellar velocity fields (\rm already shown in \rm \citetalias{krajnovic2011}) of all $D$, $R$ and $u$ objects in this sample (26 objects in total; the 10 galaxies classified as $d$ are analysed separately below). The figure reveals a diversity of \hi-stars misalignments, including corotating, polar and counterrotating gas distributions. In a number of cases the strong variation of PA(\hi) with radius indicates the warping of the gas disc.

More quantitatively, we derive PA(\hi) as a function of radius from the \hi\ velocity field using the \textsc{kinemetry} software \citep{krajnovic2006}. \rm In order to maximize the radius out to which PA(\hi) can be measured we lower the default minimum coverage of an ellipse with data points to 30 per cent. As we are only interested in the PA values this is an acceptable choice. \rm We compare PA(\hi) to the PA of the stellar kinematical major axis, PA(stars), which is given in \citetalias{krajnovic2011}. Before proceeding we highlight two aspects which should be kept in mind in order to interpret the results of this analysis correctly.

First, while PA(\hi) is a function of radius and can be measured sometimes out to tens of $R_\mathrm{e}$, PA(stars) is an average value measured over an aperture of typically $\sim1\ R_\mathrm{e}$ (see \citetalias{krajnovic2011} for a detailed discussion). Therefore, we are not comparing the \hi\ kinematics to the stellar kinematics at the same radius, but the \hi\ kinematics (out to large radius) to the inner stellar kinematics. A comparison to the stellar kinematics at larger radius would require deeper optical spectroscopy over a larger field \rm \citep[e.g.,][]{proctor2009,weijmans2009,arnold2011,arnold2013} or the study of planetary nebulae \citep[e.g.,][]{romanowsky2003,coccato2009,cortesi2013} or globular cluster kinematics \citep[e.g.,][]{kartha2013,pota2013}. A comparison between gas and stellar kinematics at a common, small radius is presented in \citetalias{davis2011b} using ionized and molecular gas data.\rm

Secondly, while PA(stars) is a well defined quantity for FRs, where it is typically constant within the observed field, its interpretation may be ambiguous in SRs, where the stellar kinematics is more complex or characterized by very low-level rotation. The SRs for which we can measure PA(\hi) are NGC~3414, NGC~3522, NGC~5631 (all KDCs) and UGC~03960 ($b_\star$). In the case of NGC~3414 PA(stars) reflects the kinematics of the region outside (and counterrotating relative to) the KDC. To obtain the PA of the KDC one would have to subtract $180\,^{\circ}$ from the value given in \citetalias{krajnovic2011}. On the contrary, in NGC~5631 PA(stars) represents the inner (i.e., the KDC's) stellar rotation. In NGC~3522 the error on PA(stars) is very large because of the overall low-level rotation. In this case we calculate the misalignment relative to the photometric major axis, which coincides with the (uncertain) kinematical one. Similarly, PA(stars) is unconstrained in UGC~03960 and we adopt the photometric PA as a reference. \rm Note that using the photometric rather than kinematical PA would make no difference for NGC~3414, while NGC~5631 is nearly round on the sky and kinematics is the only possible choice.\rm

In Fig. \ref{fig:allmisal} we show PA(\hi) $-$ PA(stars) as a function of radius for all 26 galaxies in this subsample. This figure and Fig. \ref{fig:allvelfield} show that $\sim30$ per cent of the systems host corotating \hi\ (NGC~2859, NGC~3838, NGC~3945, NGC~4026, NGC~4036, NGC~4203, NGC~5582, UGC~06176), while counterrotating and polar \hi\ distributions account each for $\sim 10$ per cent of the sample (NGC~3626, NGC~3941 and NGC~6798 in the first group, NGC~2594, NGC~3998 and NGC~4262 in the second). Note however that the observed misalignment of polar systems depends on viewing angle and it is likely that the \hi\ is on a polar configuration in a larger fraction of all ETGs (e.g., NGC~3619, NGC~5103 and UGC~09519). In most of the remaining systems the misalignment angle varies significantly with radius, indicating the presence of a warp. In the two most extreme cases the \hi\ changes from polar to corotating (NGC~2685) and from counter- to corotating (NGC~5173) with increasing radius.

The situation is somewhat different for $d$s. The \hi\ in these objects is typically very faint \rm and extends over only a couple of Westerbork beams\rm . For this reason instead of using \textsc{kinemetry} we  assume that PA(\hi) does not vary with radius and determine its value by direct analysis of the \hi\ cubes. Position-velocity diagrams along the adopted \hi\ kinematical major axes are published in \citetalias{serra2012a} (their Fig. 5). We find that most (8/10) of the $d$s are corotating. The exceptions are one counterrotator (NGC~3032) and one polar system (NGC~3499), as discussed in that paper.

Considering all 36 galaxies together we find therefore that in $\sim45$ per cent of all ETGs with an \hi\ disc or ring the gas (which can spread out to tens of $R_\mathrm{e}$) is corotating with the stars inside $\sim1\ R_\mathrm{e}$. These galaxies are all FRs where the gas accretion episode traced by the \hi\ \rm might be \rm related to the formation of the stellar disc. They are in many respects similar to spiral galaxies as they host cold gas and stars corotating on the same disc -- the only difference being the absence of bright, star-forming spiral arms owing to the low \hi\ column density \citepalias{serra2012a}.

The remaining 55 per cent of this sample is composed of $\sim10$ per cent of counterrotating \hi\ systems, $\sim10$ per cent of polar objects, and $\sim35$ per cent of ETGs with a complex, warped \hi\ distribution, some of which may in fact be polar. The link between \hi\ disc and stellar disc in these objects is less obvious.

These results show that the existence of multiple kinematical sub-components in ETGs, well documented in their inner region for the stellar and ionized gas phases, holds also on much larger scales. Following our results on the fraction of ETGs with a kpc-scale KDC ($\sim7$ per cent including counterrotating cases; \citetalias{krajnovic2011}) and misaligned ionized-gas kinematics ($\sim36$ per cent; \citetalias{davis2011b}; see also \citealt{bureau2006}) we are now finding misaligned \hi\ distributions on scales from a few to tens of kpc in $\sim12$ per cent of all ETGs ($\sim15$ per cent outside the Virgo cluster; \rm for previous studies of smaller samples of ETGs with misaligned \hi\ see, e.g., \citealt{vangorkom1987}; \citealt{bertola1992}; \citealt{bettoni2001}\rm).

Whether or not associated with the presence of a stellar disc, the \hi\ detected in ETGs provides important clues on the origin of the cold interstellar medium detected in the inner region of these objects in the form of ionized and molecular gas. This is clear in Fig. \ref{fig:allmisal}, where we show at $R=0$ the data point corresponding to the ionized gas kinematics (when detected), which was already found to be invariably aligned with the CO kinematics \citepalias{davis2011b}. The figure confirms previous claims that \hi, H$_2$ and ionized gas share the same kinematics in the region where they overlap (\citealt{morganti2006}; \citetalias{davis2011b}). The few exceptions are NGC~1023, NGC~3414, NGC~4278, NGC~5103 and NGC~5631. The first hosts an unsettled gas distribution with very complex gas kinematics, and it is unlikely that the central value of PA(\hi) derived from the velocity field is representative of the actual inner gas motion. Furthermore, \cite{morganti2006} show that in both NGC~3414 and NGC~4278 the ionized gas velocity field  twists in such a way that the outer ionized gas kinematics is aligned with the inner \hi\ kinematics. The same occurs in NGC ~5631. Therefore, these three galaxies conform to the general rule too. We do not have a good explanation for the misalignment between \hi\ and ionized gas in NGC~5103, which represents the only genuine exception to the rule. Not shown in Fig. \ref{fig:allmisal}, $d$ galaxies contribute to strengthening this result. The \hi\ is always corotating with the CO and ionized gas, including the two cases where the \hi\ is misaligned relative to the stars (NGC~3032 and NGC~3499).

\begin{figure*}
\includegraphics[width=18cm]{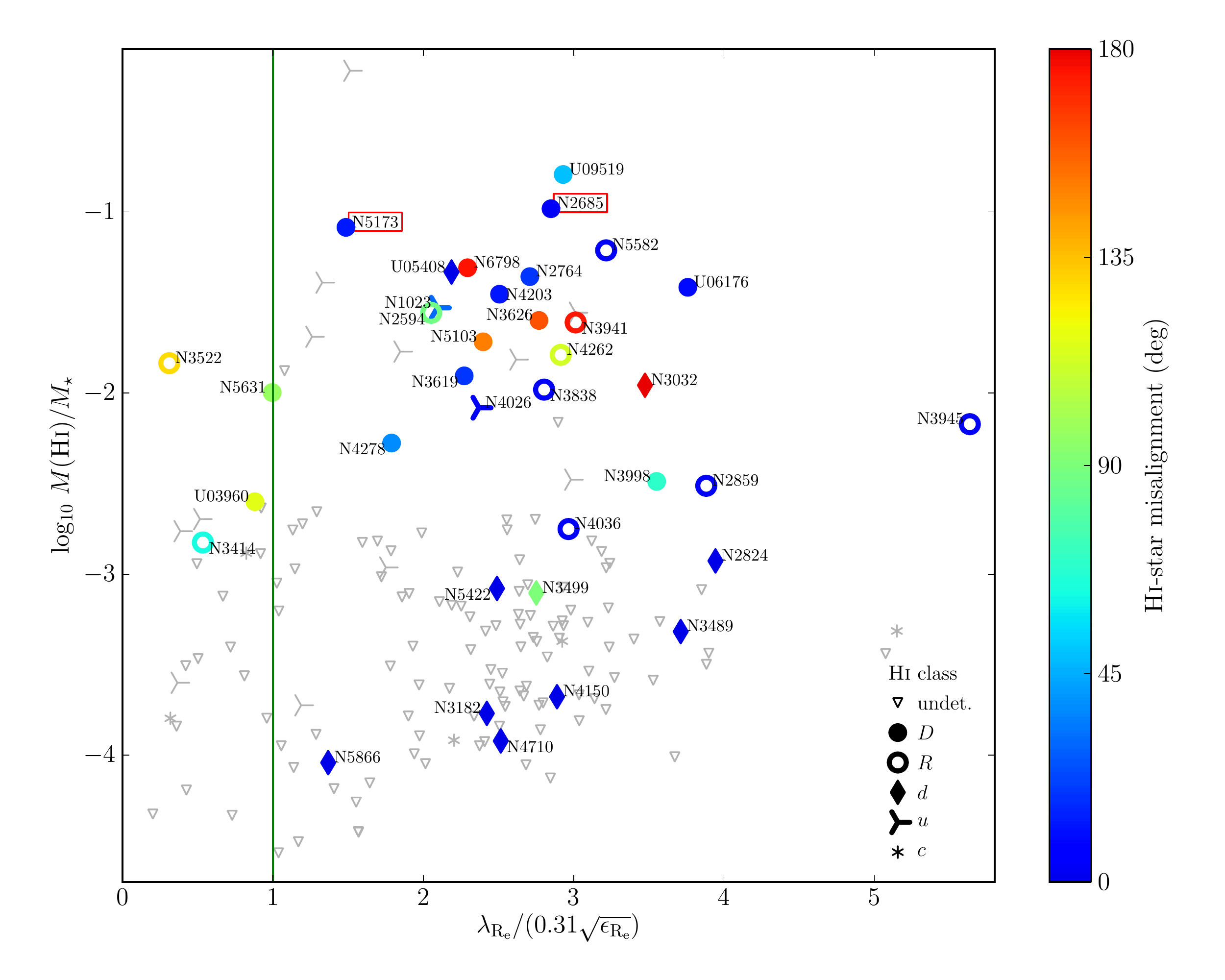}
\caption{\rm \hi\ kinematical misalignment on the \mhim\ versus $\lambda_\mathrm{R}/(0.31\sqrt{\epsilon})$ plane. This figure is the same as Fig. \ref{fig:mhille} but here we highlight only the 36 galaxies for which we are able to study the \hi\ kinematical misalignment (see Sec. \ref{sec:misal} for details on this subsample; \rm the remaining galaxies are shown with light grey markers\rm). We colour-code them according to the median misalignment angle calculated over the full radial extent of the \hi\ disc or ring. See Fig. \ref{fig:allmisal} for the variation of the misalignment angle as a function of radius in individual galaxies. Note that NGC~2685 and NGC~5173, the two most heavily warped \hi\ systems in the sample, have a small median misalignment angle which is not representative of their true complexity. For this reason we highlight them here with a red box.\rm }
\label{fig:mhille_misal}
\end{figure*}

\subsection{Misalignment in fast and slow rotators}
\label{sec:misal2}

Figure \ref{fig:mhille_misal} summarizes our findings by showing all \hi\ misalignments on the \mhim\ versus $\lambda_\mathrm{R}/(0.31\sqrt{\epsilon})$ plane of Fig. \ref{fig:mhille}. For each galaxy we adopt here the median misalignment value measured from the radial profile shown in Fig. \ref{fig:allmisal}. The median value is a fair representation of the overall misalignment for all galaxies except the heavily warped NGC~2685 and NGC~5173, which we highlight in the figure. On the one hand Fig. \ref{fig:mhille_misal} shows a significant population of FRs where \hi\ and stars are kinematically aligned (dark blue symbols, with the just mentioned exception of NGC~2685 and NGC~5173). In these galaxies gas and stars are just two different but coupled components of the same disc. On the other hand, as already emphasized above, the \hi\ discs and rings detected in FRs are not always associated with the inner stellar disc, not even for the flattest and fastest rotators (see for example NGC~3626, NGC~5103 and UGC~09519). In many FRs the large scale \hi\ kinematics is obviously not a continuation of the inner stellar kinematics to larger radius. On the contrary, counterrotating and polar \hi\ distributions account for exactly $1/4$ of all FRs analysed in this section, while another $1/4$ consists of warped discs. The detected \hi\ is therefore aligned with the inner stellar disc only in half of the cases. \rm Neutral hydrogen observations show therefore that FRs, while being a dynamically simple and homogeneous family inside $\sim1\ R_\mathrm{e}$, are much more complex and diverse in their outer regions, in agreement with results obtained using other kinematical tracers at large radius \citep[e.g.,][]{coccato2009,pota2013}.\rm

The other clear result is that in none of the four SRs the \hi\ kinematics is aligned with the stellar kinematics or morphology. In fact, all these systems exhibit a (close to) polar \hi\ distribution, although PA(\hi) typically varies with radius (Fig. \ref{fig:allmisal}). We also note that the \hi\ morphology of these systems, although disc- or ring-like, appears disturbed (Fig. \ref{fig:allvelfield}), with some cases intermediate between a $D$/$R$ and a $u$ classification. The lack of very regular, settled \hi\ discs/rings in SRs and the prevalence of polar systems may be the only difference from the FR family (\hi-wise), although even among FRs many gas discs and rings appear disturbed or not perfectly settled (e.g., NGC~2764, NGC~3998, NGC~4203, NGC~4278, NGC~5103). A larger sample of \hi-rich SRs would be needed to confirm this indication.

The above results, together with the established weak relation between \mhi\ and ETG luminosity (e.g., \citealt{knapp1985,wardle1986}; \citetalias{serra2012a}), confirm the severe disconnect between ETG \hi\ properties and their stellar component. They show, as argued in Sec. \ref{sec:mainpoint}, that \hi\ provides clues about the complex evolution of ETGs which would go unnoticed using only other observables. Confirming this complexity, we also find that the \hi\ kinematical misalignment does not depend in any clear way on ETG stellar mass, environment density (quantified in \citetalias{cappellari2011b}) or \hi\ mass. For example among the very \hi-rich galaxies in poor environments we find NGC~5582 (corotating $R$), NGC~6798 (counterrotating $D$) and UGC~09519 (warped $D$). Low-mass, isolated galaxies include NGC~2594 (polar $R$), UGC~06176 (aligned $D$) and again UGC~09519. Among the massive ETGs we find relatively small, aligned \hi\ discs/rings as in NGC~2859 and NGC~4036 but also (close to) polar \hi\ as in NGC~3414 and NGC~5631, and the warped $D$ NGC~4278. The only exception is that the lowest-\mhi\ systems are all $d$s \citepalias{serra2012a} and, as illustrated above, mostly kinematically aligned to the stellar rotation. There is of course some dependence of \hi\ properties on ETG mass and environment across the entire \atlas\ sample. \rm This is discussed in detail in \citetalias{serra2012a}, where we show that unsettled \hi\ distributions become more common, and the typical \hi\ mass decreases, as environment density and galaxy mass increase. Here we find that there is no additional trend with environment and stellar mass when studying the kinematical misalignment of settled \hi\ discs.\rm

Ever since the work of \citet{knapp1985} this decoupling between stars and \hi\ in ETGs has been attributed to the external origin of the \hi. The main point of this interpretation is that most of the present-day gas must be accreted once the formation of the (inner region of the) host ETG is basically complete. The accreted gas could come from the intergalactic medium or be brought in by small gas-rich satellites which individually do not contribute significantly to the stellar mass of the ETG but may bring a significant addition to its interstellar medium. \rm In fact, even when gas is acquired at the same time as the formation of the stellar body, e.g. in gas-rich major mergers, there does not need to be a strong coupling between the stellar body of the remnant and the larger-scale \hi\ distribution. Merging and \rm accretion events are expected to occur in a $\Lambda$CDM Universe, where galaxies grow hierarchically, and our results may be seen as a confirmation of these theories. Qualitative statements of this kind have been made by previous authors and, at this stage, do not represent an advancement in our understanding of ETG formation. For this reason we attempt to take a step forward and perform a quantitative comparison to galaxy formation simulations in the next section.

\section{Comparison to cosmological simulations}
\label{sec:sims}

We compare ETGs in the \atlas\ sample to galaxies formed in two sets of hydrodynamical simulations performed by \citet{oser2010} and \cite{hirschmann2013}. These are high-resolution re-simulations of $\sim50$ \rm dark-matter haloes \rm extracted from a large, $\Lambda$CDM, dark-matter-only simulation. In each re-simulation baryonic processes such as hydrogen and helium cooling, star formation and supernova feedback are implemented in the presence of a uniform UV background. The two sets of simulations have different implementations of supernova feedback. In the first set the feedback is purely thermal (hereafter, NoW  simulations because they do not include galactic stellar winds). These are the simulations analysed by \citetalias{naab2013} to study \rm the generic formation paths of \rm SRs and FRs in a cosmological context. \rm Simulations in the second set start from identical initial conditions but include an additional empirical model for galactic outflows driven by winds from star-forming regions (\citealt{oppenheimer2006}; we will refer to these as W  simulations). All details on the galactic wind implementation and a comparison of galaxy properties between the two models are presented in \cite{hirschmann2013}\rm . \rm Here we study for the first time the distribution and kinematics of cold gas in the simulations. \rm As we will show, these two sets give profoundly different predictions for galaxies' \hi\ properties. We refer to \cite{oser2010} and \cite{hirschmann2013} for a full description of the simulations.

For the purpose of this work it is important to first investigate whether the simulated galaxies are a reasonable match to the ETGs we want to compare them to. On the one hand, they cover the same $M_\star$ range (from $\sim10^{10}$ to a few times $10^{11}$ \msun). \rm Furthermore, the NoW simulations are in reasonable agreement with the observed mass-size relation of ETGs \citep{oser2012} and exhibit a diversity of stellar kinematical properties similar to that of real ETGs (\citetalias{naab2013}). On the other hand, not all simulated galaxies, in particular the W sample, \rm would necessarily meet the morphological selection of the \atlas\ sample, which requires the absence of large-scale spiral arms in a galaxy optical image \citepalias{cappellari2011a}. This is essentially a selection against strong star-formation on a disc while \rm about half of all simulated galaxies have relatively large specific star formation rate ($>10^{-11}$ yr$^{-1}$) at $z=0$. Similar values are measured in only a handful of the observed ETGs (Davis et al., submitted). \rm
 We note however that the uncertainty on the simulated star formation rate is large because of uncertainties in, e.g., star formation and feedback physics and their actual implementation. Additionally, neither of the simulated samples include AGNs, which \rm could \rm be an efficient way to suppress star formation. Therefore, we decide not to exclude star-forming galaxies from the comparison. Instead, we highlight this as a caveat of this work and stress that more sophisticated simulations would be needed to fully replicate the \atlas\ selection.


The simulations do not include a full treatment of different gas phases and the correspondence between simulated gas particles and \hi\ is not obvious. As a first step we define cold gas following the criterion of \cite{hirschmann2012}, which is based on a selection of gas particles on the density-temperature plane. This is preferred to a simple temperature selection since dense star-forming gas (which we want to include in the cold-gas budget) can be relatively hot in the NoW simulations. Additionally, we assume that this cold gas is made of two phases, atomic and molecular (ignoring that some of it may in fact be ionized). We separate the two phases on the gas' face-on view by applying equation 39 of \cite{krumholz2009}. In their equation $s$ is the \hi+H$_2$ column density in \msun\ pc$^{-2}$ (see \citealt{schruba2011} for a comparison between this theoretical relation and observations). This results in a relatively sharp transition from an atomic- to a molecular dominated regime at a total gas column density of $\sim10$ \msun\ pc$^{-2}$ \citep{wong2002,bigiel2008}. The resulting face-on \hi\ images are used to calculate \hi\ mass and column density distribution for all simulated galaxies (at the typical gas resolution of $\lesssim 0.5$ kpc).

\begin{figure}
\includegraphics[width=8.5cm]{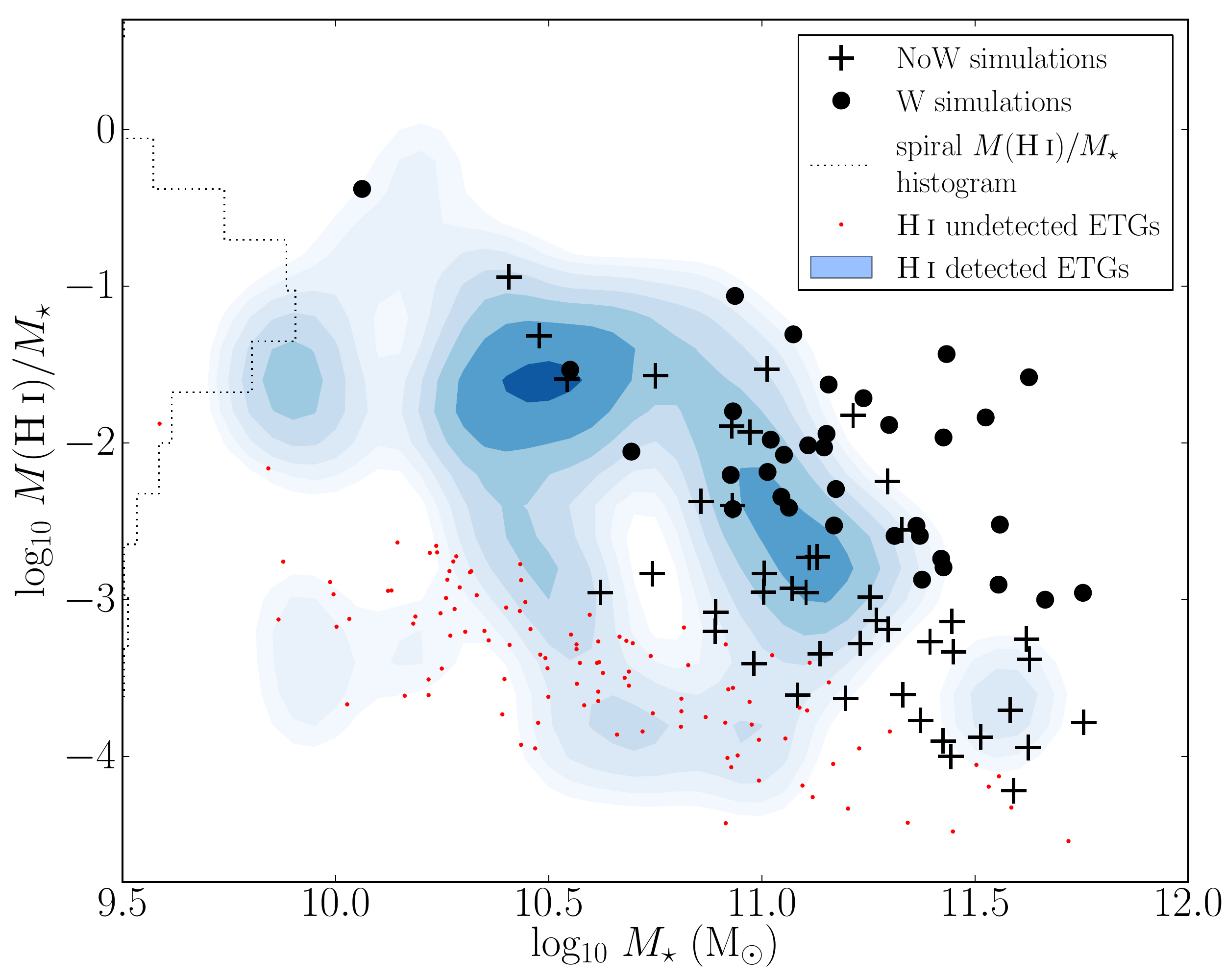}
\caption{\mhim\ plotted against $M_\star$ for simulated galaxies (black circles and crosses) and real ETGs from \citetalias{serra2012a}. For the latter, \rm the blue area shows the kernel density estimate derived using all 53 galaxies \rm with detected \hi\ while the red dots indicate \hi\ non-detections and are therefore upper limits on \mhim. The dotted histogram along the vertical axis represents the \mhim\ distribution of spiral galaxies derived as explained in the text.}
\label{fig:cosmomass}
\end{figure}

In order to study the morphology and kinematics of the \hi\ we construct seven different projections of each simulated galaxy relative to an hypothetical observer: face on (relative to the stellar kinematics); rotated by 30, 60 and $90\,^{\circ}$, respectively, about the stellar morphological major axis in the face-on view; and rotated by 30, 60 and $90\,^{\circ}$, respectively, about the stellar morphological minor axis in the face-on view. When constructing these projections we do not separate \hi\ from H$_2$ as Krumholz's prescription is strictly valid only when the gas is seen face-on. This does not affect our conclusions since we are mainly concerned with the kinematics of the gas, which is the same for all cold-gas phases.

All gas mass and surface density calculations as well the analysis of the \hi\ and stellar kinematics presented below are performed after removal of star and gas particles associated with satellite dark-matter sub-haloes according to the \textsc{subfind} algorithm of \cite{springel2001}. Furthermore, we exclude from the calculation of gas mass and surface density the discrete cold gas clouds which are a known artefact of smoothed-particle-hydrodynamics simulations and which are not part of the gas discs we wish to study \citep{sijacki2012}. These clouds, visible in some of the images discussed in Sec. \ref{sec:cosmomassmorph}, have a typical mass above $10^7$ \msun\ and may therefore have a significant impact on the value of \mhi\ if included in the calculation. 

Before showing the result of the comparison between simulated and real ETGs we note that the \atlas\ sample is volume limited and complete down to $M_\mathrm{K}=-21.5$ mag \citepalias{cappellari2011a}. On the contrary, the sample of simulated galaxies is not volume limited (it is a set of individual simulations rather than a single, large simulated volume) and does not follow the observed galaxy luminosity function. Therefore, we cannot compare the statistical distribution of real and simulated galaxy properties. Instead, we investigate whether the diversity of ETG \hi\ properties is reproduced by the simulations.

\begin{figure*}
\includegraphics[width=18cm]{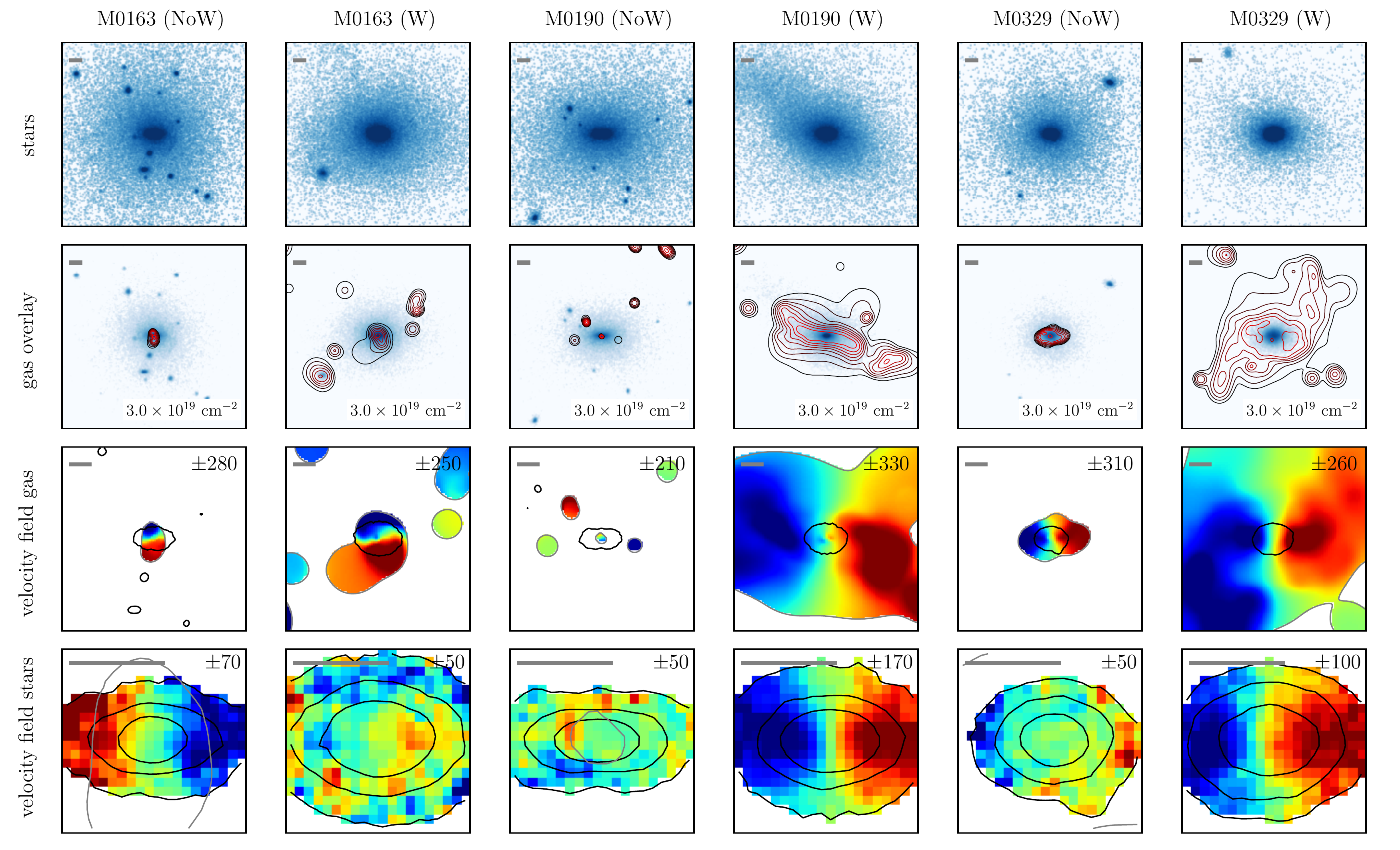}

\includegraphics[width=17.6cm]{colorbar.pdf}

\caption{Three examples of simulated galaxies. For each galaxy we show the result of the NoW and the W simulation next to each other to highlight their differences. Each column represents one simulated object -- four rows per galaxy -- viewed at an inclination of $60\,^{\circ}$. The first row corresponds to an extremely deep optical image ($\sim 29$ mag arcsec$^{-2}$ in $V$ band). The remaining rows are equivalent to those shown in Fig. \ref{fig:allvelfield} for real ETGs. Namely, the second row corresponds to a SDSS-like optical image with \hi\ contours overlaid; the third row shows the intensity-weighted \hi\ velocity field; the last row shows the intensity-weighted stellar velocity field. The linear scale of the various panels is the same as in Fig. \ref{fig:allvelfield}, with the grey scale-bar representing 5 kpc. Galaxies are sorted according to decreasing dark-matter halo mass (left to right). The outer stellar isophote corresponds to \rm a stellar mass surface density of 500 \msun\ pc$^{-2}$. \rm The colour scheme of the velocity fields is represented by the horizontal colour bar at the bottom of the figure. \rm}
\label{fig:simulated1}
\end{figure*}

\subsection{\hi\ mass and morphology}
\label{sec:cosmomassmorph}

The first step of our comparison is presented in Fig. \ref{fig:cosmomass}, where we plot \mhim\ against $M_\star$ for galaxies in the two sets of simulations.  Stellar mass values are taken from \citetalias{naab2013} for the NoW simulations and are measured in an identical way for the W simulations. For comparison we show the distribution of real ETGs distinguishing between \hi\ detections (blue density field) and non-detections (red dots). We also show the \mhim\ histogram of spiral galaxies. This is obtained from the \mhil\ histogram shown in \citetalias{serra2012a} assuming that for $M_\star>10^{10}$ \msun\ the typical spiral has $M_\star/L_\mathrm{K}=0.8$ \msun/\lsun\ \citep{bell2003}.


\begin{figure*}
\includegraphics[width=18cm]{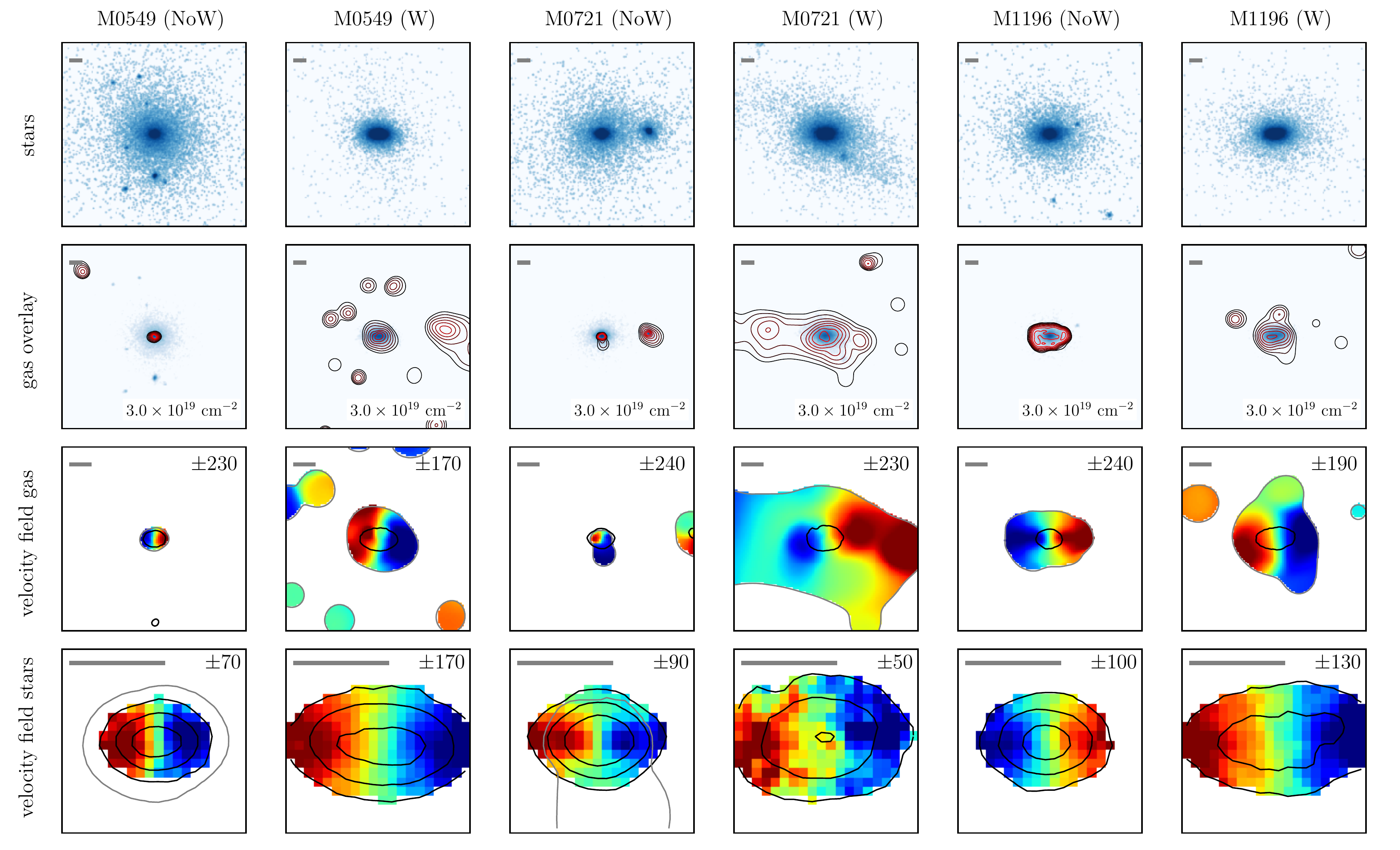}

\includegraphics[width=17.6cm]{colorbar.pdf}
\caption{Three more examples of simulated galaxies. See caption of Fig. \ref{fig:simulated1}.}
\label{fig:simulated2}
\end{figure*}

The figure shows that the NoW  simulations produce galaxies whose \hi\ content is to first order comparable to that of real ETGs and lower than that of spirals. The main discrepancy between NoW simulated and real ETGs is that extremely gas-poor galaxies with \mhim\ below a few times $10^{-4}$ are underrepresented in the simulations. This results in an overall smaller range of \mhim\ at any given $M_\star$ compared to the observations. Nevertheless, both the fact that such range is large in the simulations and that the typical \mhim\ decreases with $M_\star$ (above a stellar mass of $\sim5\times10^{10}$ \msun) are important points of agreement with the observed ETG sample.

The situation is different for the W  simulations. These galaxies are typically gas-richer than NoW  ones \citep{hirschmann2013}. They populate the high-\mhim\ end of the observed ETG distribution. At $M_\star>10^{11}$ \msun\ we find a number of objects whose \hi\ content is too large by a factor of 10 to 100 compared to the data. We conclude that W  systems are, as a family, too \hi\ rich and that NoW galaxies are a better (albeit not perfect) match to real ETGs in terms of \hi\ mass.

Having established this we can now analyse the \hi\ morphology of the simulated galaxies. We show three examples in Fig. \ref{fig:simulated1} and another three in Fig. \ref{fig:simulated2}. For each galaxy we show the NoW and the W simulation next to each other. These images and velocity fields are immediately comparable to those in Fig. \ref{fig:allvelfield} and are drawn adopting the same linear scale and gas contour levels. The only difference between observed and simulated galaxy figures is that for the latter we also show the equivalent of a deep optical image (top panel, see figure caption).

Figures \ref{fig:simulated1} and \ref{fig:simulated2} are meant to show a few cases representative of the diversity of galaxies produced by the simulations as well as the difference between the NoW and W simulation for a same halo. Indeed, it is immediately obvious that changing the star formation feedback implementation has a strong impact on the properties of a simulated galaxy. The same object can be a FR in the NoW simulation and a SR in the W one (M0163) or vice versa (M0190, M0329). Furthermore, it can be surrounded by a large gas distribution in the W simulation but have the gas disc entirely confined inside the stellar body in the NoW simulation (e.g., M0190, M0721). In fact, the larger size of the \hi\ distribution in the W simulations is a general result valid for all galaxies and a consequence of the mechanical nature of the W feedback, \rm which pushes gas to large radius while keeping it cold. \rm A visual comparison between Fig. \ref{fig:allvelfield} and Figs. \ref{fig:simulated1} and \ref{fig:simulated2} shows that these large W gas discs match reasonably well the size of the \hi\ discs and rings in gas-rich ETGs. \rm They too, like the observed systems, exhibit slight warps in the outer regions, possibly owing to their large size. \rm The NoW gas distributions are on the contrary too small.

\begin{figure}
\includegraphics[width=8.5cm]{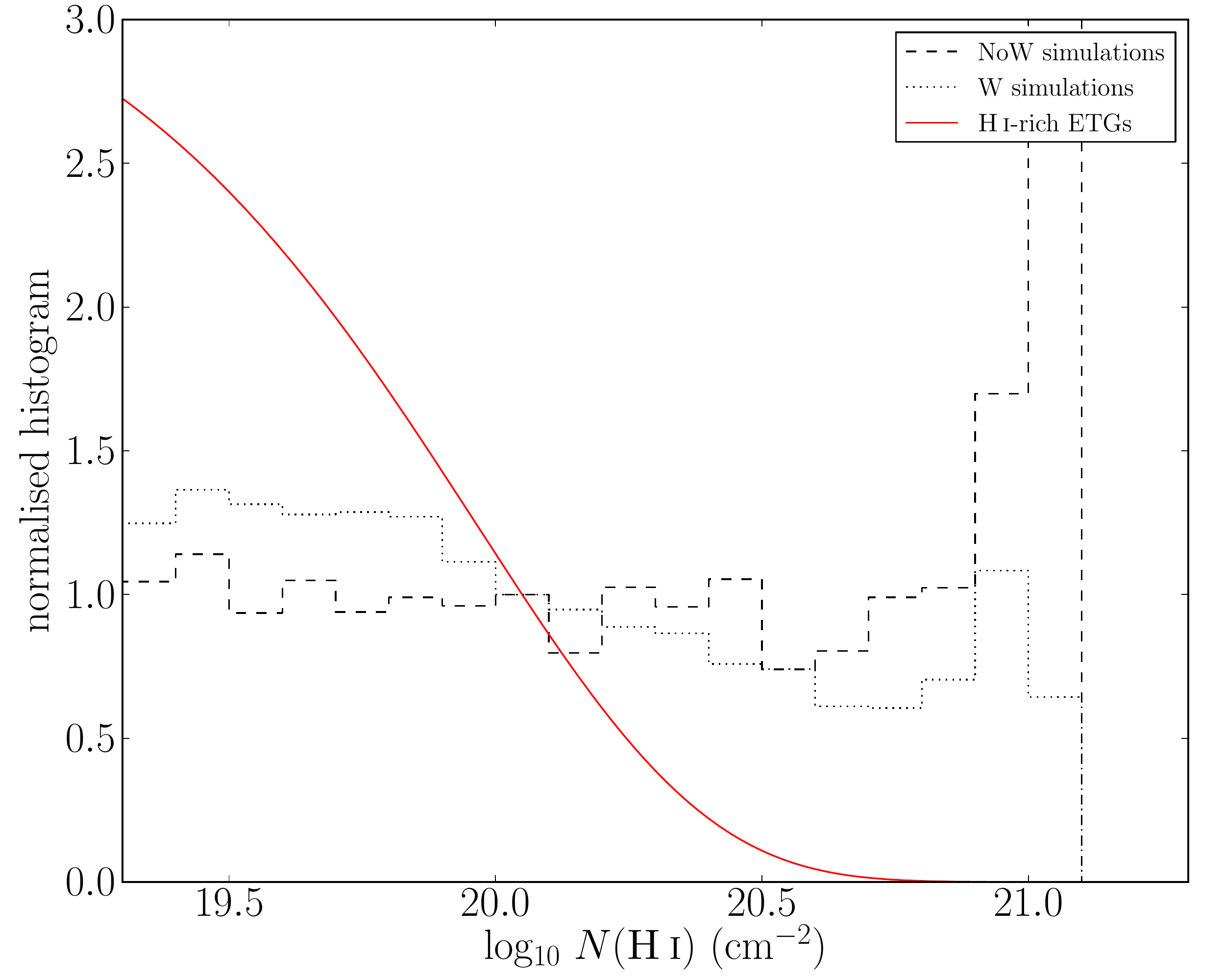}
\caption{\hi\ column density distribution in the simulations compared to that in real ETGs. \rm The latter is represented by the best-fitting Schechter function derived in \citetalias{serra2012a}\rm . For both NoW and W simulations we plot the median of the normalized distributions of all individual galaxies. The distributions are obtained for the cold-gas face-on view. Both median distributions and the Schechter function are scaled to intersect at $N$(\hi) $\sim10^{20}$ cm$^{-2}$. The peak at (and sharp drop after) $N$(\hi) $=10^{21}$ cm$^{-2}$ is due to our definition of \hi\ in the simulations (see the text). \rm This effect is more pronounced in the NoW case because the distribution of \it total \rm gas surface density extends to larger values than in the W simulations. \rm }
\label{fig:cosmodens}
\end{figure}

The fact that the NoW simulations are a better match to the observed ETG \mhi\ distribution (Fig. \ref{fig:cosmomass}) but that W objects are a better match to the observed size of the \hi\ discs may seem contradictory. In fact, this is easily understood noting that in all simulations the \hi\ column density is on average higher than observed. For this reason \hi\ discs with the right mass are too small (NoW) and \hi\ discs with the right size are too massive (W). This result, which is clear from a comparison of the \hi\ contours in Fig. \ref{fig:allvelfield} with those in Figs. \ref{fig:simulated1} and \ref{fig:simulated2}, is shown more quantitatively in Fig. \ref{fig:cosmodens}. \rm In this figure the black lines represent the \hi\ column density distribution in simulated galaxies while the red solid line shows the distribution in real galaxies (represented by the best-fitting Schechter function derived in \citetalias{serra2012a}). \rm This mismatch is consistent with the fact that simulated galaxies appear to be forming stars at a rate higher than real ETGs at $z=0$ \citep{hirschmann2013}.

Although not the focus of this study, it is worth noting in this context that all simulated galaxies host cold gas above a column density of $10^{21}$ cm$^{-2}$. This gas, which we do not include in our calculation of \hi\ mass and column density, is more centrally concentrated than the \hi\ and can be related to the molecular gas found in the central region of ETGs \citepalias{young2011}. Its mass is typically between $5\times10^7$ and $5\times10^9$ \msun\ with a handful of exceptions at lower mass and no significant difference between NoW  and W  simulations. Both sets of simulations miss therefore the large population of ETGs with $M$(H$_2$) $<5\times10^7$ \msun\ reported in \citetalias{young2011} while on the gas-rich side of the distribution the molecular gas masses are in reasonable agreement with the observations. We find that the molecular-to-total cold-gas fraction is larger in the NoW than in the W simulations. This is a consequence of their higher average cold-gas column density coupled with our method for separating atomic from molecular gas.



\subsection{\hi\ discs and misalignment in simulated fast and slow rotators}
\label{sec:misalsim}

Figures \ref{fig:simulated1} and \ref{fig:simulated2} show that not only the \hi\ mass and morphology but also the gas kinematical misalignment is affected by the feedback implementation. For example, M0549 hosts a small counterrotating gas disc in the NoW simulation and a large, corotating one in the W simulation; M0721 hosts a small corotating disc in the NoW simulation and a very large, counterrotating one in the W simulation. This kind of difference between the \hi-misalignment of NoW and W galaxies is very common within the sample of simulations analysed here. It is caused by the cumulative effect of galactic stellar winds on a galaxy and the satellites which it accretes as the simulation progresses.

Taken together, the NoW and the W simulations seem to  reproduce qualitatively the large variety of gas morphologies and kinematics of real \hi-rich ETGs. In this sense it is useful to associate some of the examples in Figs. \ref{fig:simulated1} and \ref{fig:simulated2} to ETGs in the \atlas\ sample. Bearing in mind that these matches are not exact and may not hold at a more quantitative level (as we illustrate below), M0163-NoW is for example a FR with a small polar gas disc which may be related to NGC~3499. M0163-W, a bright SR with a polar gas disc slightly larger than the stellar body, could be compared to NGC~3414. M0549-NoW hosts a small, counterrotating gas disc similar to NGC~3032. M1196-NoW and many W galaxies (e.g., M0190, M1196) represent cases of large, corotating \hi\ discs around FRs (similar for example to NGC~5582 and NGC~3838). M0329-W exhibits an \hi\ warp and tails comparable to those of NGC~4278. Counterrotating gas discs around FRs (e.g., NGC~6798) can be associated with M0721-W. The same galaxy simulated with the NoW feedback shows a small, corotating gas disc similar to the vast majority of the $d$ ETGs (e.g., NGC~4710, NGC~4150). This diversity is similar to the observed one and includes also some cases of unsettled \hi\ distributions, as observed \citepalias{serra2012a}.

Yet, the match between simulations and observations is far from perfect. In addition to the points of friction mentioned above with respect to mass and column density distributions, the \atlas\ sample includes \hi-rich ETGs not represented in the simulations such as SRs (NGC~3522) and FRs (NGC~2594, NGC~3998) with large polar rings. Furthermore, some simulated systems seem to have no corresponding real galaxy, such as the SRs with small, aligned gas discs (e.g., M0329-NoW). These objects may however be related to the SRs with aligned ionized-gas discs discussed in \citetalias{davis2011b}. This, together with the lack of central \hi\ depressions (or narrow \hi\ rings) in the simulations, is a reminder that a more sophisticated treatment of gas physics may be needed in order to relate simulated to real galaxies.

We conclude this section by comparing simulated and observed misalignments on the \mhim\ versus $\lambda_\mathrm{R}/(0.31\sqrt{\epsilon})$ diagram. The complexity of the observed ETG \hi\ properties is summarized in Fig. \ref{fig:mhille_misal}. That figure highlights that \hi\ discs around FRs are very diverse in terms of their kinematical misalignments from the stars. Furthermore, it shows that SRs too can host an \hi\ disc or ring and that this is typically polar relative to the stars. To investigate whether a similar picture emerges from the simulations we construct a figure equivalent to Fig. \ref{fig:mhille_misal}. The basis of this analysis is the simulated stellar and cold-gas velocity fields represented by the examples in Figs. \ref{fig:simulated1} and \ref{fig:simulated2}, which are obtained on a projection such that the inclination of the stellar body is $60\,^{\circ}$.

Measurements of PA(stars) are straightforward as nearly all simulated galaxies have their kinematical major axis aligned along the $X$-axis. In order to measure PA(\hi) as a function of radius we make use of the \textsc{rotcur} task part of the \textsc{gipsy} package \citep{vanderhulst1992}. This task fits tilted-ring models to a velocity field \rm and is equivalent to the software used for the analysis of real galaxies (see Sec. \ref{sec:misal})\rm . In this case we fix the centre of the rings to the stellar centre of mass and the \hi\ systemic velocity to the stellar value. We then solve for PA(\hi), inclination and rotation velocity of all rings. In what follows we analyse PA(\hi) only.

\begin{figure*}
\includegraphics[width=18cm]{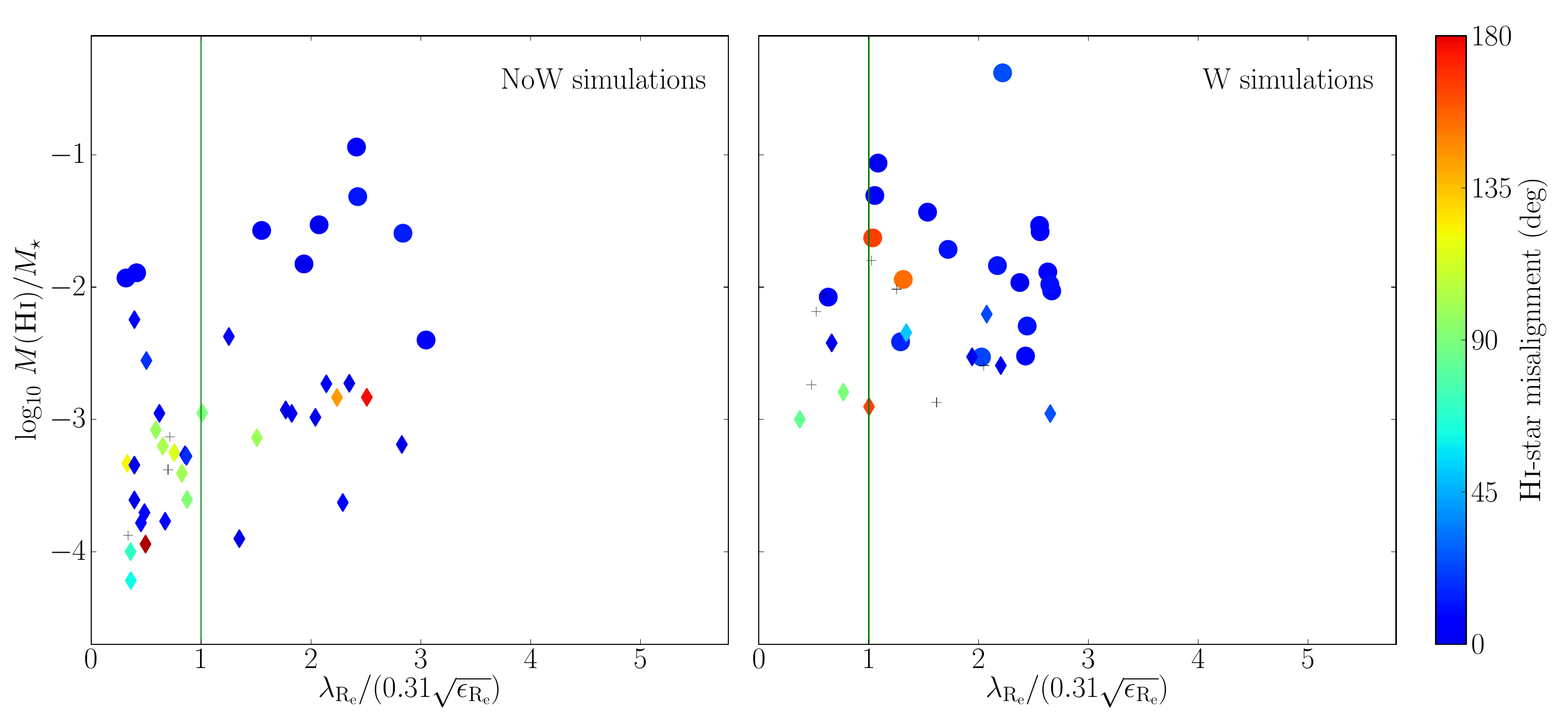}
\caption{\rm \hi\ kinematical misalignment of the simulated galaxies on the \mhim\ versus $\lambda_\mathrm{R}/(0.31\sqrt{\epsilon})$ diagram. We show NoW simulations in the left-hand panel and W simulations in the right-hand panel. The small grey crosses indicate simulations where the gas is not settled and therefore we do not measure the misalignment angle. See the caption of Fig. \ref{fig:mhille_misal} for more details on the figure. Note that we show these figures with the same axes limits as in Fig. \ref{fig:mhille_misal} to facilitate the comparison between simulated and real galaxies.\rm}
\label{fig:simulated_mhille_misal}
\end{figure*}

We summarize the result of this analysis in Fig. \ref{fig:simulated_mhille_misal}, where the values of $\lambda_\mathrm{R}$ and $\epsilon$ are measured on the edge-on projection. These values are taken from \citetalias{naab2013} for the NoW galaxies and are measured in the same way for the W objects. Galaxies with an \hi\ radius smaller than $3\ R_\mathrm{e}$ are shown as diamonds. Values of $R_\mathrm{e}$ are also taken from \citetalias{naab2013} and measured in the same way for the W models. The figure shows that, as discussed above, the simulations are able to make \hi\ discs around both FRs and SRs, in agreement with the result of Sec. \ref{sec:lambdaeps}. As noted above, it is also important that in some of these objects the \hi\ is misaligned (e.g., polar or counterrotating) relative to the stellar kinematics. However, the similarities between simulated and real ETGs on this diagram do not go much further than these qualitative statements.

A first obvious difference is that, compared to Fig. \ref{fig:mhille_misal}, the vast majority of both W and NoW FRs host \hi\ on an aligned configuration. There are a few exceptions including some of the FRs in Figs. \ref{fig:simulated1} and \ref{fig:simulated2}: M0163-NoW (polar) M0549-NoW (counterrotating) and M0721-W (counterrotating). However, strongly misaligned systems are confined to much lower levels of stellar rotation \rm and gas content \rm compared to some of the observed \hi\ counterrotators (NGC~3626, NGC~3941, NGC~5103 and NGC~6798) and polar rings (NGC~2594, NGC~3998). In the W simulations the only misaligned FRs lay, in fact, close to the empirical boundary between FRs and  SRs. The kinematical alignment between \hi\ and stars in all the other, fasterrotating simulated galaxies makes them more similar to spirals than to real FRs as a family. \rm Such stronger coupling between gas and stars compared to real ETGs may be related to the larger gas surface density and star formation rate in the simulations, as highlighted in Sec. \ref{sec:cosmomassmorph}.\rm


Rewinding the simulations reveals that the rare cases of large \hi\ misalignment in FRs at $z=0$ are usually the result of very recent events. For example, M0721-W (Fig. \ref{fig:simulated2}) spends most of its life as a FR with aligned \hi, until a merger with a relatively small but very gas-rich satellite at $z\sim0.3$ turns it into an \hi\ counterrotator. The \hi\ tail in Fig. \ref{fig:simulated2} is a remnant of that event. Another example is M0549-NoW (also in Fig. \ref{fig:simulated2}). This system is surrounded by a large, warped gas disc for a large fraction of its life. Gas in the inner part of the disc is kinematically aligned with the inner stellar rotation except for some short intervals where it appears to be close to polar. The gas disc mass and size decrease as the galaxy evolves and by $z\sim0.1$ the disc is so small that a relatively minor gas accretion event results in the current counterrotating \hi\ disc. Similarly, the FR M0163-NoW (Fig. \ref{fig:simulated1}) has been hosting a polar gas disc only for the last $\sim2$ Gyr, and in the past has been both a gas co- and counterrotator. Given that the \hi\ orbital time in \atlas\ galaxies is at most $\sim1$ Gyr it is possible that the ETGs with misaligned \hi\ have acquired their gas relatively recently.

On the contrary, FRs with an aligned (although frequently warped) gas disc at $z=0$ have typically been in such a configuration for most of their life, with occasional periods of time during which the gas is unsettled following a merger. This is true for $\sim80$ per cent of all NoW FRs and $\sim70$ per cent of all W FRs. Note that such a persistent alignment between stellar and gas kinematics is not the result of a lack of satellite accretion. On the contrary, these objects experience continuous merging (depending on their halo mass), but their gas disc is probably sufficiently massive at all times to prevent significant misalignments from developing. We do not investigate this aspect in more detail since our focus here is the comparison to observed ETGs at $z=0$ and not so much a detailed study of why simulated galaxies look the way they do.

Slightly different paths for the formation of an aligned FR are possible (although less frequent) in the simulations. \rm For example, M0721-NoW is a SR with very low-level rotation and a counterrotating gas disc at $z\sim1$. This gas lives relatively undisturbed for the next $\sim2$ Gyr and during this time forms a number of new stars sufficient to change the direction of net rotation of the stellar body and to turn the object into a FR\rm. So this is a case where rather than the gas disc changing its kinematics following gas accretion, it is the stellar body that changes its kinematics following star formation. Whether such evolutionary paths occur also in the real Universe remains to be established.

The other, already mentioned difference between Figs. \ref{fig:mhille_misal} and \ref{fig:simulated_mhille_misal} is the existence of simulated gas-rich SRs where the \hi\ is co- or counterrotating relative to the stars. These SRs have no counterpart in the \atlas\ \hi\ sample but may be related to the ionized-gas SR corotators of \citetalias{davis2011b}. Rewinding the simulations reveals that many of the SRs with a small corotating disc at $z=0$ used to have a polar disc (or sometimes a counterrotating one), which has settled only recently on an  aligned configuration. As noted above for the misaligned FRs, the \hi\ kinematical misalignment can change very frequently in the simulations, mostly following merging with satellites. Polar gas distributions like the observed ones are found too (e.g., M0163) but they are not the rule. Their history too includes frequent changes of the kinematical misalignment as a result of mergers.

The comparison presented in this section indicates that simulations are only partially successful at reproducing the observed \hi\ properties of ETGs. A careful treatment of gas physics is essential and it appears that a better match to the observations needs to be achieved before $\Lambda$CDM simulations can teach us why and how ETGs evolve into the diverse (\hi-wise) family they are today.

\section{Summary}
\label{sec:summary}

In a previous \atlas\ paper \citepalias{serra2012a} we demonstrate on a strong statistical basis that a large fraction of all ETGs outside the cluster environment host a significant mass of \hi\ gas. In particular, 1/4 of all ETGs outside the Virgo cluster host a disc or ring of low-column-density \hi\ with size from a few to tens of kpc, and mass from $\sim10^7$ to a few times $10^9$ \msun. Here we investigate the link between the \hi\ properties (mass and morphology) of these galaxies and their internal structure. The latter is determined on the basis of galaxies' optical morphology and stellar kinematics, leading to their classification as SRs or FRs. Furthermore, we study the kinematical misalignment between \hi\ (out to large radius) and stars (inside $\sim1\ R_\mathrm{e}$) in both these groups, and compare for the first time the \hi\ properties of SRs and FRs to predictions from $\Lambda$CDM hydrodynamical simulations.

We find a large diversity of \hi\ masses and morphologies within both the FR and SR families. Surprisingly, SRs are detected as often, host as much \hi\ and have a similar rate of \hi\ discs and rings as FRs. The only tentative difference between the two families is that the \hi\ discs/rings around SRs are usually not fully settled. This comparison is mostly limited by the small number of SRs in our sample (simply because these galaxies amount to just $\sim15$ per cent of the total ETG population) and a larger SR sample would be needed to improve the accuracy of some of the above conclusions. Nevertheless, our data imply that SRs can accrete a significant mass of cold gas during their life, and that they may do so at a rate similar to that of FRs. \rm This is confirmed by the frequent detection of cold dust \citep{smith2012}, kinematically-misaligned ionized gas \citepalias{davis2011b} and cuspy nuclear light profiles \citepalias{krajnovic2013b} in SRs as well by simulations, which indicate that at least some SRs may have formed during gas-rich mergers (\citetalias{bois2011}; \citetalias{naab2013}).\rm

The high detection rate of \hi\ discs/rings in galaxies with little or no evidence of an embedded stellar disc -- the SRs -- indicates that the accretion of \hi\ is not always linked to the growth of an inner stellar disc. Such weak relation between \hi\ and stellar disc is confirmed by the fact that in galaxies with a significant stellar disc component -- the FRs -- the \hi\ does not always corotate with the stars. On the contrary, we find corotation in just about half of all FRs with an \hi\ disc or ring. The remaining FRs exhibit a variety of kinematical misalignments including cases of polar and counterrotating gas. \rm The family of FRs appears therefore significantly more complex and diverse at large radius than in the inner regions, as concluded also using other kinematical tracers \citep[e.g.,][]{coccato2009,pota2013}.\rm

The \hi\ misalignments reveal another possible difference between FRs and SRs: in the latter the \hi\ disc/ring is always polar (or nearly so) relative to the stars. We do not find any SR with a co- or counterrotating \hi\ disc or ring, although we note that such gaseous discs are detected at smaller radius in the ionized gas phase \citepalias{davis2011b}.

In \citetalias{serra2012a} we show that the ETG \hi\ mass and morphology depend to some extent on galaxy stellar mass and environment density. Here we find no additional trends as the \hi\ kinematical misalignment does not seem to be related to the host stellar mass or environment in any obvious way. This complex picture highlights a diversity of ETG formation histories which may be lost in the relative simplicity of their inner structure.

We find that $\Lambda$CDM hydrodynamical simulations (both with and without galactic stellar winds) have difficulties reproducing the \hi\ properties of ETGs. We develop a simple method to define \hi\ in the simulations of \cite{oser2010} and \cite{hirschmann2013}, and find some clear inconsistencies with the observations. The decline of the \hi\ column density distribution is too shallow, resulting in an average density which is larger than in real ETGs. This causes gas discs with an \hi\ mass comparable to the observations to be too small, while discs whose size is sufficiently large are too massive. Furthermore, none of the simulations is able to produce the many gas-poor ETGs present in the observed sample.

Consistent with the observations, \hi\ discs are found in both simulated FRs and SRs. However, their kinematical misalignment from the stars match the observations only qualitatively. On the one hand, nearly all simulated FRs host corotating \hi. The few cases with polar or counterrotating gas exhibit very low levels of stellar rotation (unlike real ETGs). On the other hand, a large fraction of the simulated SRs host corotating \hi\ while none hosts a large, (nearly) polar \hi\ disc as in real SRs. We suggest that a more sophisticated treatment of gas physics and a better understanding of the corresponding feedback processes is needed in the simulations, and we conclude that a better match between observed and simulated galaxies should be achieved before cosmological simulations can be used to understand the origin of the complex \hi\ properties of ETGs.
 


\section*{Acknowledgments}

PS acknowledges support of a NWO/Veni grant. This work is based on observations obtained with the Westerbork Synthesis Radio Telescope, which is operated by ASTRON
(Netherlands Institute for Radio Astronomy) with support from the Netherlands Foundation for Scientific Research (NWO). MC acknowledges support from a Royal Society University Research Fellowship. This work was supported by the rolling grants ÔAstrophysics at OxfordÕ PP/E001114/1 and ST/H002456/1 and visitors grants PPA/V/S/2002/00553, PP/E001564/1 and ST/H504862/1 from the UK Research Councils. RLD acknowledges travel and computer grants from Christ Church, Oxford and support from the Royal Society in the form of a Wolfson Merit Award 502011.K502/jd. RLD is also grateful for support from the Australian Astronomical Observatory Distinguished Visitors programme, the ARC Centre of Excellence for All Sky Astrophysics, and the University of Sydney during a sabbatical visit. SK acknowledges support from the Royal Society Joint Projects Grant JP0869822. RMcD is supported by the Gemini Observatory, which is operated by the Association of Universities for Research in Astronomy, Inc., on behalf of the international Gemini partnership of Argentina, Australia, Brazil, Canada, Chile, the United Kingdom, and the United States of America. TN and MBois acknowledge support from the DFG Cluster of Excellence `Origin and Structure of the Universe'. MS acknowledges support from a STFC Advanced Fellowship ST/F009186/1. (TAD) The research leading to these results has received funding from the European Community's Seventh Framework Programme (/FP7/2007-2013/) under grant agreement no. 229517. MBois has received, during this research, funding from the European Research Council under the Advanced Grant Program no. 267399-Momentum. LY acknowledges support from NSF grant AST-1109803.  MH acknowledges financial support from the European Research Council under the European CommunityÕs Seventh Framework Program (FP7/2007-2013)/ERC grant agreement no. 202781. The authors acknowledge financial support from ESO.

\appendix

\section{\hi\ misalignment curves}

\rm 
Figure \ref{fig:allmisal} shows the variation of the kinematical misalignment PA(\hi) $-$ PA(stars) as a function of radius for the 26 galaxies with extended \hi\ distributions analysed using the \textsc{kinemetry} software \citep{krajnovic2006}. This includes all $D$s in the \atlas\ sample, all $R$s and the two $u$ galaxies with some settled gas, NGC~1023 and NGC~4026 (see Sec. \ref{sec:misal}). Note that PA(\hi) varies as a function of radius while PA(stars) is an average central value, as explained in detail in Sec. \ref{sec:misal}. The curves in the figure were used to determine the median misalignment values shown in Fig. \ref{fig:mhille_misal}. Note the severe warps of NGC~2685 and NGC~5173, which are discussed also in Sec. \ref{sec:misal}. For these two galaxies the small median misalignment angle is not representative of the true complexity of the gas system. The figure also shows the kinematical misalignment between ionized/molecular gas and stars presented in \citetalias{davis2011b}. The kinematical alignment between inner \hi\ and other gas phases is discussed in Sec. \ref{sec:misal1}. We refer to the caption of Fig. \ref{fig:allmisal} for more details on its content. \rm

\begin{figure*}
\includegraphics[width=5.5cm]{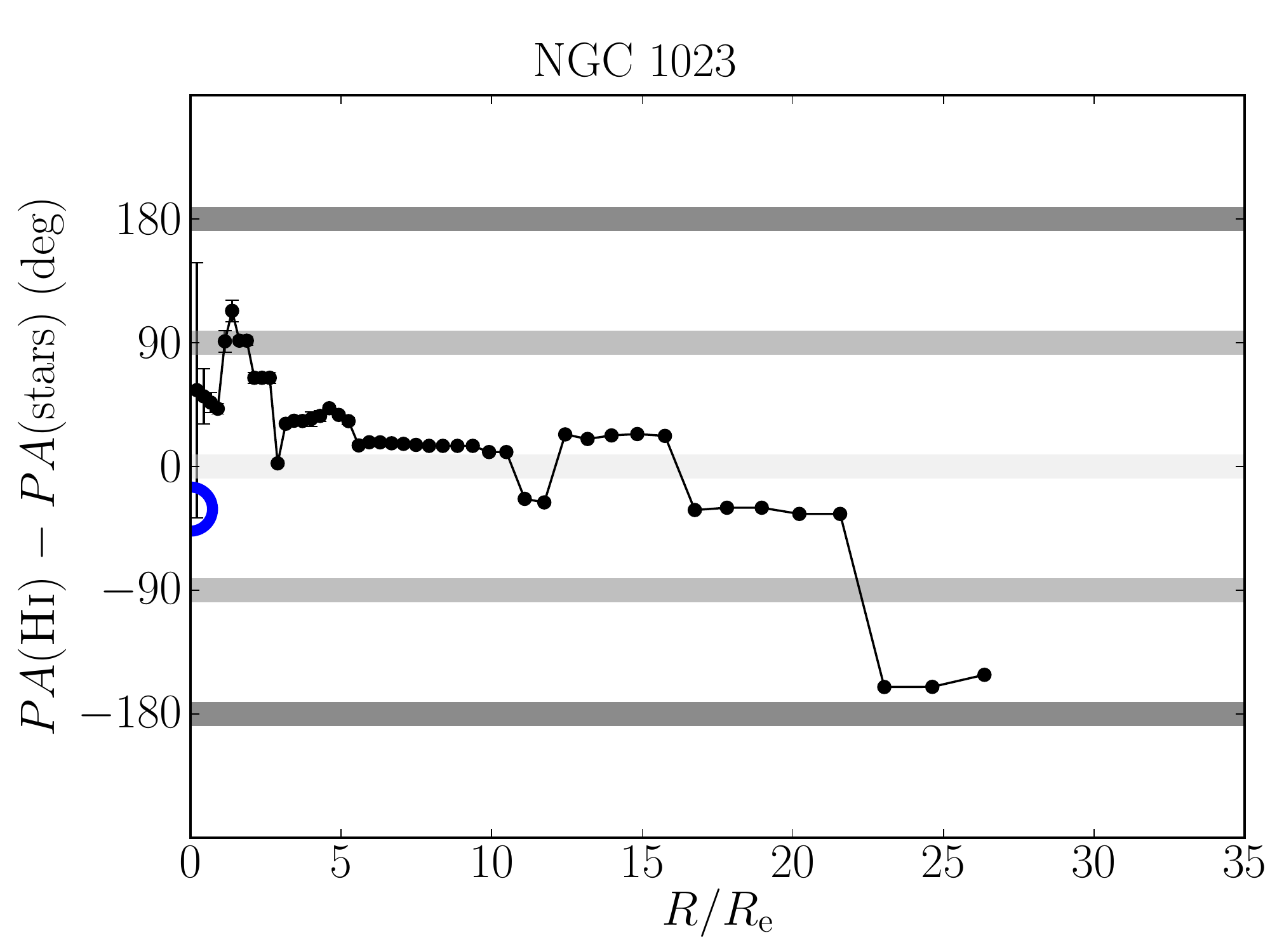}
\includegraphics[width=5.5cm]{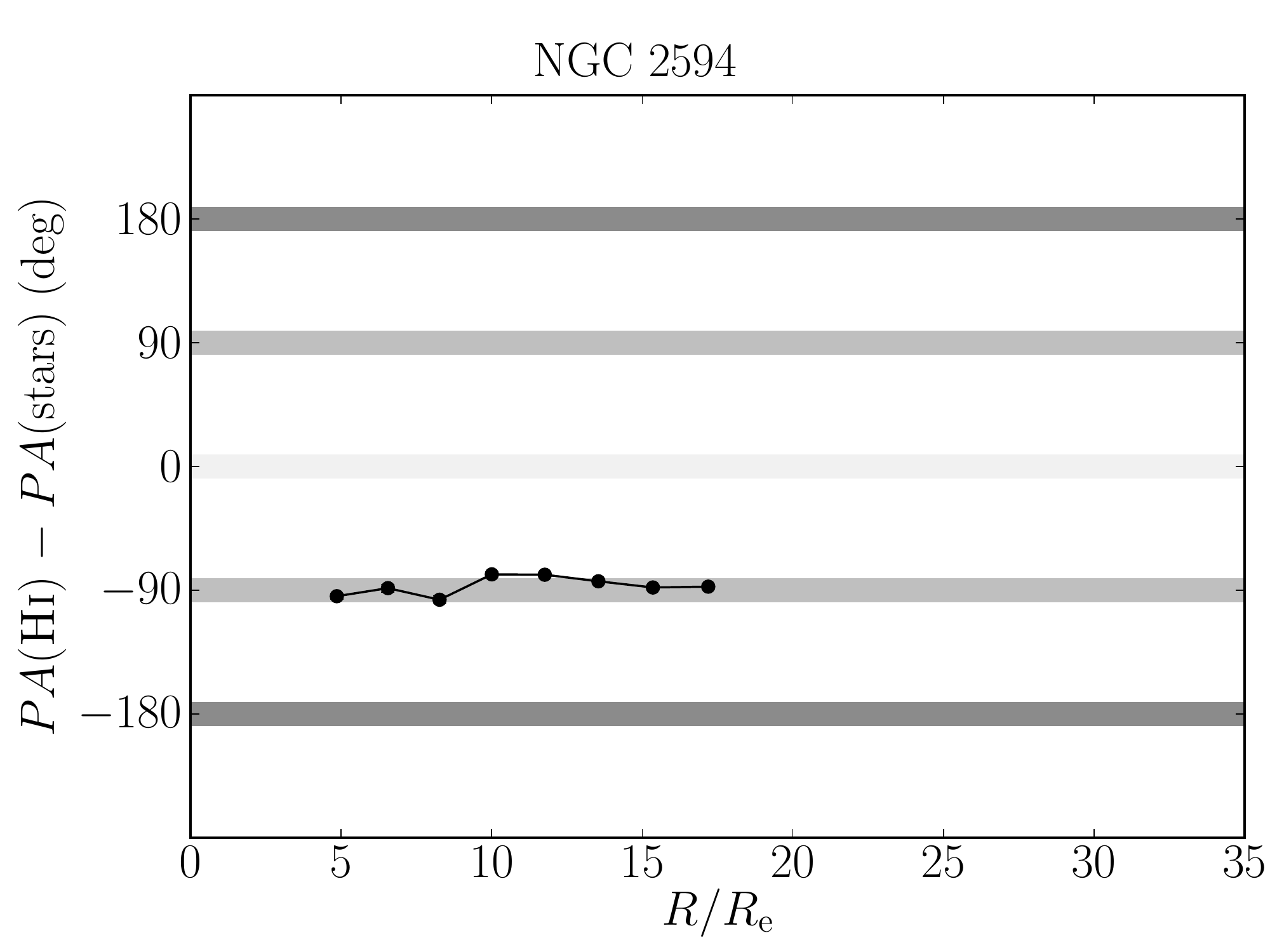}
\includegraphics[width=5.5cm]{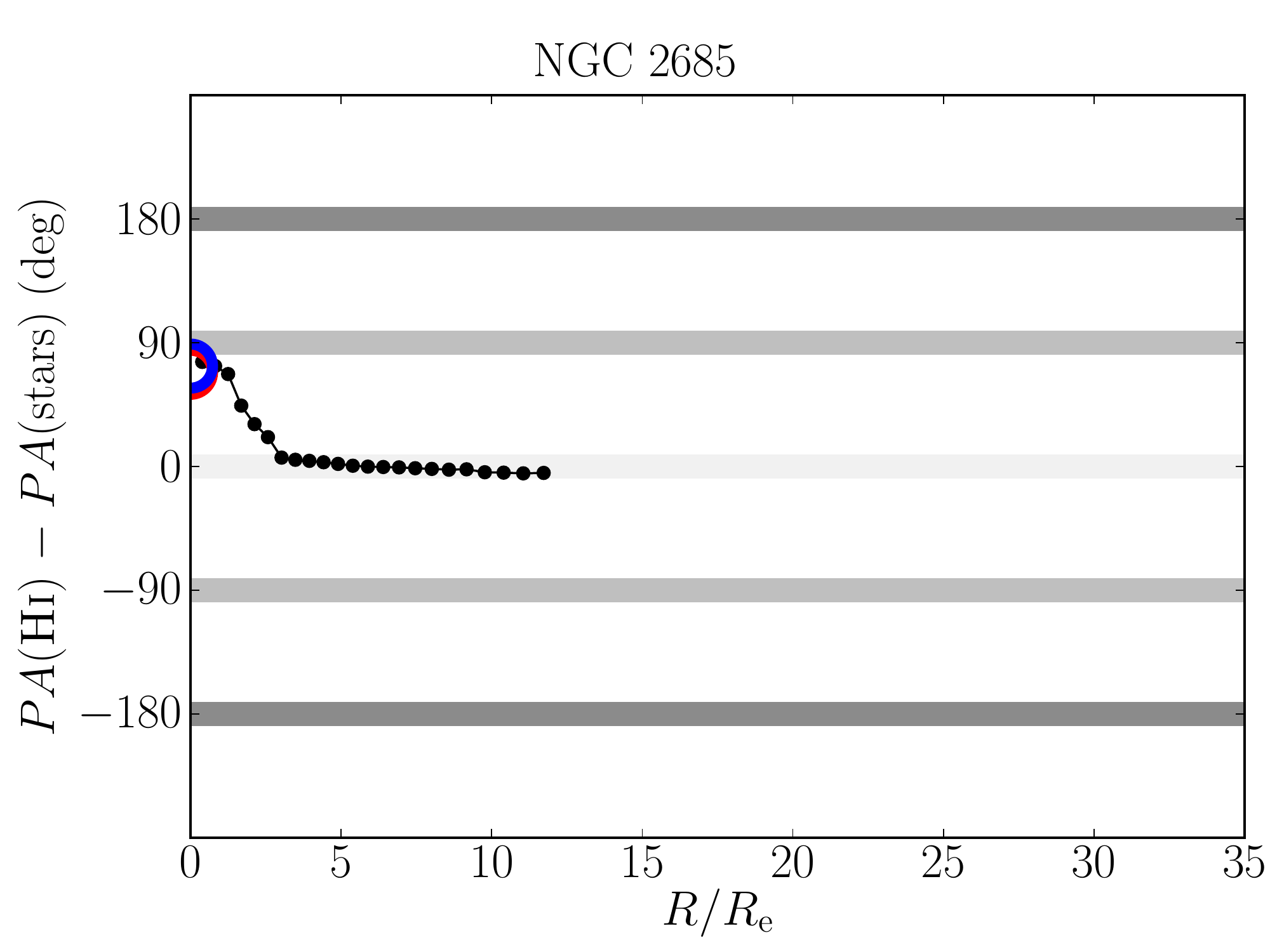}

\includegraphics[width=5.5cm]{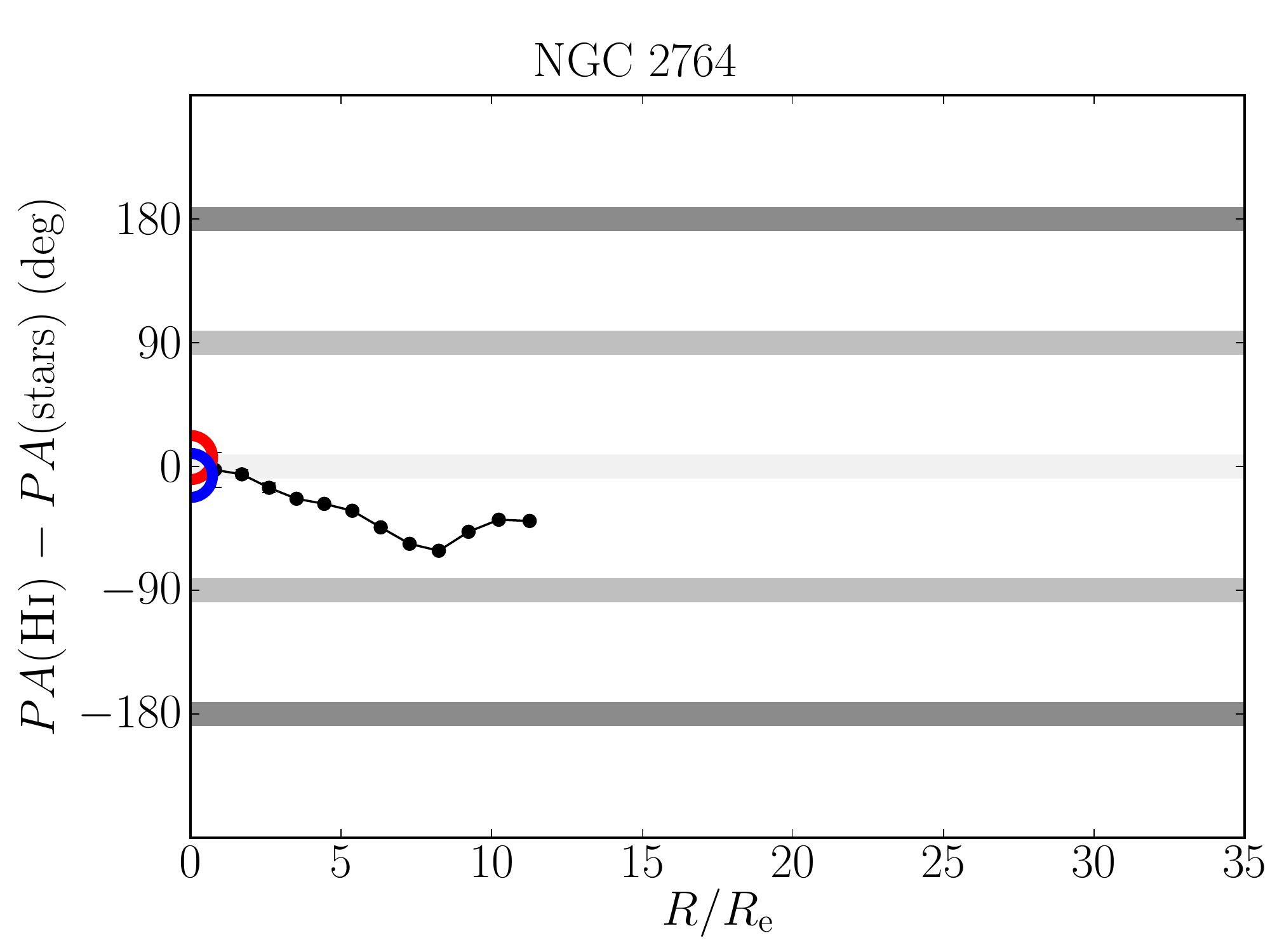}
\includegraphics[width=5.5cm]{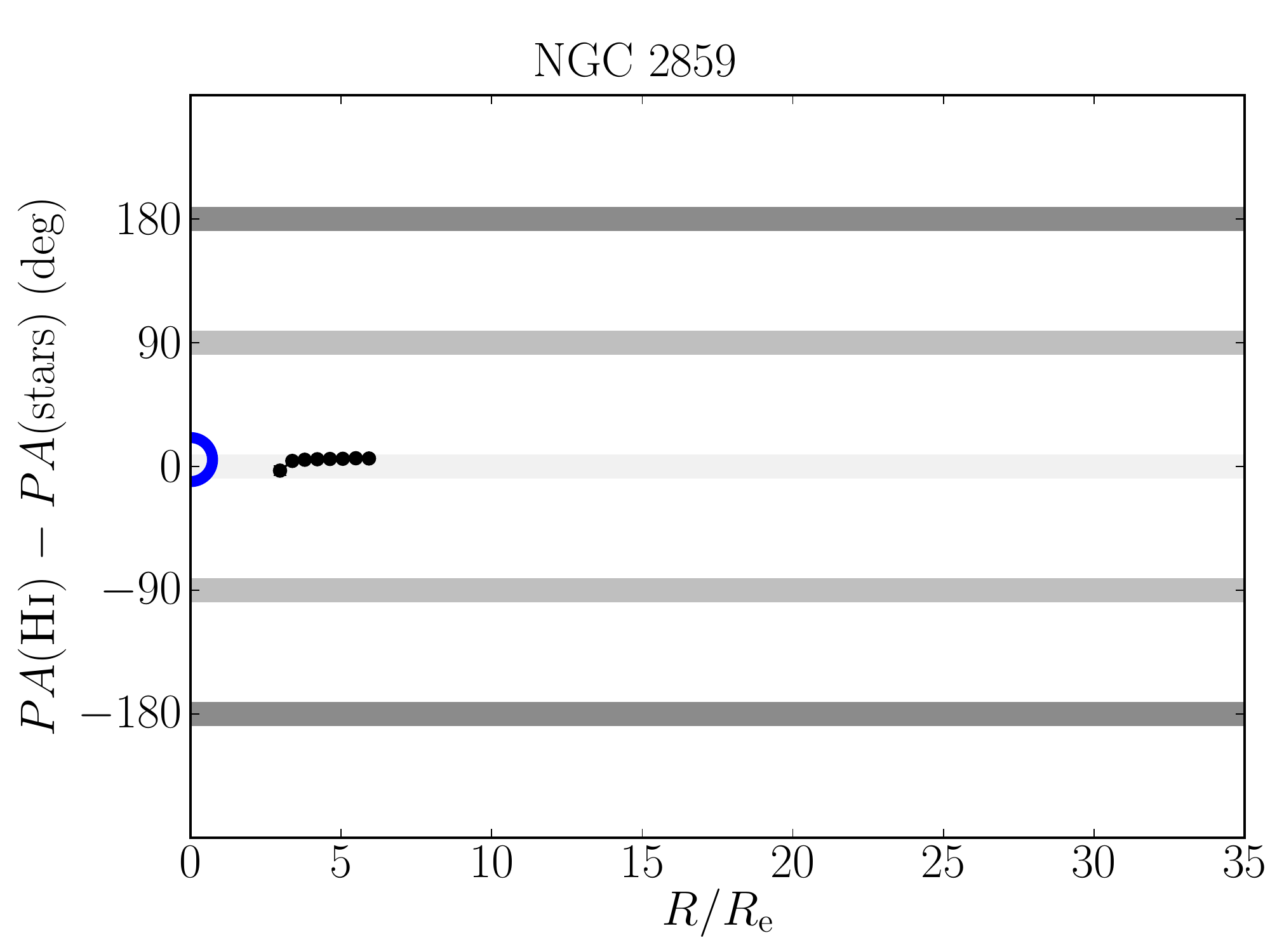}
\includegraphics[width=5.5cm]{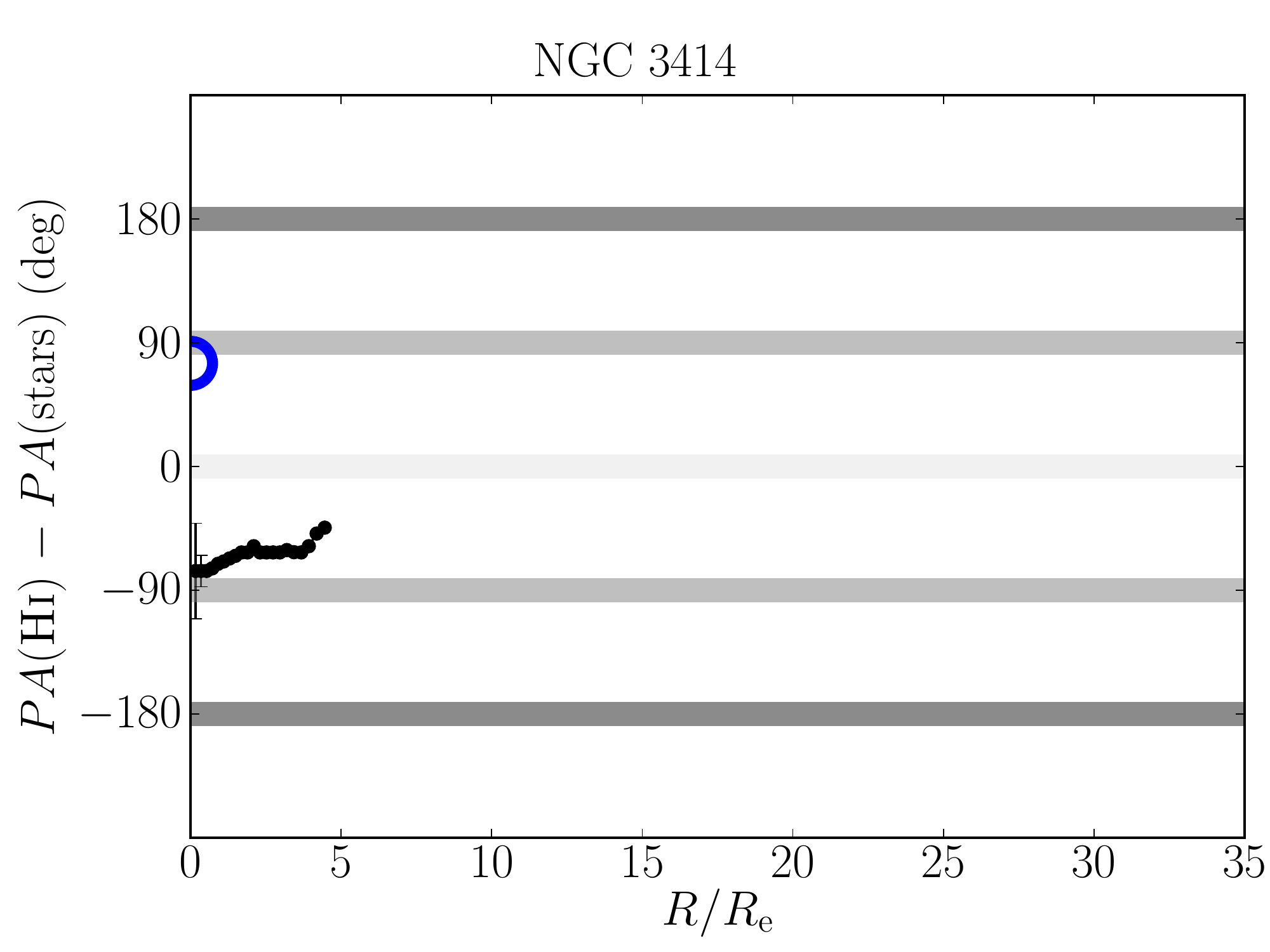}

\includegraphics[width=5.5cm]{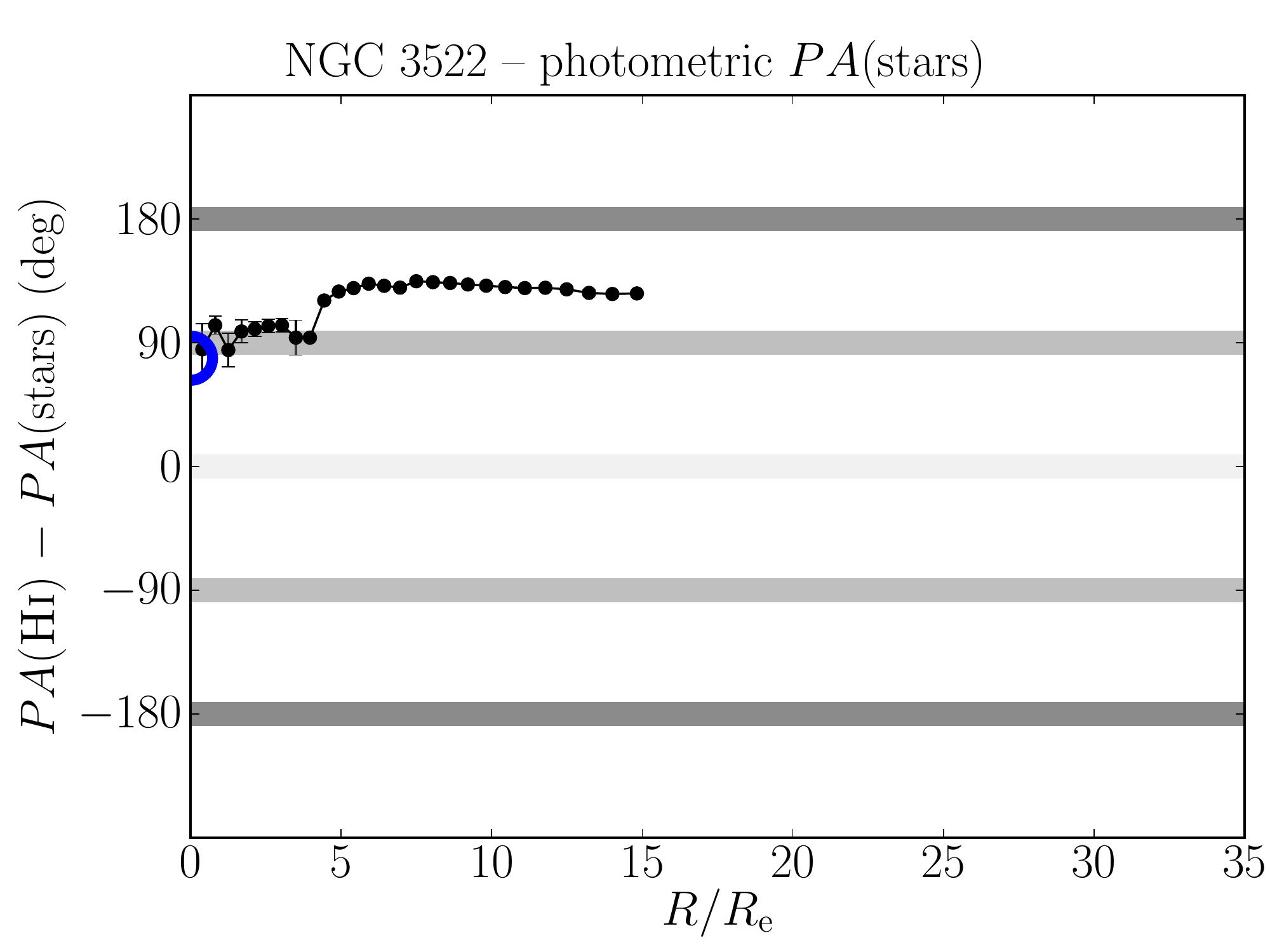}
\includegraphics[width=5.5cm]{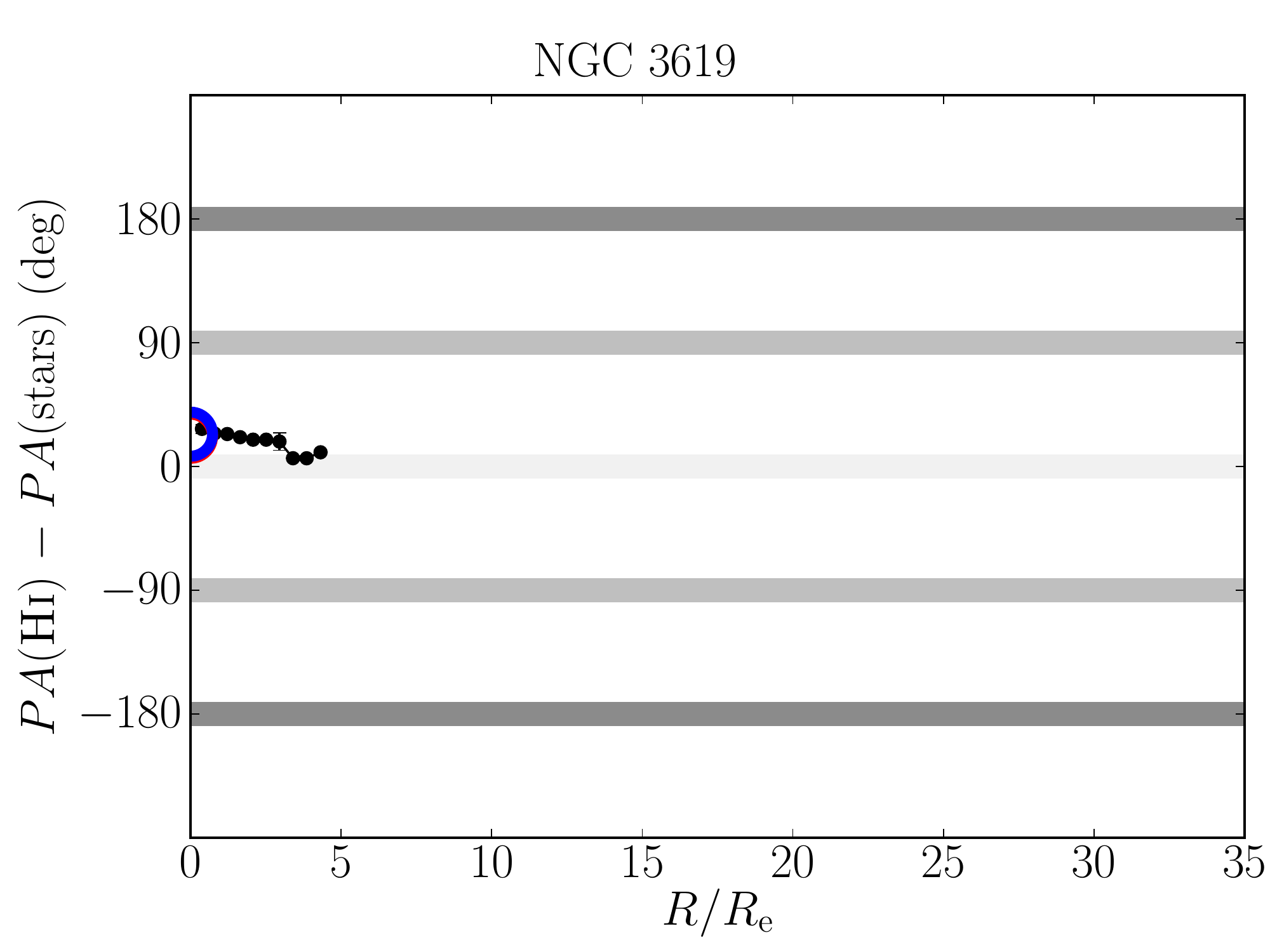}
\includegraphics[width=5.5cm]{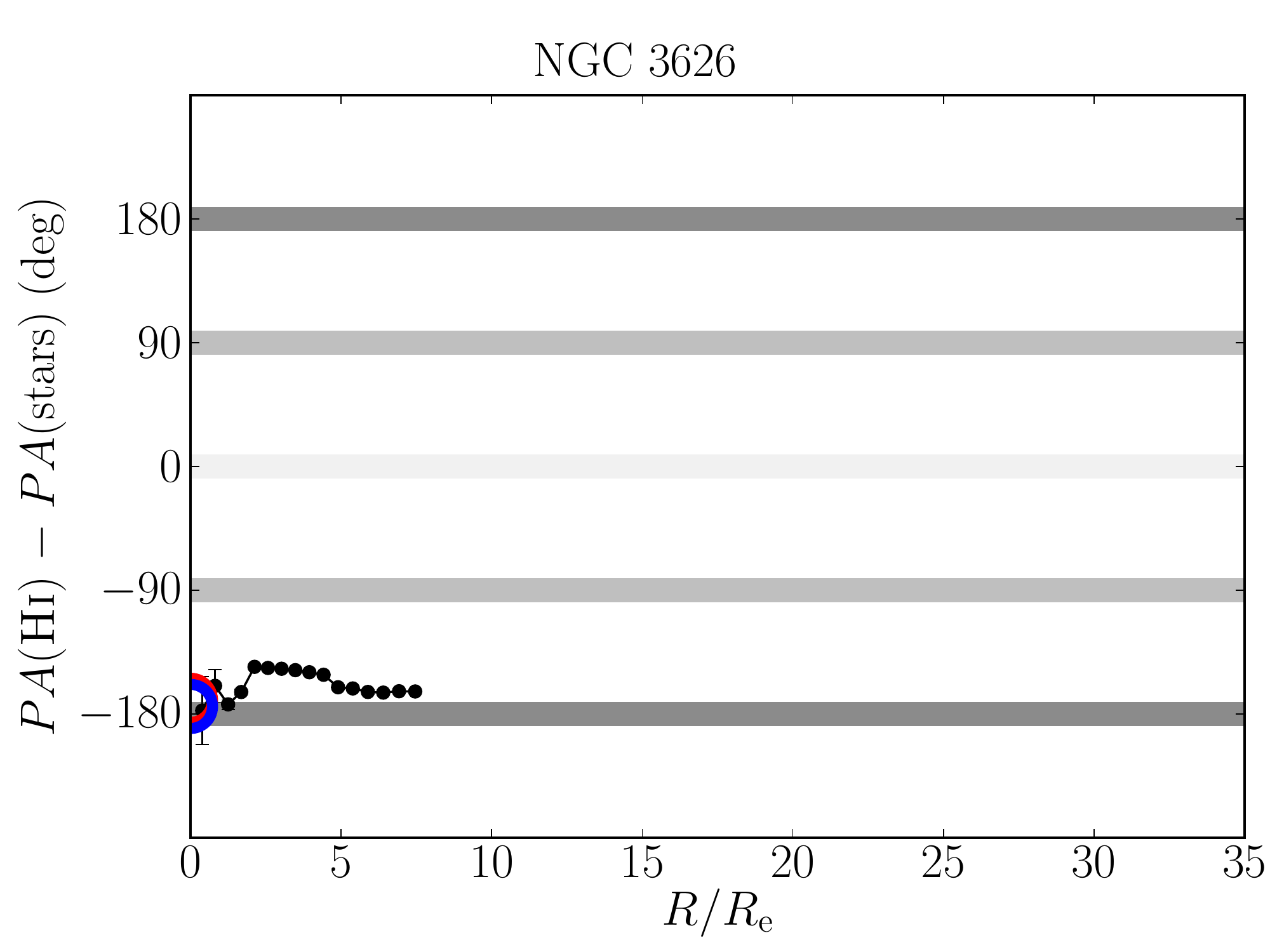}

\includegraphics[width=5.5cm]{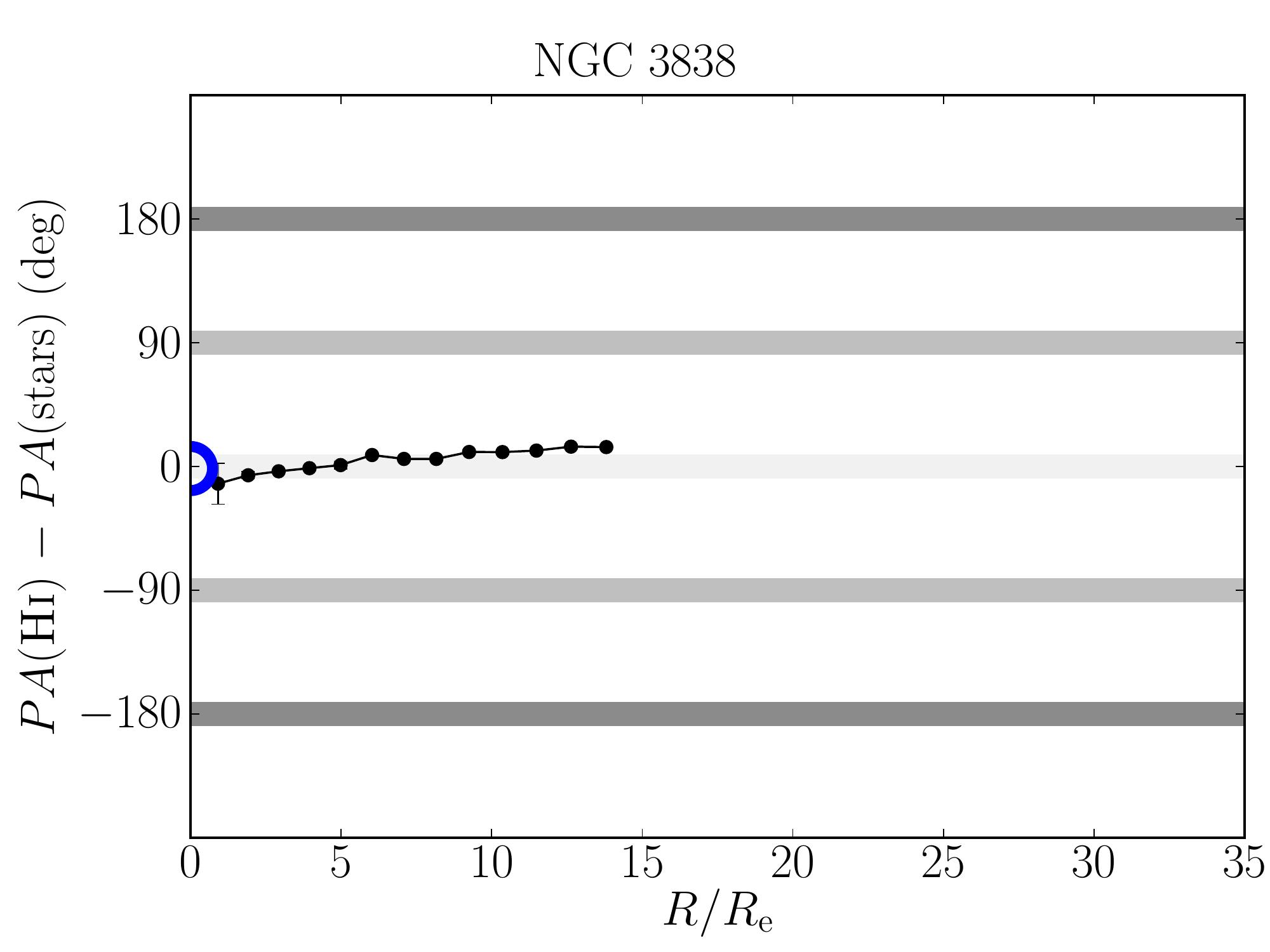}
\includegraphics[width=5.5cm]{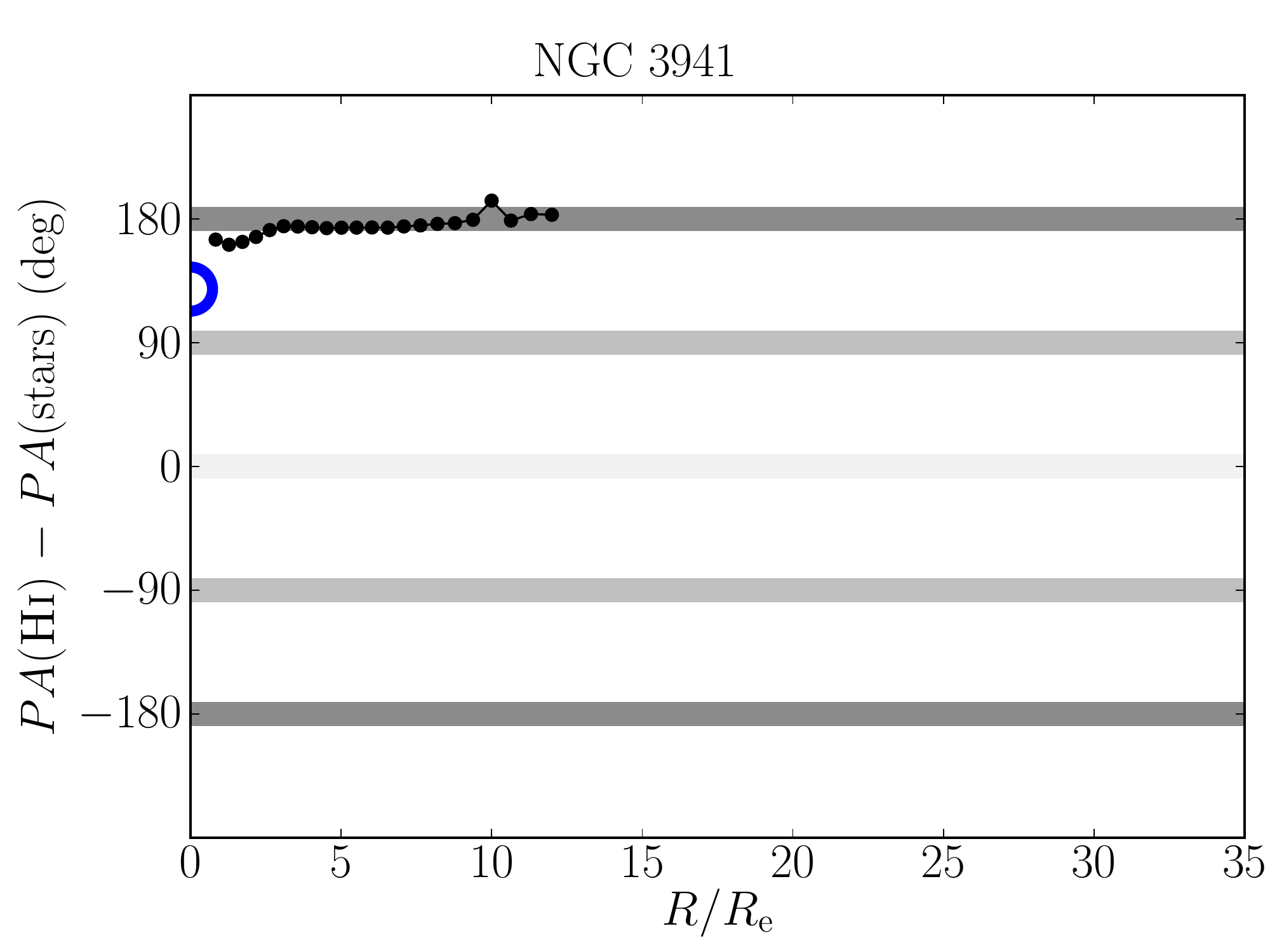}
\includegraphics[width=5.5cm]{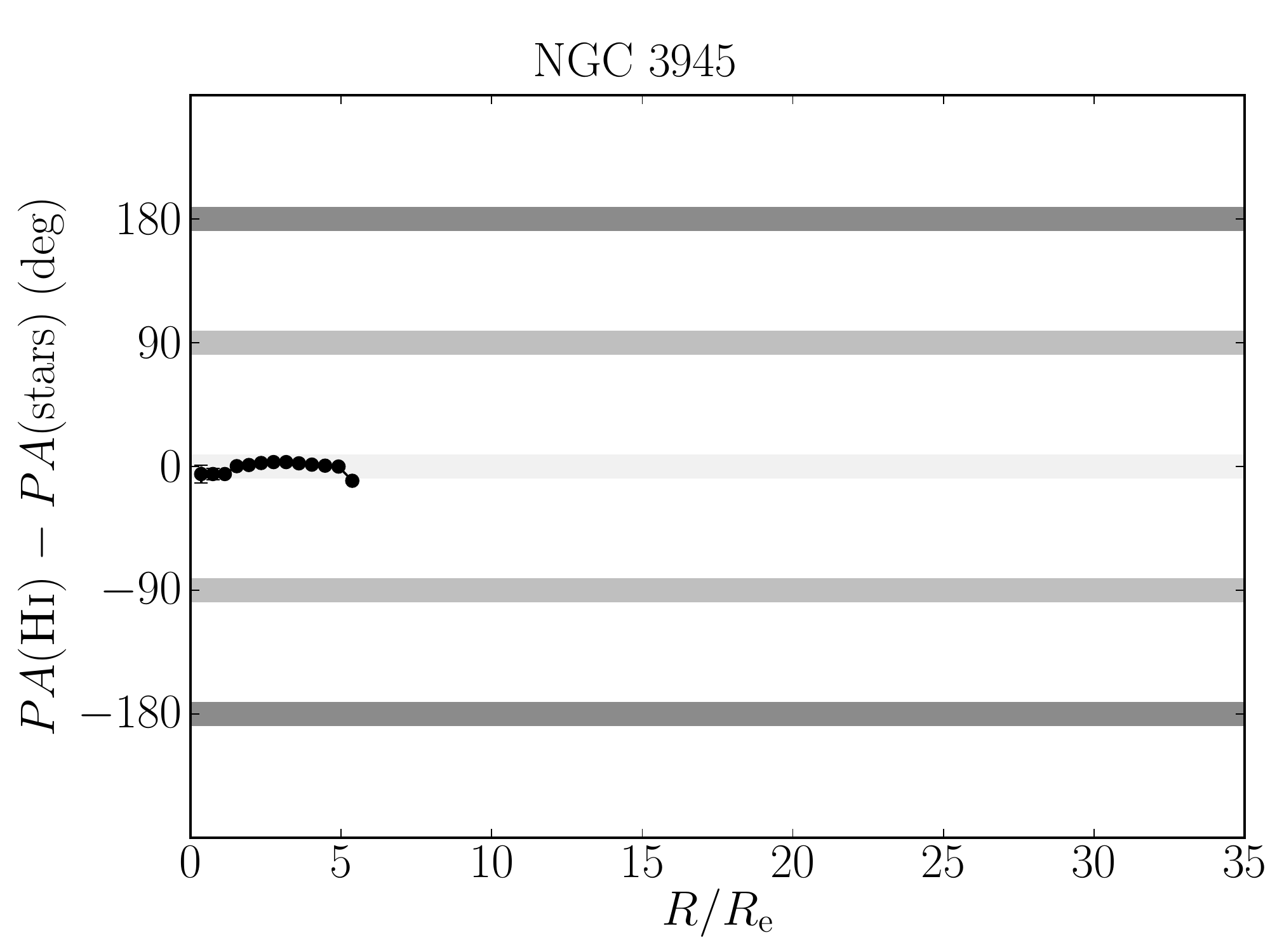}

\includegraphics[width=5.5cm]{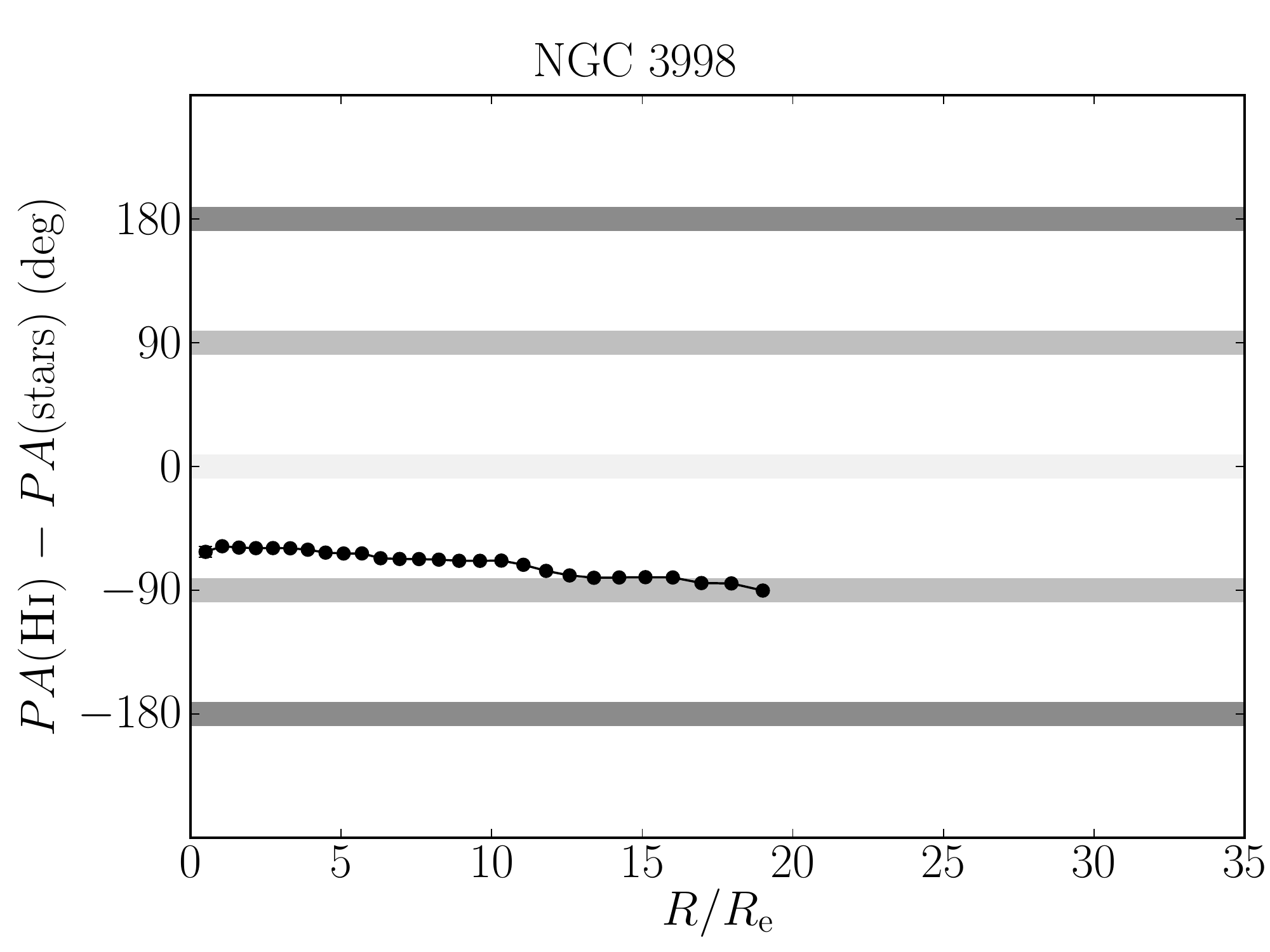}
\includegraphics[width=5.5cm]{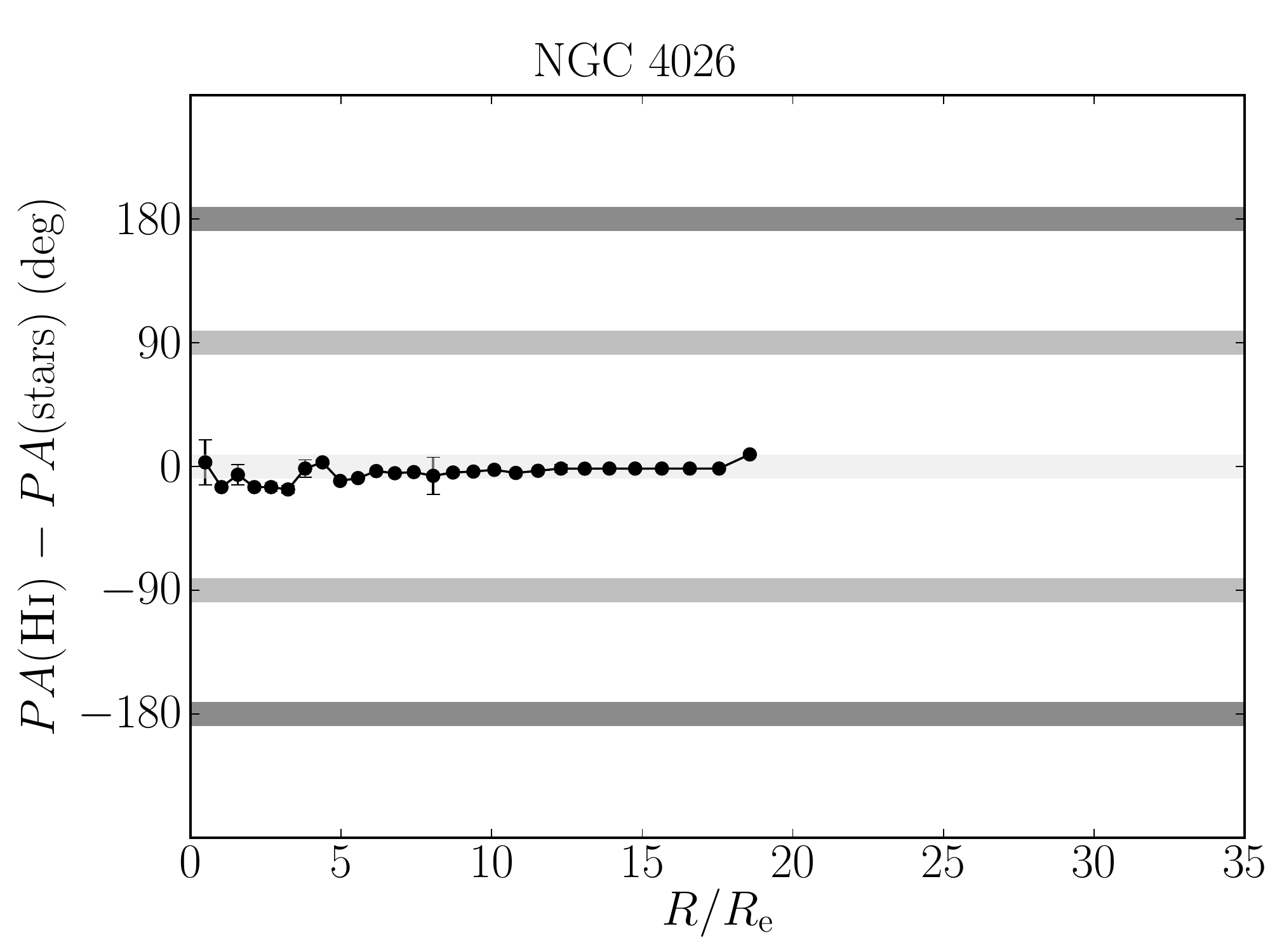}
\includegraphics[width=5.5cm]{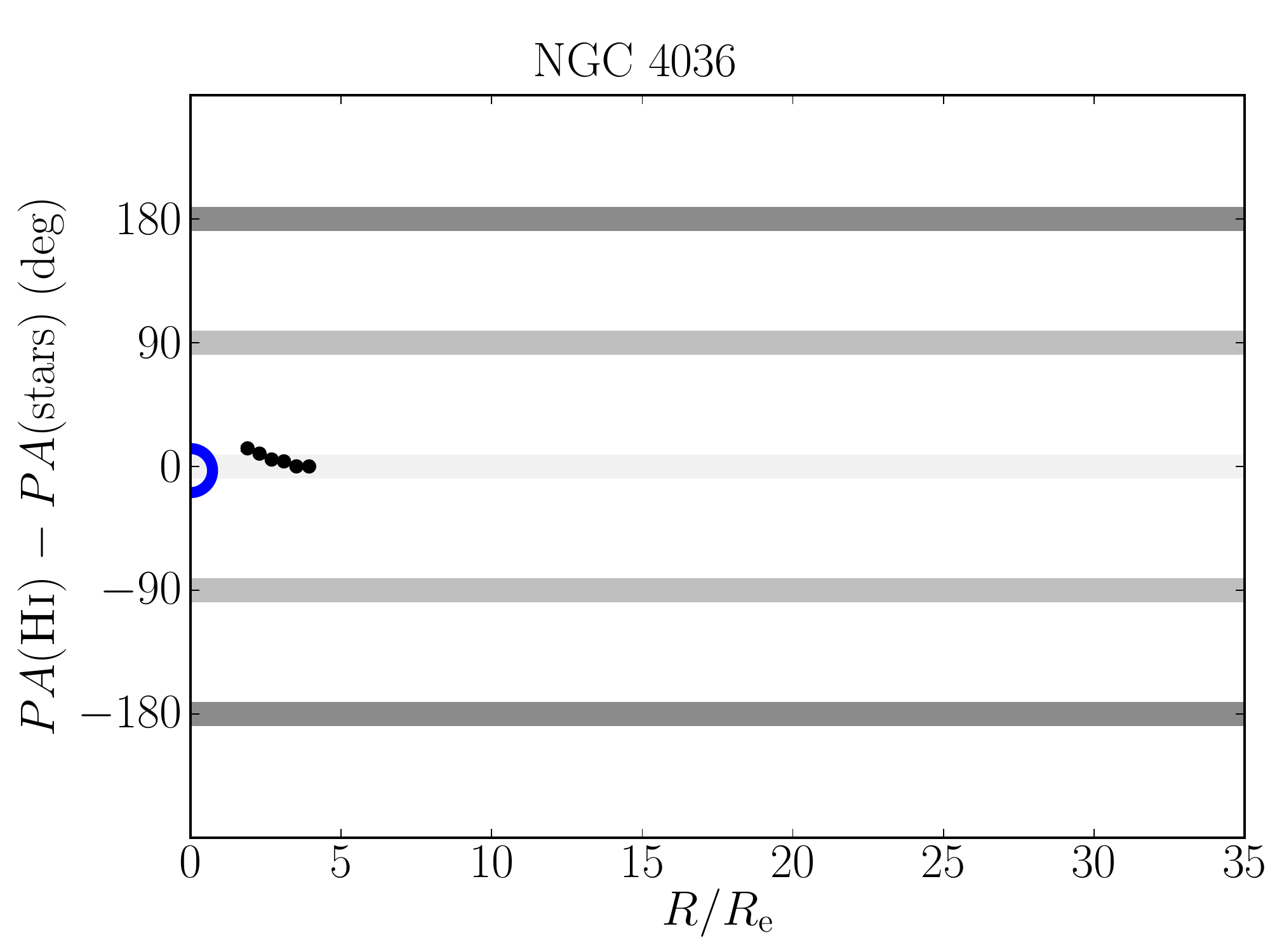}
\caption{\rm \hi\ misalignment as a function of radius in units of $R_\mathrm{e}$ \citepalias{cappellari2011a}. The black dots represent the measurements. The thick, horizontal grey lines serve as a reference to identify gas close to corotation, polar distribution and counterrotation, respectively. We highlight the few cases where PA(stars) is the photometric rather than kinematical PA. Red and blue semi-circles at $R=0$ show molecular and ionized gas misalignment, respectively.\rm }
\label{fig:allmisal}
\end{figure*}

\addtocounter{figure}{-1}
\begin{figure*}
\includegraphics[width=5.5cm]{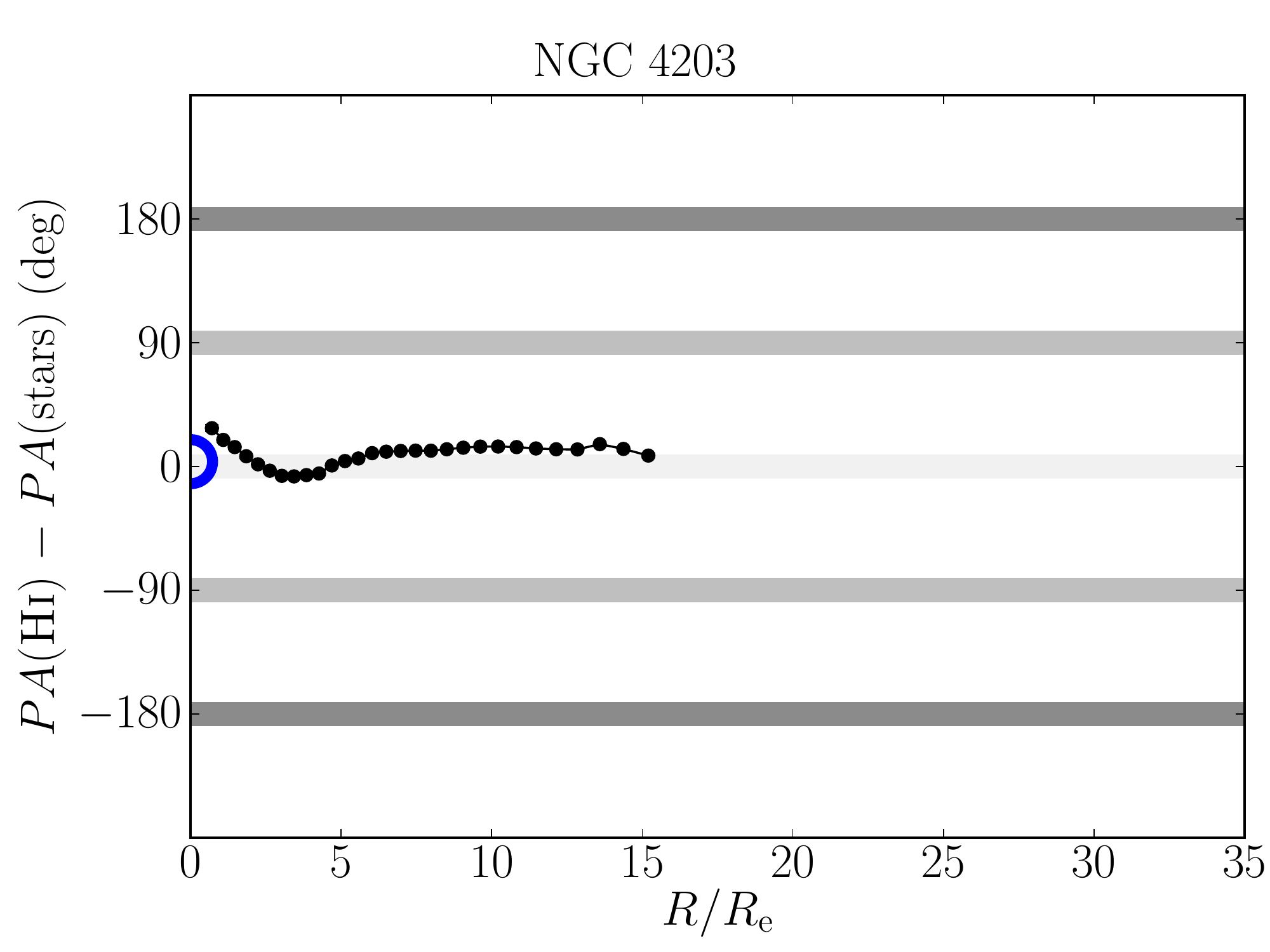}
\includegraphics[width=5.5cm]{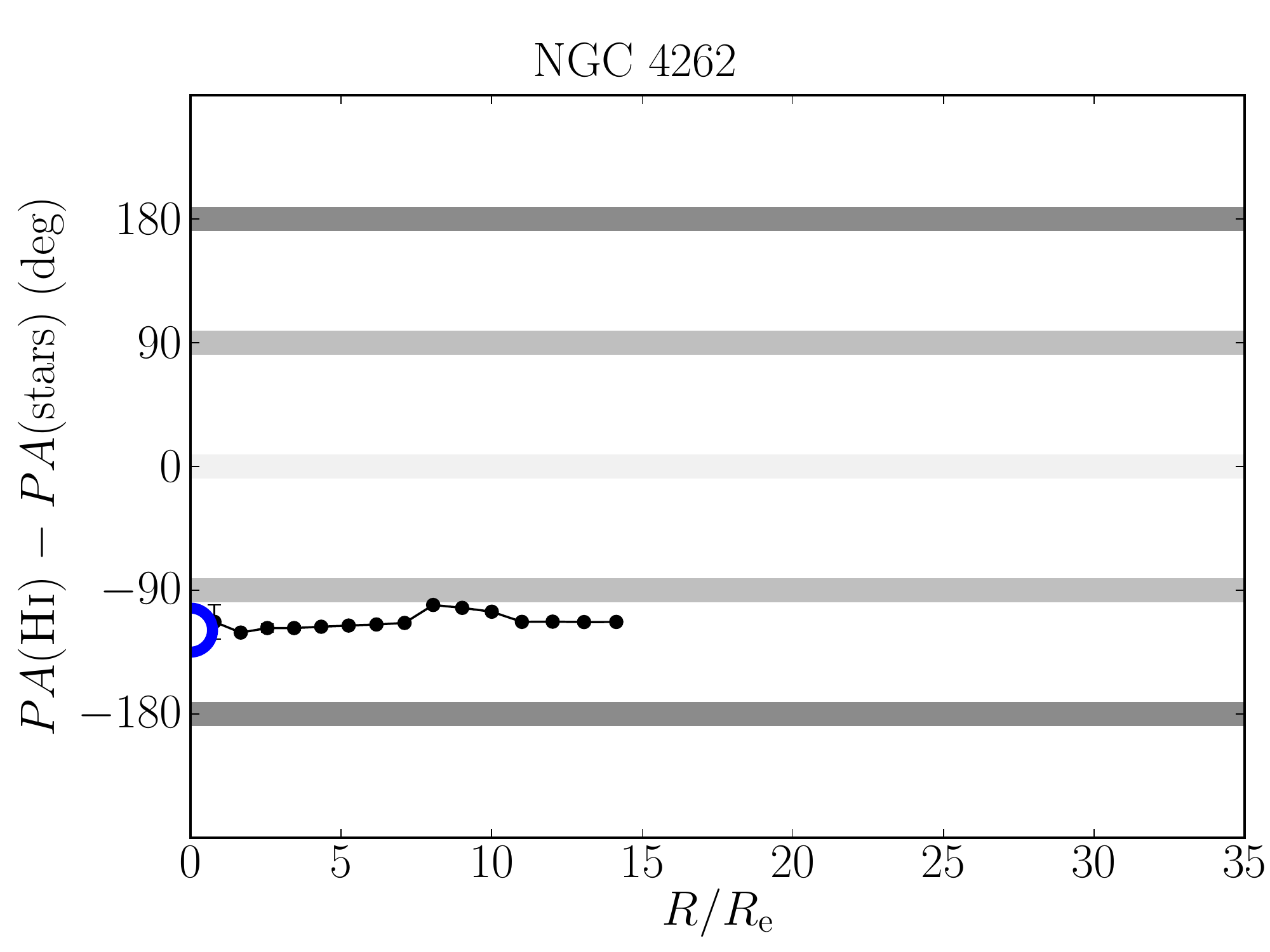}
\includegraphics[width=5.5cm]{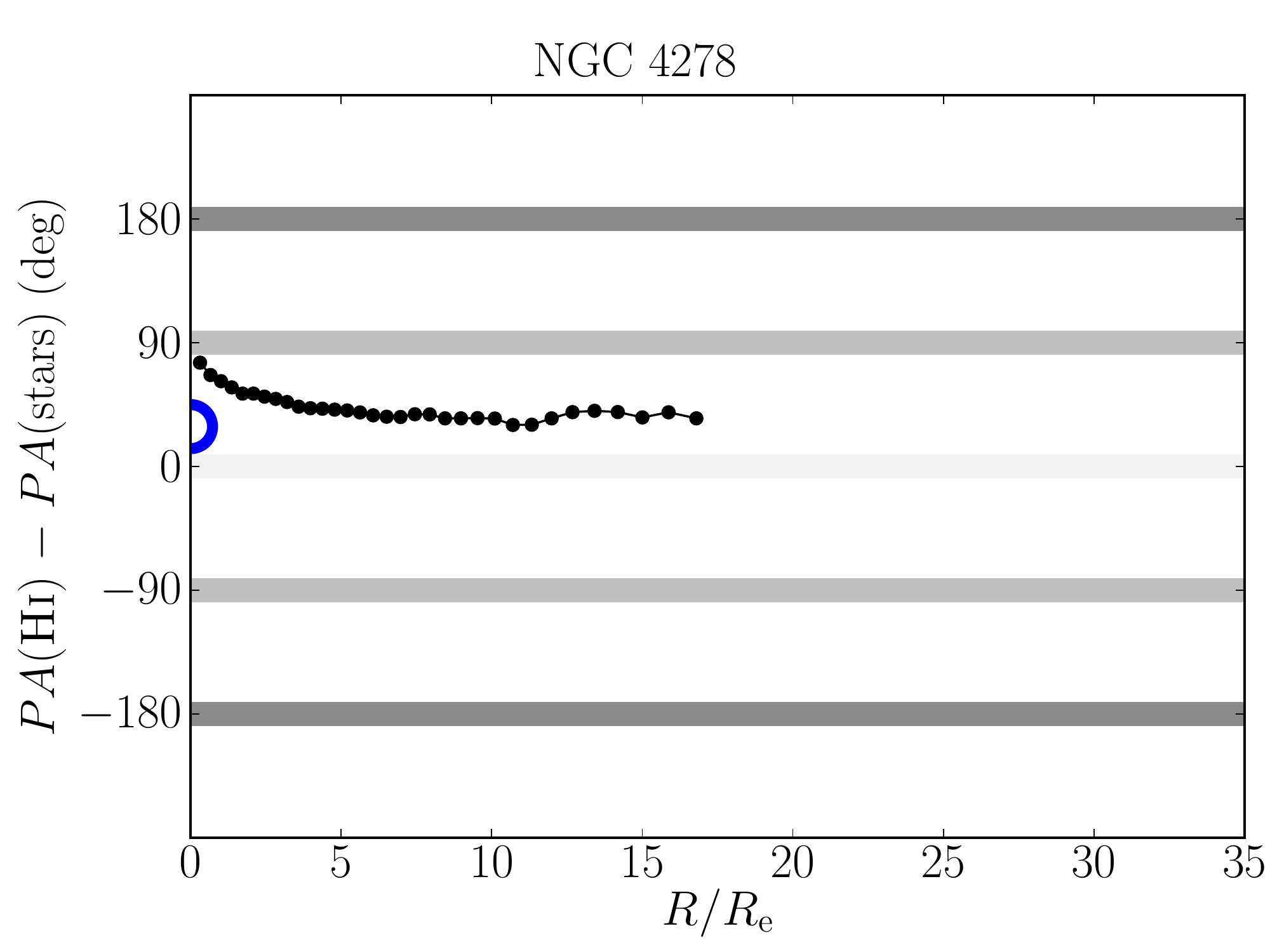}

\includegraphics[width=5.5cm]{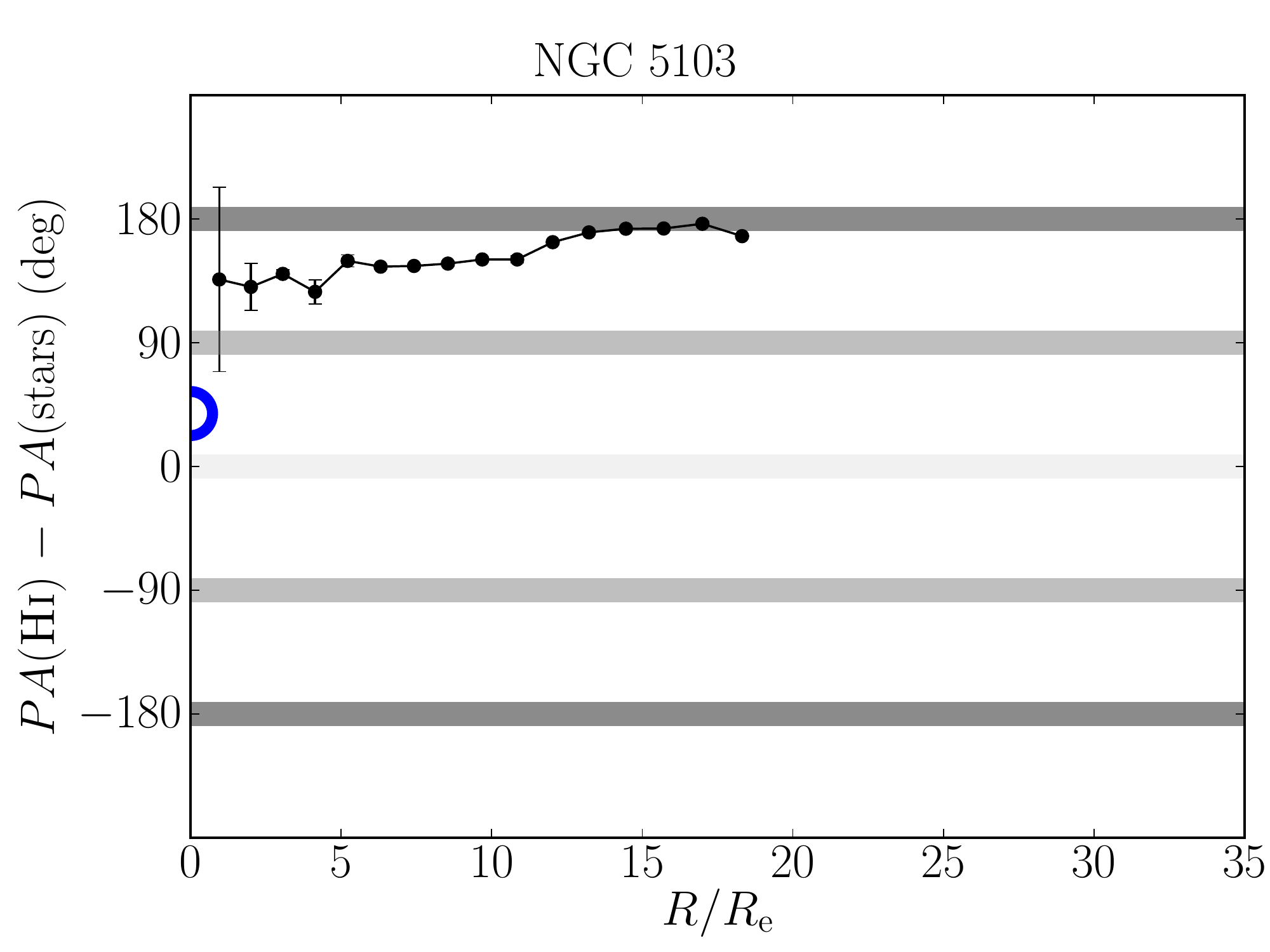}
\includegraphics[width=5.5cm]{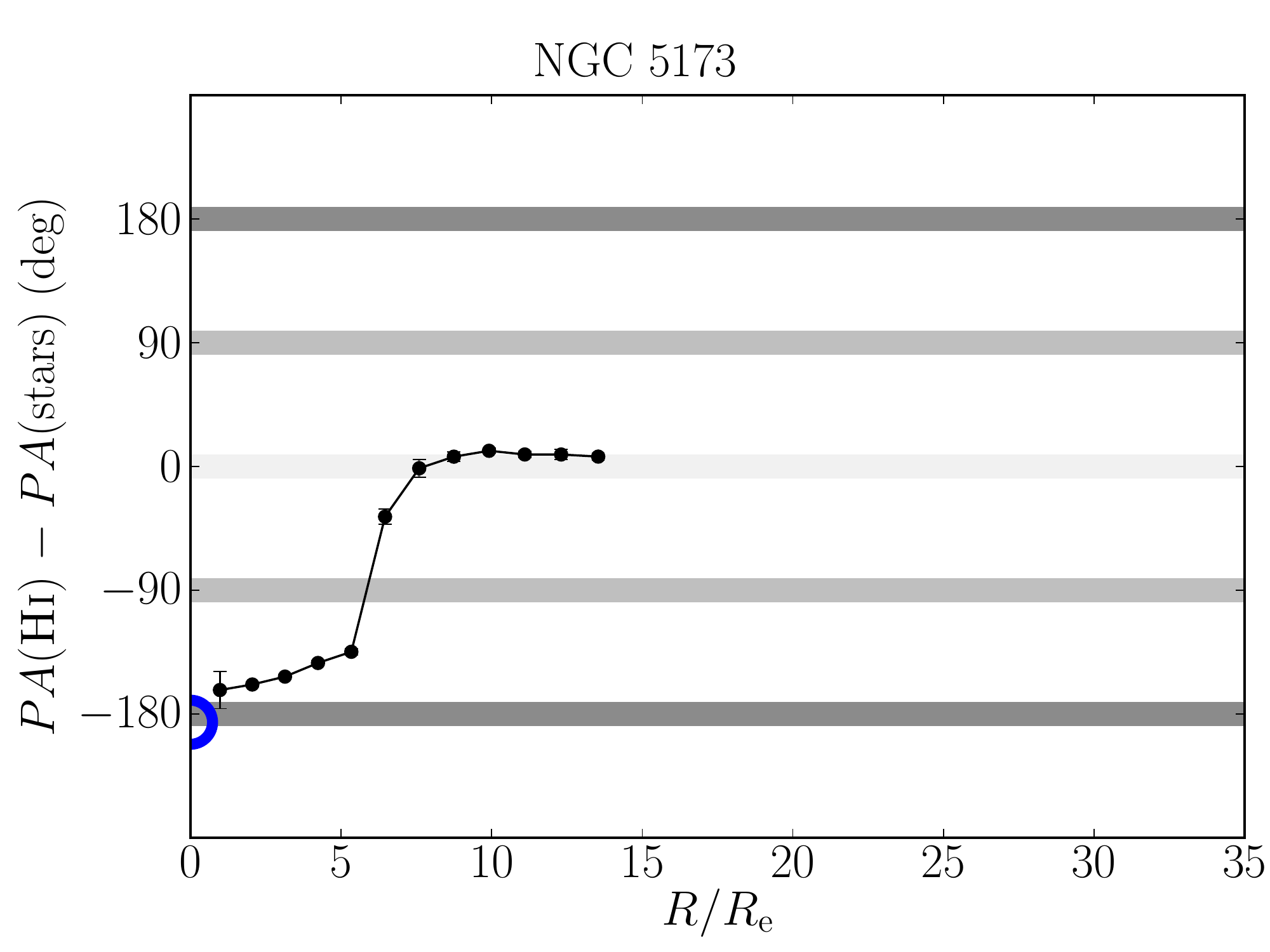}
\includegraphics[width=5.5cm]{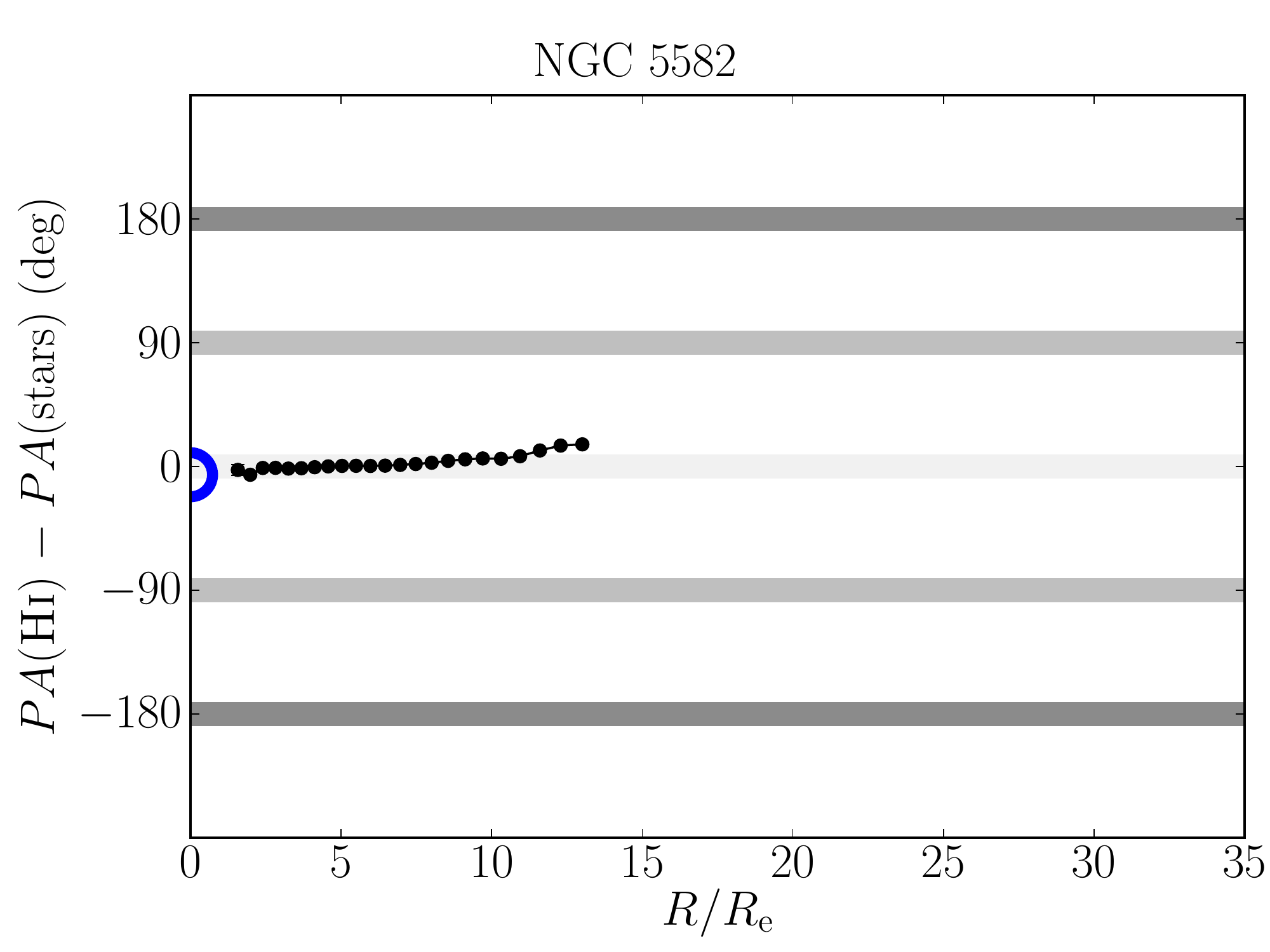}

\includegraphics[width=5.5cm]{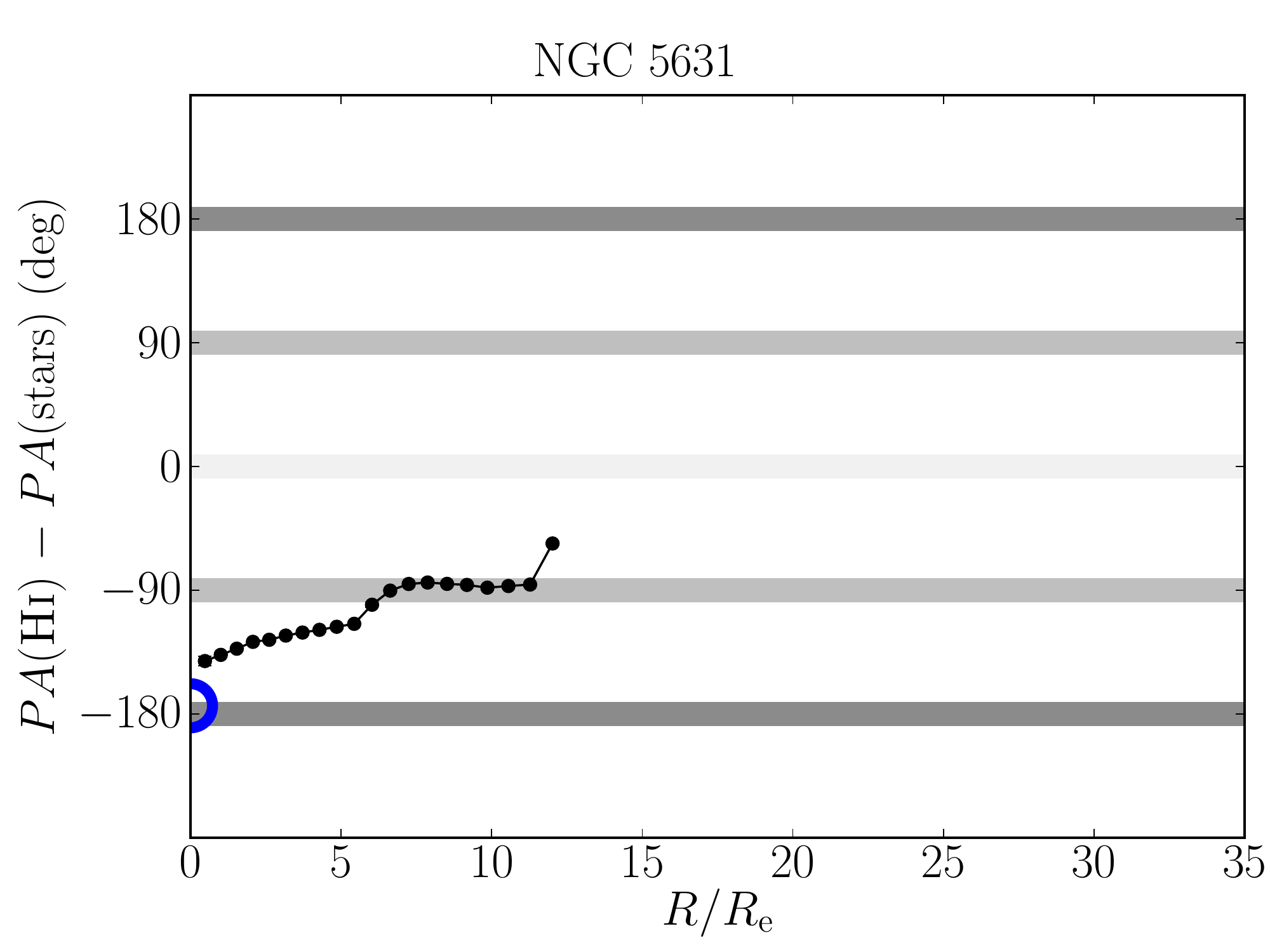}
\includegraphics[width=5.5cm]{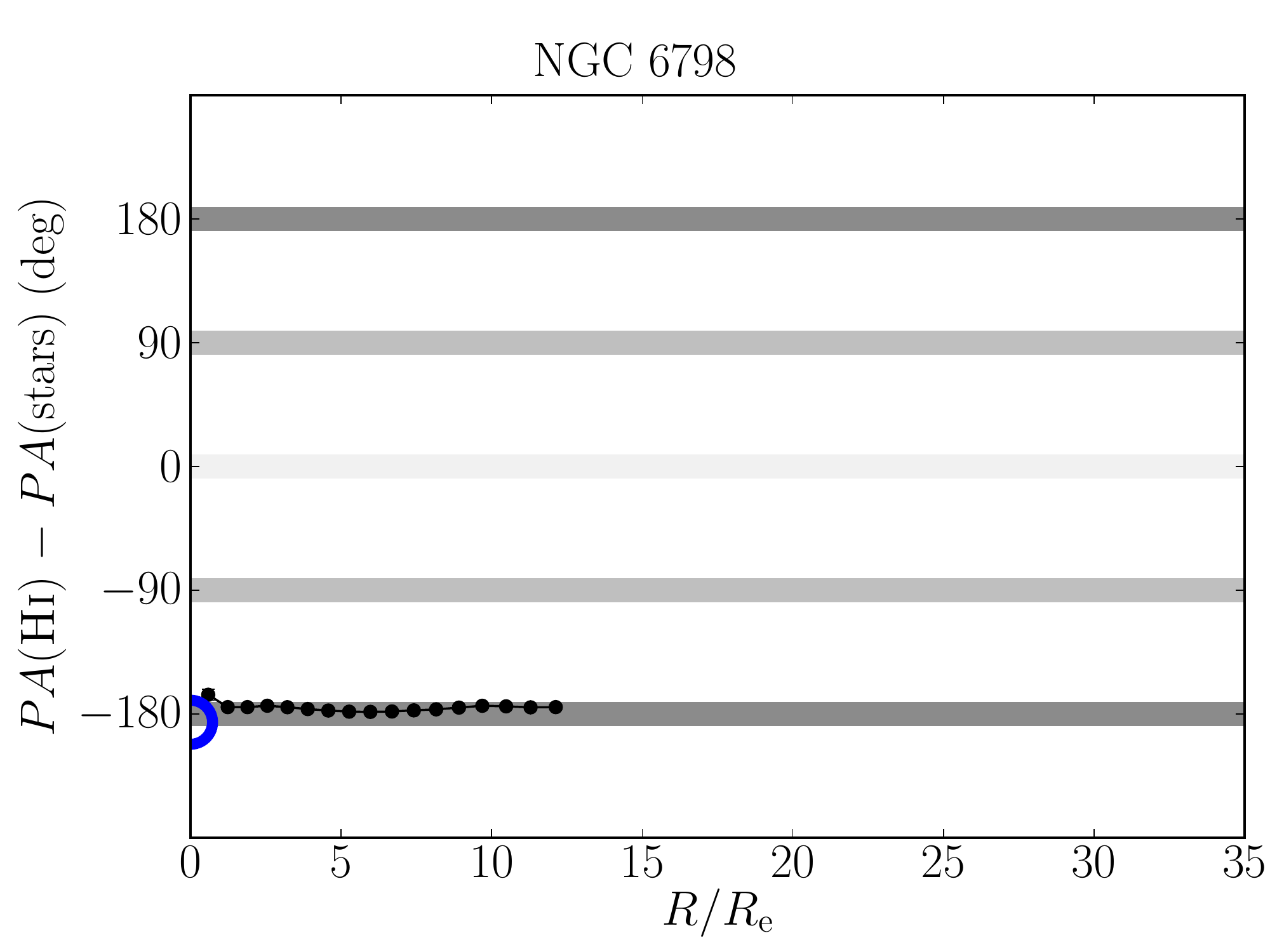}
\includegraphics[width=5.5cm]{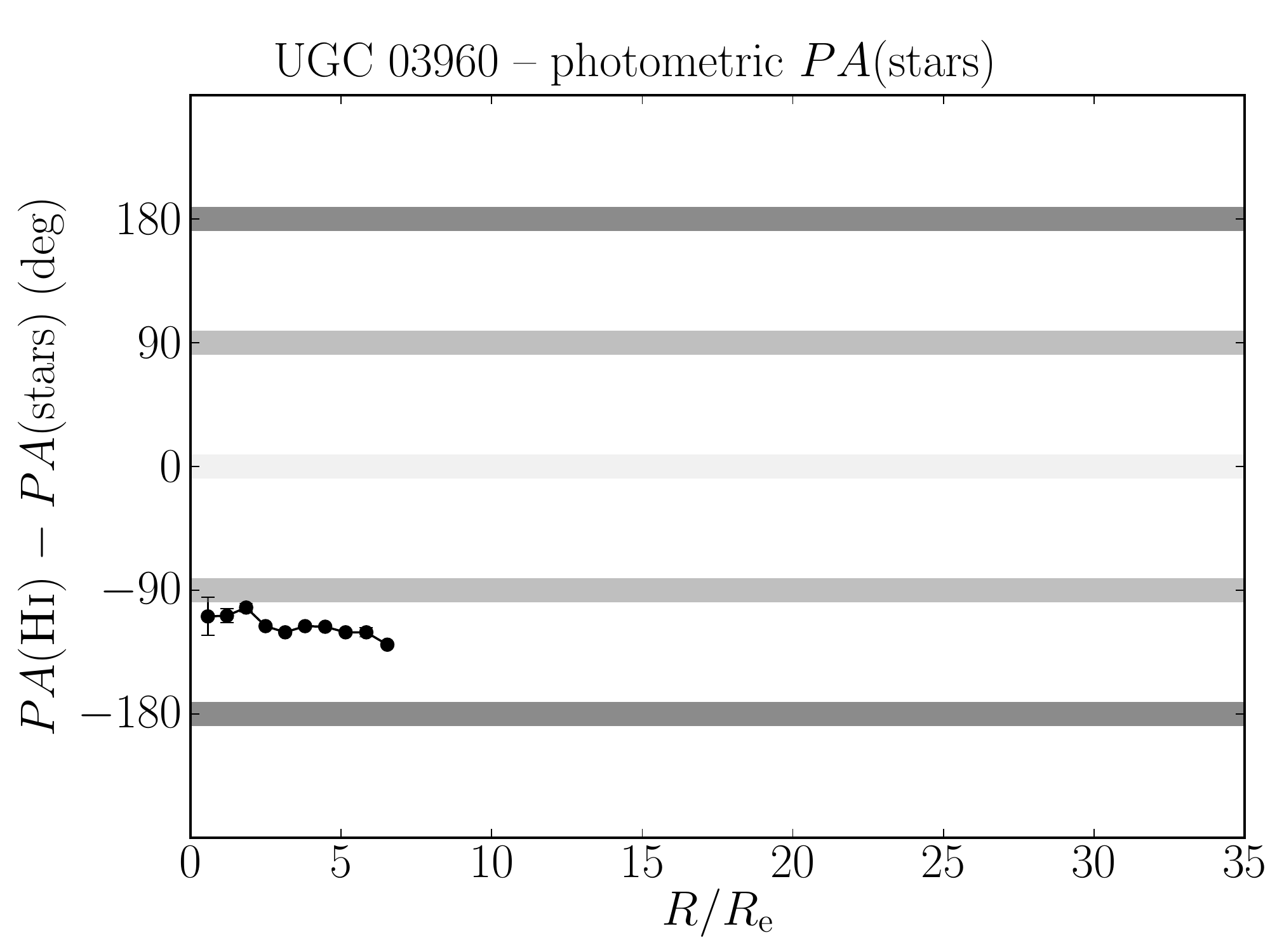}

\includegraphics[width=5.5cm]{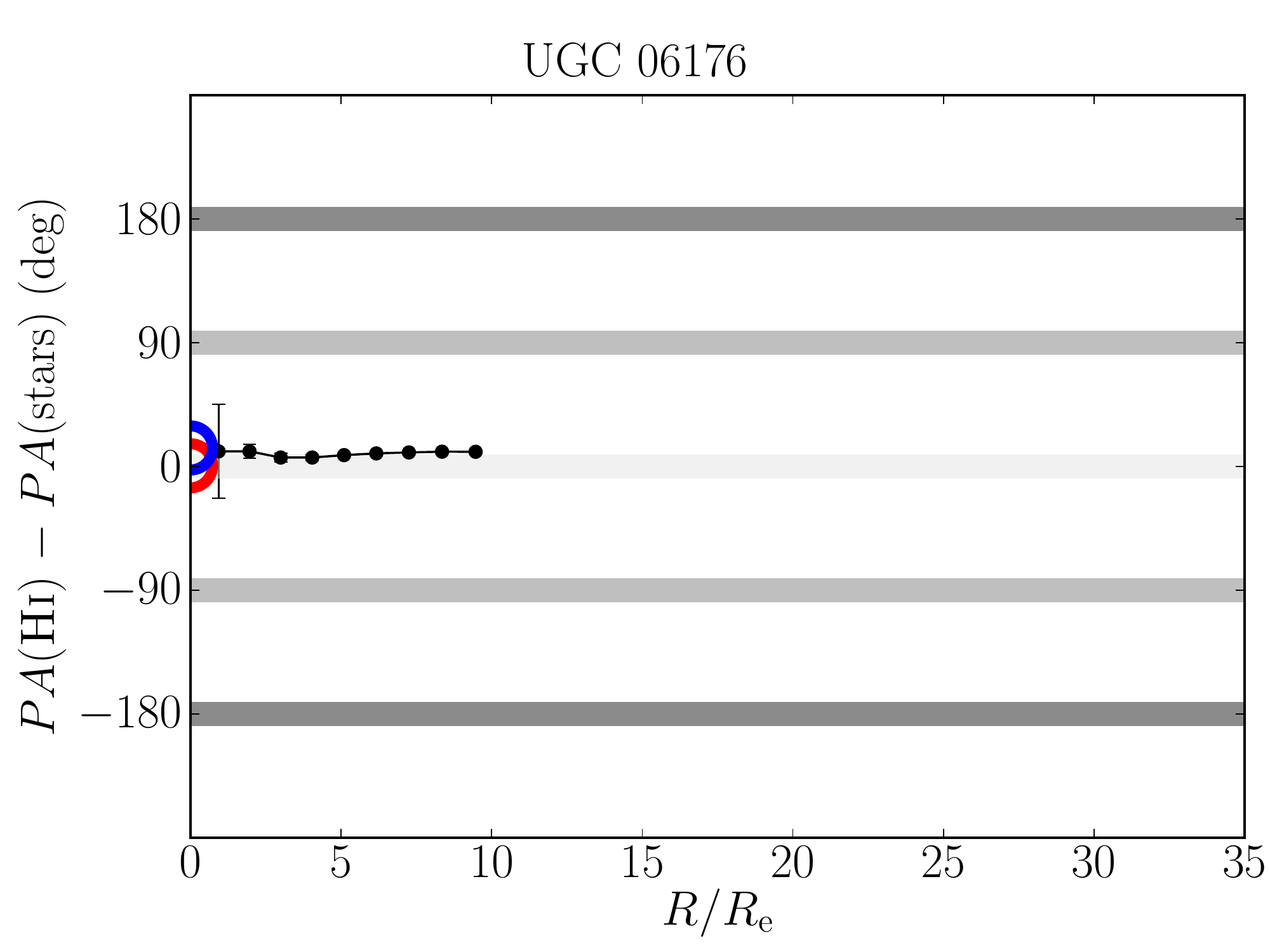}
\includegraphics[width=5.5cm]{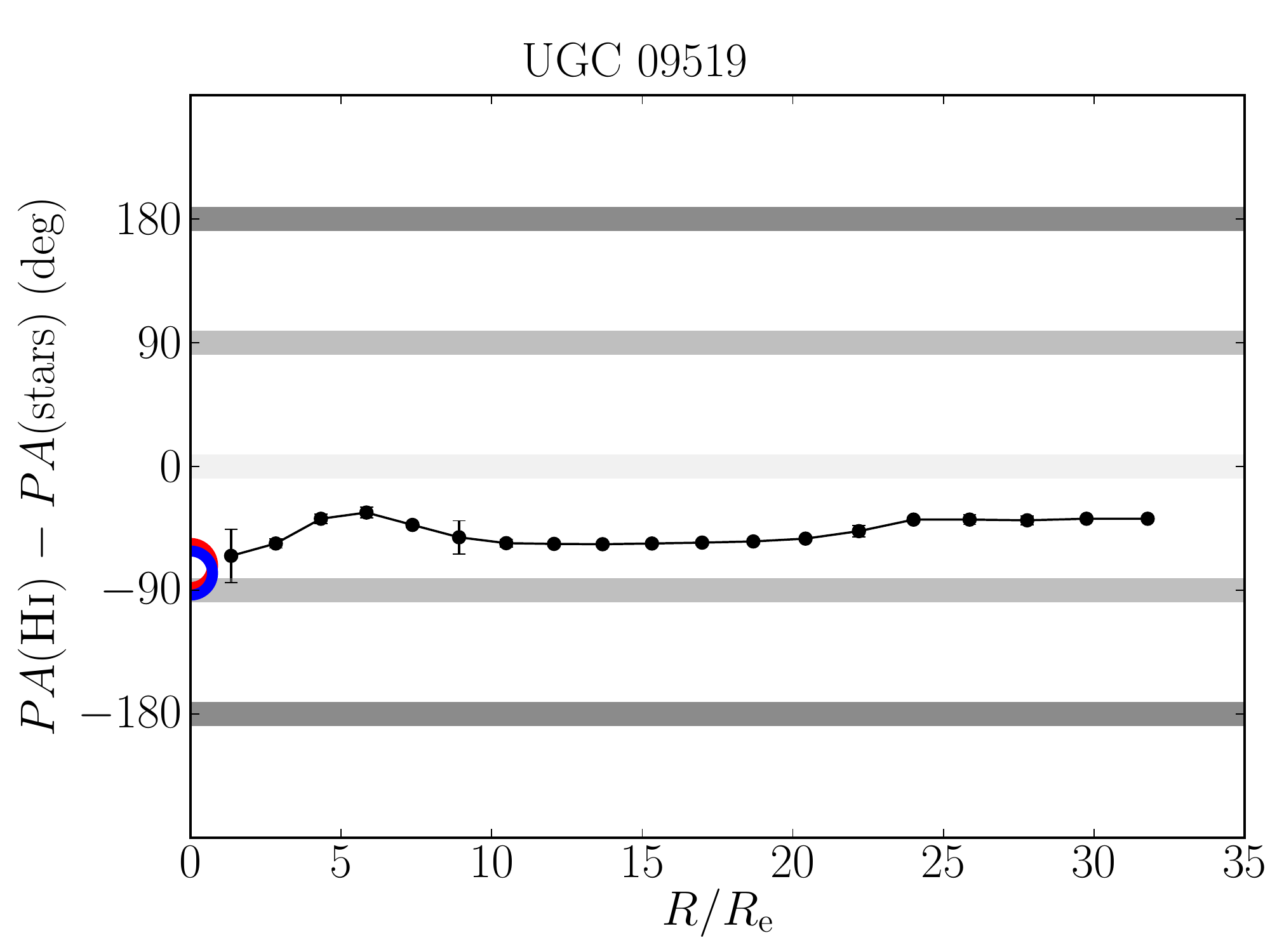}
\caption{\it Continued \rm}
\end{figure*}

\label{lastpage}

\end{document}